\documentclass[11pt,a4paper]{article}
\usepackage{jheppub,color,amsfonts,mathrsfs}
\usepackage[caption=false]{subfig}
\usepackage[titletoc,title]{appendix}

\newcommand{\ignore}[1]{}

\newcommand{\cd}{\partial}
\newcommand{\nid}{\noindent}
\newcommand{\id}{{\mathrm{Id}}}
\newcommand{\Id}{{\mathrm{Id}}}
\newcommand{\R}{\mathbb{R}}
\newcommand{\N}{\mathbb{N}}
\newcommand{\I}{\mathbb{I}}
\newcommand{\ra}{\rightarrow}
\newcommand{\ip}[1]{\left\langle #1 \right\rangle}
\newcommand{\eps}{{\boldsymbol{\varepsilon}}}
\newcommand{\nvec}{{\boldsymbol{n}}}
\newcommand{\Yvec}{{\boldsymbol{Y}}}
\newcommand{\volm}{{\rm vol}_{\mathcal{M}}}
\renewcommand{\d}{{\mathrm{d}}}
\renewcommand{\div}{{\mathrm{div}}}
\newcommand{\wt}{\widetilde}
\newcommand{\wh}{\widehat}
\newcommand{\Hess}{{\rm Hess}}
\newcommand{\pvec}{{\mathbf{p}}}
\newcommand{\zerovec}{{\mathbf{0}}}
\newcommand{\xvec}{{\mathbf{x}}}
\newcommand{\evec}{{\mathbf{e}}}
\newcommand{\Xvec}{{\mathbf{X}}}
\newcommand{\Rvec}{{\mathbf{R}}}
\newcommand{\Rrot}{{\mathscr{R}}}
\newcommand{\Orot}{{\mathscr{O}}}

\newcommand{\beq}{\begin{eqnarray}}
\newcommand{\eeq}{\end{eqnarray}}
\newcommand{\non}{\nonumber\\}

\newcommand{\p}{\partial}

\newcommand{\tr}{\qopname\relax o{tr}}
\newcommand{\diag}{\qopname\relax o{diag}}

\newcommand{\bphi}{{\boldsymbol{\phi}}}
\newcommand{\bvarphi}{{\boldsymbol{\varphi}}}

\newcommand{\bpi}{{\boldsymbol{\pi}}}

\newcommand{\grad}{\qopname\relax o{grad}}

\renewcommand{\i}{\mathrm{i}}

\newtheorem{theorem}{Theorem}

\newtheorem{prop}[theorem]{Proposition}
\newtheorem{lemma}[theorem]{Lemma}

\title{Realistic classical binding energies in the $\omega$-Skyrme model} 

\author{Sven Bjarke Gudnason$^1$,}
\affiliation{$^1$Institute of Contemporary Mathematics, School of
  Mathematics and Statistics, Henan University, Kaifeng, Henan 475004,
  P.~R.~China}
\emailAdd{gudnason(at)henu.edu.cn}
\author{James Martin Speight$^2$}
\affiliation{$^2$School of Mathematics, University of Leeds, Leeds LS2
  9JT, England}
\emailAdd{speight(at)maths.leeds.ac.uk}

\abstract{
An omega-meson extension of the Skyrme model -- without the Skyrme
term but including the pion mass -- first considered by Adkins and
Nappi is studied in detail for baryon numbers $1$ to $8$. The static
problem is reformulated as a constrained energy minimisation problem
within a natural geometric framework and studied analytically on
compact domains, and numerically on Euclidean space.  
Using a constrained second-order Newton flow algorithm, classical
energy minimisers are constructed for various values of the omega-pion
coupling. At high coupling, these Skyrmion solutions are qualitatively
similar to the Skyrmions of the standard Skyrme model with
\emph{massless} pions. At sufficiently low coupling, they show 
similarities with those in the lightly bound Skyrme model: the
Skyrmions of low baryon number dissociate into lightly bound clusters
of distinct 1-Skyrmions, and the classical binding energies for baryon 
numbers 2 through 8 have realistic values.  
}

\begin{document}
\maketitle

\section{Introduction}

Skyrmions were used to model nuclei even before the birth of
Quantum ChromoDynamics (QCD) \cite{Skyrme:1962vh}.
The symmetries of hadronic physics at low energies were understood
before QCD was an accepted theory of the strong interactions.
In fact, QCD contains an extra $U(1)$ symmetry compared to the
low-energy chiral Lagrangian, and this caused scepticism until the
so-called $U(1)$-problem was solved by 't Hooft \cite{tHooft:1976rip}.
As
a consequence of Derrick's theorem \cite{Derrick:1964ww}, the topological solitons of the Skyrme model \cite{Skyrme:1962vh}
need something more than the kinetic term to be stabilised.
There is, however, little -- if any -- phenomenological support for
adding the Skyrme term\footnote{The Skyrme term can be viewed as a
  specific combination of 2 higher-order terms in the chiral
  Lagrangian for which the 4 time derivatives exactly cancel. The two
  terms naturally appear in such an expansion, but there is no
  phenomenological reason for the cancellation. Nevertheless, it
  simplifies the quantisation of the zero modes in the model.}.
Starting from just the symmetries of the low-energy hadronic physics,
it is possible to include just one more particle into the theory to
stabilise the topological solitons, namely the omega vector meson.
This was understood already in the seminal paper by Adkins and Nappi
\cite{Adkins:1983nw}.
In a full-fledged hadronic physics model, several vector mesons would
have to be incorporated. However, if the scope is simply the
low-energy effective nuclear spectrum, perhaps a few -- or just one --
vector meson could be sufficient.
The alternative option of including the rho meson instead of the omega
meson was considered in a series of papers
\cite{Meissner:1986vu,Bando:1987br,Harada:2003jx} and recently also by
Sutcliffe and Naya
\cite{Sutcliffe:2010et,Sutcliffe:2011ig,Naya:2018mpt,Naya:2018kyi}.\footnote{There
  exists an alternative approach to Skyrmions which is relevant for
  nuclei at finite or high density. In such approach only a single
  Skyrmion is calculated, but with periodic boundary conditions. The
  size of the box is then related to the density of nucleons. In this
  setup, the $\omega$ meson has been considered (together with the
  $\rho$ meson) in the literature to quite some extent
  \cite{Ma:2012kb,Ma:2012zm,Ma:2013ooa,Ma:2013ela,Ma:2016nki}. } 

In the past 36 years, the omega vector meson extension of the chiral
Lagrangian as a model for nuclei has not received much attention. Sutcliffe considered the model
\cite{Sutcliffe:2008sk}, but only constructed solutions of degree 1 to 4 within the
rational map approximation, which approximates the field equations by ODEs.
Recently, Speight considered the model with the addition of an
(explicit) isospin symmetry breaking term in the form of a derivative
coupling of the omega meson field and the pion field
\cite{Speight:2018zgc}, but considered only the degree 1 sector where, again, only ODEs need be solved. 

There is a good reason for this relative paucity of results: the static field equations in this model are {\em not} the Euler-Lagrange equations for the theory's static energy functional, so standard energy minimisation algorithms (based on gradient descent or simulated annealing) do not solve the static problem. The underlying cause for this difficulty is that the vector
field representing the omega meson enters the Lagrangian with the ``wrong sign.'' 
We overcome this obstacle by observing that static solutions solve a {\em constrained} energy minimisation problem in which $\omega^0$ (the temporal component of the omega field) is uniquely determined by the Skyrme field. We solve this constrained energy minimisation problem by arrested Newton flow for the Skyrme field, updating $\omega^0$ after each time step by solving the constraint equation. This equation is a linear inhomogeneous PDE which can be efficiently solved via a standard conjugate gradient method. 
The resulting algorithm, being based on a second order flow, is much faster than comparable heat-flow methods \cite{Foster:2009rw}, allowing, for the first time, extensive simulation of a wide selection of topological sectors for a range of coupling values.

We find that this omega extended Skyrme model, although very simple --
with only two parameters to dial -- has regions in parameter space with
extremely low binding energies.
This addresses one of the usual problems with Skyrme-type models -- that they
are too strongly bound.
In this model, we have a line of vanishing classical binding energy
and beyond that even ``negatively'' bound solutions (that is, they are
metastable\footnote{By metastable we mean a solution which is only a
  local minimum of the energy functional. The metastability implies a
  quantum mechanical thinking, that by quantum fluctuations, the
  solution may tunnel over to the global minimum in a finite time,
  which is exponentially prolonged by the barrier between the two
  minima. }).
A vanishing classical binding energy means that the multisoliton
 -- although metastable -- can be broken up and will possess
the same energy with all the constituent $B=1$ Skyrmions indefinitely
separated. 
The weakly bound multi-Skyrmions in turn provide a larger number of
metastable solutions (local minimisers of the energy functional).

The paper is organised as follows.
Section \ref{sec:model} sets up the model and the notation of the
paper.
The second variation of the energy functional for the model is derived,
and its implications for stability of the
model on compact domains discussed, in sec.~\ref{sec:secondvar}. 
The numerical method is introduced in sec.~\ref{sec:nummet}. Classical solutions at the coupling proposed by Sutcliffe are found and compared
with the approximate solutions he found within a rational map approximation \cite{Sutcliffe:2008sk}.
A semi-classical quantisation scheme is proposed, and 
applied to the 1-Skyrmion, in
sec.~\ref{sec:collcoordquant}.
Then an attempt to find the optimal calibration of the model is made in
sec.~\ref{sec:calibration}. This optimal calibration has radically lower coupling than that proposed by Adkins and Nappi \cite{Adkins:1983nw} or Sutcliffe \cite{Sutcliffe:2008sk} and the classical solutions display new qualitative behaviour.
These solutions are illustrated and discussed in
sec.~\ref{sec:numsols}.
Inter-Skyrmion forces are studied in sec.~\ref{sec:scatpot} and an asymptotic formula for the interaction energy between well-separated Skyrmions derived using a point source formalism.
Finally the paper is concluded with a discussion in
sec.~\ref{sec:conc}.

Since the paper is somewhat lengthy and contains many topics, we will
suggest shortened routes through it for two contrasting types of
reader. 
The reader primarily interested in the application of the Skyrme model
to nuclear physics could start at subsection \ref{sec:explicit_stuff}
then, omitting section \ref{sec:secondvar} and its associated appendix
entirely, skip directly to section \ref{sec:nummet} and proceed
through to section \ref{sec:conc}. By contrast, the reader primarily
interested in the differential geometry of generalized sigma models
could read sections 
\ref{sec:model} (skipping \ref{sec:explicit_stuff}),
\ref{sec:secondvar}, the associated appendix \ref{app:secondvar} and
\ref{sec:nummet}, take a look at figures \ref{fig:g347M0176} and
\ref{fig:g1434M0176}, then skip to section \ref{sec:conc}.

\section{The model}\label{sec:model}

We will find it convenient to give a coordinate free, geometric
formulation of the field theory. This is both economical and flexible,
providing field equations which work in arbitrary dimension, on any
background geometry, for any target space. It also allows us to
emphasise certain conceptual points which are important for our
numerical method.  
The reader wishing to see a formulation of the model and its static
field equations in the case of most direct interest, expressed in
explicit coordinates, can skip to section \ref{sec:explicit_stuff}.

Let $(\mathcal{M},\eta)$ be a Lorentzian $d+1$ manifold with pseudo-metric $\eta$, representing spacetime, $(N,h)$ be a compact Riemannian manifold (metric $h$) equipped with a closed $d$-form $\Omega$, and $V$ be a smooth function on $N$. The fields consist of a smooth map
$\bvarphi:\mathcal{M}\ra N$ (the Skyrme field) and a $1$-form $\omega$ on $\mathcal{M}$ (the omega meson). The action of the model is
\begin{equation}\label{action}
  S(\bvarphi,\omega)=\frac{1}{8}\ip{\d\bvarphi,\d\bvarphi}_{L^2}
  -\int_{\mathcal{M}}V\circ\bvarphi\,\volm
  -\frac12\ip{\d\omega,\d\omega}_{L^2}
  +\frac12\ip{\omega,\omega}_{L^2}
  +g\int_{\mathcal{M}}\omega\wedge\bvarphi^*\Omega,
\end{equation}
where $g$ is (without loss of generality) a positive coupling constant, $\volm$ denotes the volume form on $(\mathcal{M},\eta)$, and $\ip{\cdot,\cdot}_{L^2}$ denotes the $L^2$ pseudo-inner-product on $\mathcal{M}$ defined by its Lorentzian metric (and the metric on $N$ for the first term). The case of direct interest has $\mathcal{M}=\R^{1,3}$ (Minkowski space), $N=S^3$, the unit sphere in $\R^4$ and $\Omega$ the normalised volume form on $S^3$ (normalised so that $\int_{S^3}\Omega=1$). Note, in particular, that the {\em baryon current} in this formulation is the vector field on $\mathcal{M}$ metrically dual to the $1$-form $B=\star\bvarphi^*\Omega$, where $\star$ denotes the Hodge isomorphism on $(\mathcal{M},\eta)$, and that this vector field is divergenceless by closure of $\Omega$. We may identify $\bvarphi=(\varphi_0,\varphi_1,\varphi_2,\varphi_3)$ whose components are
traditionally named $\sigma=\varphi_0$ and $\pi_i=\varphi_i$, $i=1,2,3$ (the pions). A standard choice of potential is 
\begin{equation}\label{Vdef}
V(\bvarphi)=\frac{m^2}{4}(1-\varphi_0),
\end{equation}
which gives the pions mass $m$ (in units of the omega mass). With these choices, the action \eqref{action} coincides with that introduced by Adkins and Nappi, in the normalisation used by Sutcliffe \cite{Sutcliffe:2008sk}. 

Returning to the general case, the field equations are obtained by demanding that $(\bvarphi,\omega)$ is a formal critical point of $S$: for all
smooth variations $(\bvarphi_s,\omega_s)$ of $(\bvarphi,\omega)=(\bvarphi_s,\omega_s)|_{s=0}$ of compact support in $\mathcal{M}$,
\begin{equation}
\frac{\d{\:}}{\d{s}}S(\bvarphi_s,\omega_s)\bigg|_{s=0}=0.
\end{equation}
To proceed further, it is convenient to choose an isometric embedding $N\subset\R^k$ (such an embedding certainly exists; for $N=S^3$ we may
choose the canonical embedding in $\R^4$) and to associate to any smooth map $\bvarphi:\mathcal{M}\ra N$ the $(d-1)$-form $\Xi_\bvarphi$ on $\mathcal{M}$ valued in
$\bvarphi^{-1}TN$ defined so that
\begin{equation}
h(Y,\Xi_\bvarphi(X_1,X_2,\ldots,X_{d-1}))=\Omega(Y,\d\bvarphi(X_1),\d\bvarphi(X_2),\ldots,\d\bvarphi(X_{d-1})), 
\end{equation}
for all $X_1,\ldots,X_{d-1}\in T_p\mathcal{M}$ and $Y\in T_{\bvarphi(p)}N$.
Recall that $\bvarphi^{-1}TN$ is the vector bundle over $\mathcal{M}$
whose fibre over $p\in\mathcal{M}$ is the vector space
$T_{\bvarphi(p)}N$. 
This bundle will be of some significance in the following. A
comprehensive description of it, and the geometric structures it
canonically possesses, may be found in ref.~\cite{Urakawa:1993}.  

Given a smooth variation $(\bvarphi_s,\omega_s)$, we define
$\eps=\cd_s\bvarphi_s|_{s=0}$ and $\alpha=\cd_s\omega_s|_{s=0}$. Note
that $\eps$ is a section of $\bvarphi^{-1}TN$ while $\alpha$ is a
$1$-form on $\mathcal{M}$ and that both, by assumption, have support
in some compact set $K\subset\mathcal{M}$. It follows immediately from
eq.~\eqref{action} and the Homotopy Lemma (see, for example
ref.~\cite{Speight:2006dn}) that 
\beq
\frac{\d{\:}}{\d{s}}S(\bvarphi_s,\omega_s)\bigg|_{s=0}&=&\frac14\ip{\d\bvarphi,\d\eps}_{L^2}-\ip{(\grad V)\circ\bvarphi,\eps}_{L^2}
-\ip{\d\omega,\d\alpha}_{L^2}+\ip{\omega,\alpha}\nonumber\\
&&\quad+g\int_{K}\left(\omega\wedge \d(\bvarphi^*\iota_{\eps}\Omega)+\alpha\wedge\bvarphi^*\Omega\right)\nonumber\\
&=&\ip{\eps,\frac{(-1)^{d+1}}{4}\star\d\star\d\bvarphi-(\grad V)\circ\bvarphi}_{L^2}
-g\int_{\partial K}\omega\wedge \bvarphi^*\iota_\eps\Omega\nonumber \\ &&\quad+\int_K\d\omega\wedge\bvarphi^*\iota_\eps\Omega
+\ip{\alpha,-\star\d\star\d\omega+\omega+g\star\bvarphi^*\Omega}_{L^2}\nonumber \\
&=&\ip{\eps,\frac{(-1)^{d+1}}{4}\star\d\star\d\bvarphi-(\grad V)\circ\bvarphi+g(-1)^d\star(\d\omega\wedge\Xi_\bvarphi)}_{L^2}\nonumber \\
&&\quad +\ip{\alpha,-\star\d\star\d\omega+\omega+g\star\bvarphi^*\Omega}_{L^2},
\eeq
where we have used Stokes's Theorem and the facts that, on a
Lorentzian $(d+1)$-manifold, the coderivative  
$\Omega^p(\mathcal{M})\ra\Omega^{p-1}(\mathcal{M})$ adjoint to ${\d}$
is $(-1)^{p(d+1)}\star\d\star$, and $\star\star=(-1)^{d(p+1)}$ \cite{Willmore:1987}.  
This should vanish for all $\eps\in\Gamma(\bvarphi^{-1}TN)$ and all
$\alpha\in\Omega^1(\mathcal{M})$. Hence
\beq
\frac{(-1)^{d+1}}{4}P_{\bvarphi}(\star\d\star\d\bvarphi)-(\grad
V)\circ\bvarphi+(-1)^d g\star(\d\omega\wedge\Xi_\bvarphi)&=&0,\label{eq:eomphi}\\
-\star\d\star\d\omega+\omega+g\star\bvarphi^*\Omega&=&0,\label{eq:eomomega}
\eeq
where $P_{\bvarphi}:\R^k\ra T_{\bvarphi}N$ denotes\footnote{The term
$P_\bvarphi(\star\d\star\d\bvarphi)$ is, up to sign, the {\em tension
    field} of the map $\bvarphi$. It can be defined without reference
  to an embedding 
$N\subset\R^k$ using the natural connexion on the bundle
  $\bvarphi^{-1}TN$ \cite{Urakawa:1993}. The extrinsic formulation is more
  convenient for our purposes.} the orthogonal projection defined by
the isometric embedding $N\subset\R^k$. 
These are the field equations for the action $S$. Note that each term
on the left hand side of eq.~\eqref{eq:eomphi}, and hence the
left-hand side itself, is a section of $\bvarphi^{-1}TN$.  

So far, $\mathcal{M}$ was an arbitrary Lorentzian
manifold. Henceforth, we assume that $\mathcal{M}=\R\times X$ with a
product metric 
$\eta=\d{t}^2-\zeta$, where $(X,\zeta)$ is a Riemannian $d$-manifold. We shall
denote the Hodge isomorphism on $X$ by $*$, to distinguish it from
the  
isomorphism on $\mathcal{M}$. Now $\omega=\omega_0\d{t}+\omega_{X}$ where
$\omega_0$ and $\omega_{X}$ are curves (parametrised by $t$) in
$\Omega^0(X)$ and $\Omega^1(X)$ respectively. We shall denote by
$\d\omega_0$ and $\d\omega_X$ the curves in $\Omega^1(X)$ and
$\Omega^2(X)$ obtained by applying  
$\d\Omega^p(X)\ra\Omega^{p+1}(X)$ at each fixed $t$, and
$\dot\omega_0=\cd_t\omega_0\in\Omega^0(X)$,
$\dot\omega_X=\cd_t\omega_X\in\Omega^1(X)$. 
Similar conventions apply to $\d\bvarphi$ and $\dot\bvarphi$, having
interpreted $\bvarphi$ as a curve in $C^\infty(X,N)$.  
In this case, the theory enjoys time translation symmetry and hence,
by Noether's Theorem, has a conserved energy functional 
\begin{align}
  E&=\int_X*\bigg(
  \frac18|\dot\bvarphi|^2
  +\frac12|\dot\omega_X|^2
  +\frac18|\d\bvarphi|^2
  +V(\bvarphi)
  -\frac12|\d\omega_0|^2
  -\frac12\omega_0^2
  +\frac12|\d\omega_X|^2
  +\frac12|\omega_X|^2 \non
  &\phantom{=\int_X*\bigg(\ }
  -g\omega_0B_0\bigg),
\end{align}
where $B_0=*\bvarphi^*\Omega\in\Omega^0(X)$. Note that the quantity
\begin{equation}
B=\int_X B_0*1=\int_X\bvarphi^*\Omega,
\end{equation}
is a homotopy invariant of the map $\bvarphi(t,.):X\ra N$ since
$\Omega$ is closed, and hence is independent of $t$. For suitable $X$
and $N$ it may be interpreted as the baryon number of the field
$\bvarphi$.  

Our aim is to find {\em static} solutions of the field equations, so
let us assume that all fields are independent of $t$. Then
$\bvarphi=\bphi\circ\bpi$, where $\bphi:X\ra N$ is a fixed map and
$\bpi:\R\times X\ra X$ is projection.
Furthermore,
$\star\bvarphi^*\Omega=B_0\, \d{t}=(*\bphi^*\Omega)\d{t}$. Hence,
eqs.~\eqref{eq:eomphi}, \eqref{eq:eomomega} are satisfied by
$\omega=f\,\d{t}$ and $\bvarphi=\bphi\circ\bpi$, where $f:X\ra\R$, provided 
\beq
\frac14P_\bphi(\triangle\bphi)+(\grad V)\circ\bphi+g*(\d f\wedge\Xi_\bphi)&=&0,\label{static1}\\
(\triangle+1) f&=&-g*\bphi^*\Omega,\label{static2}
\eeq
where $\triangle$ is the usual\footnote{We use the geometer's sign
  convention, so $\triangle=-\cd_1^2-\cd_2^2-\cd_3^2$ on $\R^3$.}
Laplacian on $(X,\zeta)$. 
This is the coupled pair of PDEs we seek to solve.
Note that any solution of them has, by virtue of eq.~\eqref{static2}
(and, if $X$ is noncompact, a suitable decaying boundary condition on
$\omega_0=f$), 
\begin{equation}
-g\int_X\omega_0B_0*1=\ip{f,(\triangle+1)f}_{L^2(X)}=\|\d f\|_{L^2(X)}^2+\|f\|_{L^2(X)}^2,
\end{equation}
and hence energy
\begin{equation}\label{Edef}
E(\bphi,f)=\int_X*\left(\frac18|\d\bphi|^2+V\circ\bphi+\frac12|\d f|^2+\frac12 f^2\right).
\end{equation}

We claim that eq.~\eqref{static1} is precisely the Euler-Lagrange equation
for the energy functional $E(\bphi,f)$ subject to the constraint
\eqref{static2}. To verify this, let $(\bphi_s,f_s)$ be a smooth
variation of $(\bphi,f)$ satisfying \eqref{static2} for all $s$. Once
again let 
$\eps=\cd_s\bphi_s|_{s=0}\in\Gamma(\bphi^{-1}TN)$ and
$\alpha=\cd_sf_s|_{s=0}\in C^\infty(X)$. Then, differentiating
eq.~\eqref{static2} with respect to the variation parameter yields 
\begin{equation}
(\triangle+1)\alpha=-g*\d(\bphi^*\iota_\eps\Omega),\label{eq:dconstraint}
\end{equation}
and hence
\beq
\frac{\d\:}{\d{s}}E(\bphi_s,f_s)\bigg|_{s=0}&=&\frac14\ip{\d\bphi,\d\eps}_{L^2(X)}+\ip{\eps,(\grad V)\circ\bphi}_{L^2(X)}
+\ip{\d f,\d \alpha}_{L^2(X)}+\ip{f,\alpha}_{L^2(X)}\nonumber\\
&=&\ip{\eps,\frac14\triangle\bphi+(\grad V)\circ\bphi}_{L^2(X)}+\ip{f,(\triangle+1)\alpha}_{L^2(X)}\nonumber\\
&=&\ip{\eps,\frac14P_\bphi(\triangle\bphi)+(\grad V)\circ\bphi}-g\ip{f,*\d(\bphi^*\iota_\eps\Omega)}_{L^2(X)},
\eeq
where we have used eq.~\eqref{eq:dconstraint} in the last line.
Now
\beq
\ip{f,*\d(\bphi^*\iota_\eps\Omega)}_{L^2(X)}&=&\int_Xf\d(\bphi^*\iota_\eps\Omega)
=-\int_X \d f\wedge\bphi^*\iota_\eps\Omega\nonumber\\
&=&-\ip{\eps,*(\d f\wedge\Xi_\bphi)}_{L^2(X)},
\eeq
where, once again, decaying boundary conditions were imposed if $X$ is noncompact. Hence
\begin{equation}\label{Egrad}
\frac{\d\:}{\d{s}}E(\bphi_s,f_s)\bigg|_{s=0}=\ip{\eps,\frac14P_\bphi(\triangle\bphi)+(\grad V)\circ\bphi+g*(\d f\wedge\Xi_\bphi)}_{L^2(X)},
\end{equation}
that is, $E(\bphi,f)$ is stationary for all variations preserving the
constraint \eqref{static2} if and only if eq.~\eqref{static1} holds.   

Equation \eqref{Egrad} has a useful reinterpretation. Given any smooth
map $\bphi:X\ra N$, the constraint equation \eqref{static2} uniquely
determines the smooth function $f:X\ra\R$, so we may think of $E$ as a
function $C^\infty(X,N)\ra\R$, that is, as a functional of $\bphi$
only. Formally, $C^\infty(X,N)$ is an infinite dimensional manifold
whose tangent space at a map $\bphi$ is $\Gamma(\bphi^{-1}TN)$, the
vector space of smooth sections of the bundle 
$\bphi^{-1}TN$. This space carries a natural inner product called the
$L^2$ metric, so that, formally, $C^\infty(X,N)$ is a Riemannian
manifold. In this picture, eq.~\eqref{Egrad} states that the
\emph{gradient} of the function $E:C^\infty(X,N)\ra\R$ with respect to
the $L^2$ metric is 
\begin{equation}\label{gradE}
\grad E_\bphi=\frac14P_\bphi(\triangle\bphi)+(\grad V)\circ\bphi+g*(\d f\wedge\Xi_\bphi).
\end{equation}
Note that this is, at each fixed $\bphi$, a section of $\bphi^{-1}TN$,
and hence defines a vector field on $C^\infty(X,N)$.

\subsection{Summary in explicit coordinates}\label{sec:explicit_expressions}\label{sec:explicit_stuff}

Let us summarize what we have found so far in the special case of most
direct interest, where spacetime, $\mathcal{M}$, is $3+1$ dimensional
Minkowski space and the target space $N=S^3$, expressing all
quantities in a standard choice of explicit coordinates. The Skyrme
field is 
$\bphi=(\phi_0,\phi_1,\phi_2,\phi_3)$ 
subject to the constraint $\bphi\cdot\bphi=1$. The action functional \eqref{action} is
\begin{equation}\label{eq:actionexplicit}
S(\bphi,\omega_\mu)=\int_{\R^{3,1}}\left\{\frac18\cd_\mu\bphi\cdot\cd^\mu\bphi-\frac14m^2(1-\phi_3)-\frac14\omega_{\mu\nu}\omega^{\mu\nu}+\frac12\omega_\mu\omega^\mu+g\omega_\mu B^\mu\right\}\d{}^4x,
\end{equation}
where $\omega_\mu$ is the omega meson vector field,
$\omega_{\mu\nu}=\cd_\mu\omega_\nu-\cd_\nu\omega_\mu$ is the field
strength for the omega meson, $m$ is
the pion mass, $g$ is a coupling between the omega meson and the
baryon current, which reads
\beq
B_\mu=\frac{1}{12\pi^2}\epsilon_{\mu\nu\rho\sigma}\epsilon_{abcd}\phi_a\cd_\nu\phi_b\cd_\rho\phi_c\cd_\sigma\phi_d,
\eeq
having adopted the conventions that $\epsilon_{0123}=+1$ and that repeated spacetime indices $\mu,\nu,\ldots$ and field space indices $a,b,\ldots$ are summed over $\{0,1,2,3\}$. 

We have found that a static field configuration $\bphi(x^1,x^2,x^3)$, $\omega_0=f(x^1,x^2,x^3)$,
$\omega_i=0$, satisfies the Euler-Lagrange equations for the action \eqref{eq:actionexplicit} if and only if it is a critical point of the static energy functional
  \beq \label{eq:Eexplicit}
  E(\bphi,f) = \int_{\R^3} \left(
  \frac18\p_i\bphi\cdot\p_i\bphi + \frac14m^2(1 - \phi_3) +
  \frac12\p_if\p_if + \frac12f^2
  \right)\d{}^3x\;,
  \eeq
subject to the constraint\footnote{The constraint equation \eqref{eq:static2explicit} can be interpreted as a variant of Gauss's law of electrostatics:  if we think of baryon density $B_0$ as a kind of ``electric" charge density, then \eqref{eq:static2explicit} is the equation for the
``electrostatic" potential $f$ induced by $B_0$ in the unusual case where the ``photon" has unit mass. Of course, this is merely an analogy.}
\begin{equation}\label{eq:static2explicit}
(-\cd_i\cd_i+1)f=-gB_0.
\end{equation}
It follows that $(\bphi,f dx^0)$ is a static solution of the model if and only if the functions $(\phi_0,\phi_2,\phi_2,\phi_3,f)$ satisfy eq.~\eqref{eq:static2explicit} and 
\begin{equation}\label{eq:gradEexplicit}
-\frac14(\cd_i\cd_i\phi_b-\phi_b\phi_a\cd_i\cd_i\phi_a)+\frac{m^2}{4}(\phi_0\phi_b-\delta_{b0})
-\frac{g}{2\pi^2}\frac{1}{2!}\epsilon_{ijk}\epsilon_{abcd}\phi_a\cd_if\cd_j\phi_c\cd_k\phi_d=0.
\end{equation}
It is important to note that the equations \eqref{eq:gradEexplicit}
and \eqref{eq:static2explicit} are \emph{not} the Euler-Lagrange
equations for the functional $E(\bphi,f)$, but are the correct
equations for finding static solutions in this model.

Since eq.~\eqref{eq:static2explicit} uniquely determines $f$ for any
given $\bphi$, we may formally use it to eliminate $f$ from the energy 
functional $E(\bphi,f)$, which is thus reinterpreted as a functional
$E(\bphi)$ of the Skyrme field only. The left hand side of
eq.~\eqref{eq:gradEexplicit} can then be identified with
$(\grad E_\bphi)_b$,  the component of the \emph{gradient} of the
functional $E$ at the configuration $\bphi$ in the field space
direction $b$. This interpretation will be central to the numerical
method we develop for solving eqs.~\eqref{eq:static2explicit},
\eqref{eq:gradEexplicit} in practice.

\section{Stability and the second variation formula}\label{sec:secondvar}

As just observed, since eq.~\eqref{static2} uniquely determines $f$ for each given $\bphi$,
we may interpret $E(\bphi,f)$ as a functional of $\bphi$ only which, in
a slight abuse of notation, we will denote $E(\bphi)$. The static
$\omega$-Skyrme model thus defines a natural geometric variational
problem for maps $\bphi:(X,\zeta)\ra (N,h)$ between Riemannian
manifolds -- to minimise $E(\bphi)$ in a given homotopy class of maps
-- analogous to the classical harmonic map problem, where the energy
to be extremised is simply the Dirichlet energy, 
\begin{equation}
E_D(\bphi)=\frac12\int_X|\d\bphi|^2*1.
\end{equation}
Equation \eqref{static1} is the condition for $\bphi$ to be a
\emph{critical point} of $E(\bphi)$, but its solutions are not
necessarily local minima: they could be 
saddle points instead. To distinguish between minima and saddle points
of $E(\bphi)$ we must consider its \emph{second} variation. The goal of
this section is to compute and apply this second variation, exploiting
the close analogy with the well-established setting of harmonic
maps. To avoid technical issues with boundary conditions, we will
assume throughout this section that $(X,\zeta)$ is closed. 

We begin by briefly recalling the first and second variation formulae
for $E_D(\bphi)$.  
Associated to any smooth map $\bphi:(X,\zeta)\ra (N,h)$ is a smooth
section of $\bphi^{-1}TN$ called the \emph{tension field}, 
\begin{equation}
\tau(\bphi):=\sum_{i}(\nabla^\bphi_{e_i} \d\bphi(e_i)-\d\bphi(\nabla_{e_i}e_i)),
\end{equation}
where $\{e_i\}$ is a local orthonormal frame on $(X,\zeta)$, and
$\nabla^\bphi$ denotes the pullback of the Levi-Civita connexion
$\nabla^N$ on $TN$ to $\bphi^{-1}TN$. This is the natural connexion on
$\bphi^{-1}TN$ (recall, this vector bundle over $X$ whose fibre above
$x\in X$ is $T_{\bphi(x)}N$), constructed from $\nabla^N$. A thorough
treatment of its definition and properties is presented in
ref.~\cite[ch.\ 4]{Urakawa:1993}. In the extrinsic formulation used in
section \ref{sec:model}, $\tau(\bphi)=-P_\bphi\triangle\bphi$. 
Given a smooth one-parameter variation $\bphi_t$ of $\bphi=\bphi_0:X\ra N$,
with infinitesimal generator
$\eps:=\cd_t\bphi_t|_{t=0}\in\Gamma(\bphi^{-1}TN)$, the associated
variation of $E_D(\bphi)$ is
\begin{equation}
\frac{\d\: }{\d{t}}E_D(\bphi_t)\bigg|_{t=0}=-\int_X h(\tau(\bphi),\eps)*1, 
\end{equation}
so $\bphi$ is a critical point of $E_D$ (a harmonic map) if and only if
$\tau(\bphi)=0$. Consider now an arbitrary \emph{two}-parameter
variation $\bphi_{s,t}$ of a harmonic map $\bphi=\bphi_{0,0}$, with
infinitesimal generators $\eps:=\cd_s\bphi_{s,t}|_{s=t=0}$ and
$\wh\eps:=\cd_t\bphi_{s,t}|_{s=t=0}$. Then 
\begin{equation}
\frac{\cd^2 E_D(\bphi_{s,t})}{\cd s\cd t}\bigg|_{s=t=0}=\int_Xh(\eps,J_\bphi\wh\eps)*1,
\end{equation}
where $J_\bphi:\Gamma(\bphi^{-1}TN)\ra\Gamma(\bphi^{-1}TN)$ is a certain
second-order linear self-adjoint elliptic differential operator,
constructed from $\nabla^\bphi$ and the curvature 
tensor $R$ of $(N,h)$, called the \emph{Jacobi operator}. Explicitly,
\begin{equation}
J_\bphi\eps:=-\sum_{i}\left(\nabla^\bphi_{e_i}\nabla^\bphi_{e_i}\eps-\nabla^\bphi_{\nabla_{e_i}e_i}\eps+R\big(\eps,\d\bphi(e_i)\big)\d\bphi(e_i)\right).
\end{equation}
The second variation thus defines a symmetric bilinear form on $\Gamma(\bphi^{-1}TN)$
\begin{equation}
\Hess^D_\bphi(\eps,\wh\eps):=\int_Xh(\eps,J_\bphi\wh\eps)*1,
\end{equation}
called the \emph{Hessian}. We say that the harmonic map $\bphi$ is
\emph{stable} if $\Hess^D_\bphi(\eps,\eps)\geq 0$ for all $\eps$, and
\emph{unstable} otherwise. Determining the stability of a harmonic map
thus reduces to a question about the eigenvalues of its Jacobi
operator. 

How does this generalise to our variational problem? We have already computed the first variation, \eqref{Egrad}, 
\begin{equation}
\frac{\d\: }{\d{t}}E(\bphi_t)\bigg|_{t=0}=\int_X h\left(-\frac14\tau(\bphi)+(\grad V)\circ \bphi+g*(\d f\wedge\Xi_\bphi),\eps\right)*1, 
\end{equation}
in the notation just introduced.
To state the second variation formula requires two more preliminary
definitions. First, given a smooth map $\bphi:X\ra N$, we define the
linear first-order differential operator
$\dot{\Xi}_\bphi:\Gamma(\bphi^{-1}TN)\ra\Gamma(\bigwedge^{d-1}T^*X\otimes\bphi^{-1}TN)$
which maps a section 
$\eps$ of $\bphi^{-1}TN$ to the $(d-1)$-form on $X$ valued in
$\bphi^{-1}TN$ satisfying 
\beq
h\big(Y,\dot{\Xi}_\bphi(\eps)(X_1,X_2,\ldots,X_{d-1})\big)&=&\Omega\big(Y,\nabla^\bphi_{X_1}\eps,\d\bphi(X_2),\ldots,\d\bphi(X_{d-1})\big)\nonumber \\&& \mathop+
\Omega\big(Y,\d\bphi(X_1),\nabla^\bphi_{X_2}\eps,\ldots,\d\bphi(X_{d-1})\big)+\cdots\nonumber \\&&\cdots+\Omega\big(Y,\d\bphi(X_1),\d\bphi(X_2),\ldots,\nabla^\bphi_{X_{d-1}}\eps\big),
\label{linbar}
\eeq
for all $x\in X$, $Y\in T_{\bphi(x)}N$, $X_1,\ldots,X_{d-1}\in T_xX$.
Second, given a smooth map $\bphi:X\ra N$, we define the linear
integral operator 
$\alpha_\bphi:\Gamma(\bphi^{-1}TN)\ra C^\infty(X)$ which maps a section
$\eps$ of $\bphi^{-1}TN$ to the solution $\alpha$ of the linear PDE 
\begin{equation}\label{liba}
(\triangle+1)\alpha=-*\d(\bphi^*\iota_\eps\Omega),
\end{equation}
which exists and is smooth and unique by standard elliptic PDE theory.
The linear operator $\alpha_\bphi$ maps infinitesimal variations of
$\bphi$ to the corresponding infinitesimal variations of $f$. That is, 
given a variation $\bphi_t$ of $\bphi$, generated by
$\eps=\cd_t\bphi_t|_{t=0}$, the corresponding variation $f_t$ of the
solutions of eq.~\eqref{static2} has
$\cd_t f_t|_{t=0}=g\alpha_\bphi(\eps)$. We may now state the second
variation formula (the proof, which is rather involved, is presented
in Appendix \ref{app:secondvar}): 

\begin{prop} \label{prop1}
Let $\bphi:X\ra N$ and $f:X\ra\R$ satisfy \eqref{static1}, \eqref{static2}. Let $\bphi_{s,t}$ be any smooth two-parameter variation of $\bphi=\bphi_{0,0}$, $f_{s,t}$ be the
corresponding variation of $f=f_{0,0}$, preserving \eqref{static2}, $\eps=\cd_{s}\bphi_{s,t}|_{s=t=0}$ and $\wh\eps=\cd_t\bphi_{s,t}|_{s=t=0}$. Then
\begin{eqnarray*}
\Hess_\bphi(\eps,\wh\eps)&:=&\frac{\cd^2 E(\bphi_{s,t},f_{s,t})}{\cd s\cd t}\bigg|_{s=t=0}\\
&=&\int_X h\left(\eps,\frac14J_\bphi\wh\eps+\big(\nabla^N_{\wh\eps}\grad V\big)\circ\bphi+g*\big(\d f\wedge\dot{\Xi}_\bphi(\wh\eps)\big)\right)*1\\
&&\mathop+g\int_X\d f\wedge\bphi^*\big(\iota_\eps\nabla^N_{\wh\eps}\Omega\big)+g^2\int_X \alpha_\bphi(\eps)(\triangle+1)\alpha_\bphi(\wh\eps)*1.
\end{eqnarray*}
\end{prop}

In direct analogy with harmonic map theory, a critical point is stable if $\Hess_\bphi(\eps,\eps)\geq 0$ for all $\eps$, and unstable otherwise. Since $\alpha_\bphi$ is
not invertible, the stability question does not easily reduce to a
spectral problem. Nonetheless, in an interesting family of special
cases we can make significant progress.

\subsection{Stability of the identity map}\label{sec:stability_compact}

Consider the case that $(X,\zeta)=(N,h)$, $\Omega$ is the volume form
on $(N,h)$, $V=0$ and $\bphi=\id$, the identity map, that is,
$\bphi(x)=x$. If $N=S^3$, this is a simple model of dense nuclear
matter with uniform baryon density, whose stability in the
conventional Skyrme model was studied by Manton
\cite{Manton:1987xt}. We will, for the time being, leave $X=N$
general, however. It is well known that $\id:X\ra X$ is harmonic, so
$\tau(\id)=0$ \cite{Smith:1975}. Furthermore,
$*\Id^*\Omega=*\Omega=1$, since $\Omega$ was chosen to be the volume
form. Hence the function $f$ determined by eq.~\eqref{static2} is
simply the constant function $f=-g$, so $\d f=0$, and it follows
immediately that $\bphi=\id$ satisfies the Euler-Lagrange equation
\eqref{static1}: $\id$ is a critical point of $E(\bphi)$ for all
$g$. As we will see, the \emph{stability} of $\id$ depends, in general,
on the coupling $g$, however. 

The formula for the Hessian given by Proposition \ref{prop1}
simplifies radically in this case. First, since $\d f=0$, the
difficult terms involving $\dot\Xi_{\id}$ and $\nabla^N\Omega$ vanish
(actually $\nabla^N\Omega\equiv 0$ since the volume form is parallel,
so the latter term vanishes even for critical points with nonconstant
$f$). So, noting that $V=0$, 
\begin{equation}
\Hess_\id(\eps,\eps)=\frac14\ip{\eps,J_\id\eps}+g^2\left(\|\d\alpha_\id(\eps)\|_{L^2}^2+\|\alpha_\id(\eps)\|_{L^2}^2\right)\geq\frac14\ip{\eps,J_\id\eps},
\end{equation}
and it follows that if $\id$ is stable as a harmonic map, it is also a
stable critical point of $E(\bphi)$. Hence, $\id$ is stable for all $g$
in dimensions $d=1,2$, or if $(X,\zeta)$ is K\"ahler, or if
$(X,\zeta)$ is Ricci negative, for example \cite{Smith:1975}. If $\id$
is \emph{unstable} as a harmonic map (for example, if $X=S^d$, $d\geq
3$), things are more interesting: it is an unstable critical point of
$E(\bphi)$ for $g\geq 0$ small, but may exhibit a stability transition,
as $g$ increases. 

To proceed further, we note that the variation section
$\eps$ is now a section of $\id^{-1}TN\equiv TN\equiv TX$, that is a
\emph{vector field} on $(X,\zeta)$, which greatly simplifies the
Jacobi operator. In fact \cite{Smith:1975}
\begin{equation}
J_\id\eps=\sharp\triangle\flat\eps-2\rho\eps,
\end{equation}
where $\triangle$ is the usual Hodge Laplacian on one-forms, $\flat$
is the metric isomorphism $TX\ra T^*X$ defined by $\zeta$
(i.e.\ $(\flat\eps)(u):=\zeta(\eps,u)$ for all $u\in T_xX$), $\sharp$
is its inverse, and $\rho$ is the Ricci endomorphism of $(X,\zeta)$
(the linear map $\rho:T_xX\ra T_xX$ satisfying
$\zeta(u,\rho v)={\rm Ric}(u,v)$, where ${\rm Ric}$ is the usual Ricci
curvature tensor). Hence, 
\begin{align}
\Hess_\id(\eps,\eps)&=\frac14\ip{\eps, \sharp\triangle\flat\eps-2\rho\eps}+g^2\ip{\alpha_\id(\wh\eps),(\triangle+1)\alpha_\id(\eps)}\nonumber \\
&=\frac14\|\d\flat\eps\|_{L^2}^2+\frac14\|\delta\flat\eps\|_{L^2}^2-\frac12\ip{\eps,\rho\eps}+g^2\left(\|\d\alpha_\id(\eps)\|_{L^2}^2+\|\alpha_\id(\eps)\|_{L^2}^2\right),
\end{align}
where $\eps$ is an arbitrary smooth vector field on $X$. Every term in
this, except the curvature term, $-\ip{\eps,\rho\eps}/2$, is
manifestly non-negative, so the question of stability of $\id$ is
nontrivial only if the Ricci curvature of $(X,\zeta)$ is positive
somewhere. We shall prove that $\Hess_\id$ is non-negative when
evaluated on the subspace of divergenceless vector fields, and is, for
large enough $g$, also non-negative on the subspace of pure
gradients. From this, we can deduce that $\id$ is stable, for $g$
sufficiently large, if $(X,\zeta)$ is Einstein. 

\begin{lemma} \label{lem1}
For any divergenceless vector field $\eps_0$ on $(X,\zeta)$,
$\Hess_{\id}(\eps_0,\eps_0)\geq 0$. 
\end{lemma}

\nid {\it Proof:}
For any vector field $\eps$ on $X$,
$*\id^*\iota_\eps\Omega=*\iota_\eps\Omega=\div\eps$, so
$\alpha_\id(\eps)$ satisfies the PDE 
\begin{equation}
(\triangle+1)\alpha_\id(\eps)=-\div\eps.
\end{equation}
Hence, for all divergenceless vector fields $\eps_0$,
$\alpha_\id(\eps_0)=0$. Further, by a formula of Bochner and Yano
\cite{Smith:1975}, 
\begin{equation}
\ip{\eps,J_\id\eps}=\frac12\|{\cal L}_\eps\zeta\|_{L^2}^2-\|\div\eps\|_{L^2},
\end{equation}
where ${\cal L}$ denotes the Lie derivative, so for all divergenceless
vector fields $\eps_0$, 
\begin{equation}
\Hess_\id(\eps_0,\eps_0)=\frac18\|{\cal L}_{\eps_0}\zeta\|_{L^2}^2\geq 0.
\end{equation}
\hfill$\Box$

\begin{lemma} \label{lem2}
There exists $g_0\geq 0$ such that, for all $g\geq g_0$ and all smooth
functions $\ell:X\ra \R$, $\Hess_{\id}(\nabla\ell,\nabla\ell)\geq 0$.
\end{lemma}

\nid {\it Proof:}
Since $X$ is compact, there exists a constant $c>0$ such that, for all
$u\in T_xX$, ${\rm Ric}(u,u)\leq c\zeta(u,u)$, and hence, for all
vector fields $\eps$, $\ip{\eps,\rho\eps}\leq c\|\eps\|_{L^2}^2$. Let
$0=\lambda_0<\lambda_1\leq\lambda_2\leq\lambda_3\leq\cdots$ be the
eigenvalues of the Laplacian (on functions) on $(X,\zeta)$ and
$\{f_n\}$ be a corresponding $L^2$ orthonormal basis of
eigenfunctions, so $\triangle f_n=\lambda_n f_n$. Since the sequence
$(\lambda_n)$ diverges to infinity, there exists $q\in\N$ such that,
for all $n>q$, $\lambda_n\geq 2c$. Any function $\ell\in C^\infty(X)$
has a unique expansion $\ell=\sum_{n=0}^\infty a_n f_n$ in the
harmonics $\{f_n\}$. Now 
\begin{equation}
(\triangle+1)\alpha_\id(\nabla \ell)=-\div\nabla\ell=\triangle\ell=\sum_{n=1}^\infty\lambda_na_nf_n,
\end{equation}
so
\begin{equation}
\alpha_\id(\nabla\ell)=\sum_{n=1}^\infty\frac{\lambda_na_n}{1+\lambda_n}f_n.
\end{equation}
Hence
\beq
\Hess_{\id}(\nabla\ell,\nabla\ell)&\geq&\frac14\ip{\nabla\ell,\sharp\triangle\d\ell}-\frac{c}{2}\|\nabla\ell\|_{L^2}^2+g^2\ip{\alpha_\id(\nabla\ell),(\triangle+1)\alpha_\id(\nabla\ell)}\nonumber \\
&=&\frac14\ip{\d\ell,\d\delta\d\ell}-\frac{c}{2}\ip{\ell,\triangle\ell}+g^2\ip{\alpha_\id(\nabla\ell),\triangle\ell}\\
&=&\frac14\ip{\ell,\triangle^2\ell-2c\triangle\ell}+g^2\ip{\alpha_\id(\nabla\ell),\triangle\ell}\nonumber \\
&=&\frac14\sum_{n=1}^\infty\left(\lambda_n^2-2c\lambda_n+\frac{4g^2\lambda_n^2}{1+\lambda_n}\right)a_n^2\nonumber \\
&\geq&\frac14\sum_{n=1}^q\left(\lambda_n^2-2c\lambda_n+\frac{4g^2\lambda_n^2}{1+\lambda_n}\right)a_n^2
\eeq
since $\lambda_n^2\geq 2c\lambda_n$ for all $n>q$. If $g$ is chosen so
that that $4g^2$ exceeds
$$
4g_0^2:=\max\{|2c-\lambda_n|(1+\lambda_n^{-1}):1\leq n\leq q\},
$$
all the terms in this finite sum are non-negative, and the claim
immediately follows.\hfill$\Box$ 

Every smooth vector field $\eps$ on $X$ uniquely decomposes into
gradient and divergenceless components (just apply the Hodge
decomposition to the one-form $\flat\eps$), and we have just shown
that, for $g$ sufficiently large, $\Hess_\id$ is non-negative on both
the gradient and divergenceless subspaces of $\Gamma(TX)$.
If $\Hess_\id$ is diagonal with respect to the Hodge decomposition, it
follows immediately that $\id$ is stable for $g$ sufficiently
large. In particular: 

\begin{prop}\label{prop2}
Let $(X,\zeta)$ be a closed Einstein manifold. Then there exists
$g_0\geq 0$ such that, for all $g>g_0$, $\id:(X,\zeta)\ra (X,\zeta)$
is a stable critical point of $E(\bphi)$.
\end{prop}

\nid {\it Proof:}
By Lemmas \ref{lem1} and \ref{lem2}, there exists $g_0$ such that, for
all $g\geq g_0$ and all $\eps$, 
\beq
\Hess_\id(\eps,\eps)&=&\Hess_\id(\eps_0,\eps_0)+\Hess_\id(\nabla\ell,\nabla\ell)+2\Hess_\id(\nabla\ell,\eps_0)\nonumber \\
&\geq& 2\Hess_\id(\nabla\ell,\eps_0),
\eeq
where $\eps=\eps_0+\nabla\ell$ is the Hodge decomposition of $\eps$
into divergenceless and gradient parts (obtained by decomposing the
one-form $\flat\eps$ into coclosed and exact parts). Since $(X,\zeta)$
is Einstein, $\rho=c\id$ where $c$ is a constant. Hence 
\beq
\Hess_\id(\nabla\ell,\eps_0)&=&\frac14\ip{\nabla\ell,\sharp\triangle\flat\eps_0}-\frac{c}{2}\ip{\nabla\ell,\eps_0}+g^2\ip{\alpha_\id(\nabla\ell),-\div\eps_0}\nonumber\\
&=&\frac14\ip{\ell,\delta(\d\delta+\delta\d)\flat\eps_0}-\frac{c}{2}\ip{\ell,\div\eps_0}+g^2\ip{\alpha_\id(\nabla\ell),0}\nonumber\\
&=&0,
\eeq
since $\div\eps_0=-\delta\flat\eps_0=0$. The claim immediately
follows. \hfill $\Box$ 

Proposition \ref{prop2} covers, in particular, the case of most
interest, $X=S^3$. A careful recapitulation of the proof of Lemma
\ref{lem2} using the spectrum of the Laplacian for the unit $d$-sphere
reveals that the critical coupling for $X=S^d$, above which $\id$ is
stable, is 
\begin{equation}
g_0(S^d)=\frac12\sqrt{(d-2)(d+1)}.
\end{equation}

\subsection{A topological lower energy bound}\label{sec:bound}

We conclude by establishing a topological lower
bound for $E(\bphi)$. 
We now revert to the case of general $(N,h)$, $\Omega$ and $V$ while
maintaining the assumption that $X$ is compact and without boundary.  

\begin{prop} For all smooth maps $\bphi:X\ra N$,
$$
E(\bphi)\geq \frac{g^2}{2{\rm Vol}(X,\zeta)}\left(\int_X\bphi^*\Omega\right)^2.
$$
\end{prop}

\nid {\it Proof:}
By the Cauchy-Schwartz inequality and eq.~\eqref{static2},
\begin{equation}
\|-1\|_{L^2}\|f\|_{L^2}\geq\ip{-1,f}=g\int_X\bphi^*\Omega,
\end{equation}
and hence
\begin{equation}
E(\bphi)\geq\frac12\|f\|_{L^2}^2\geq\frac{g^2}{2\|-1\|_{L^2}^2}\left(\int_X\bphi^*\Omega\right)^2.
\end{equation}
\hfill$\Box$

Note that this bound is \emph{quadratic} in the topological invariant
$\int_X\bphi^*\Omega$. So, if $N=S^3$ and $\Omega$ is the (normalised)
volume form on $N$, we see that the $\omega$-Skyrme energy grows at
least quadratically with the baryon number, $E\geq {\rm const}\times B^2$.
This contrasts with the conventional Skyrme model, where the analogous
bound on compact domains is $E\geq {\rm const} \times |B|^{4/3}$
\cite{Harland:2013rxa}.
On the other hand, our bound coincides precisely with the energy bound
found by Adam and Wereszczynski \cite{Adam:2013tga} for the
so-called sextic Skyrme model 
\beq
E_{{\scriptscriptstyle \rm sextic}}(\bphi)=
\int_X\left(\frac18|\d\bphi|^2+\frac{g^2}{2}|\bphi^*\Omega|^2+V\circ\phi\right)*1,
\eeq
on a compact three manifold (with target $N=S^3$). This is one of several striking similarities between these two models, a theme to which we will return in section \ref{sec:conc}.
It is interesting to note that the sextic model on $X=\R^3$ can easily be shown \cite{Adam:2013tga} to have a linear topological energy bound,
\beq
E_{{\scriptscriptstyle \rm sextic}}(\bphi)\geq\sqrt{\frac{g}{2}}|B|,
\eeq
similar to the Faddeev bound on the standard Skyrme energy. It is natural to conjecture that the same bound holds for the $\omega$-Skyrme model on
$\R^3$, but we have been unable to prove this. Note that on any domain
$X$, for all smooth maps $\bphi:X\ra N$,
$E(\bphi)\leq E{{\scriptscriptstyle \rm sextic}}(\bphi)$, so lower
bounds on $E_{{\scriptscriptstyle \rm sextic}}$ do not imply lower bounds on
$E$.

\section{The numerical method}\label{sec:nummet}

We seek to find, within a given topological sector, the minimum of
$E(\bphi)$ as defined in eq.~\eqref{Edef}, $f$ being determined by $\bphi$
using eq.~\eqref{static2}. To do this, we choose an initial configuration
$\bphi(0)$ and solve Newton's equation for the motion of $\bphi(t)$ in
$C^\infty(X,N)$ subject to the potential function
$E:C^\infty(X,N)\ra\R$, that is 
\begin{equation}\label{newton}
P_\bphi\big(\ddot\bphi\big)=-\grad E_\bphi,
\end{equation}
starting at rest, $\dot\bphi(0)=0$.
In practice, we discretise space on a cubic grid and approximate
$\grad E_\bphi$ using finite differences, then use a 4th order
Runge-Kutta scheme to perform the time stepping. This flow will start
to roll ``downhill'', that is, reduce $E$, but will not, as it stands,
converge to a minimum of $E$. To achieve this, we compare, after each
time step, the energies of the new and old configurations. If
$E(t+\delta t)>E(t)$, we arrest the flow, restarting it with
$\dot\bphi=0$. This strategy\footnote{Introduced to one of us by Paul
  Sutcliffe.} is quite widely used in the study of topological
solitons, but does not appear to have received a commonly accepted
name.  We propose to call it ``arrested Newton flow''.  

In the present case, at each time step, to evaluate $\grad E_\bphi$
(and $E$) we must construct the function $f$ satisfying the constraint
equation \eqref{static2}. This is a linear inhomogeneous PDE, or
rather, having discretised space, a high-dimensional linear system of
algebraic equations, so the obvious strategy is to use an off-the-peg
linear algebra solver to compute $f$. This turns out to be
inefficient, since such solvers are not iterative, in the sense that
they start from scratch, making no use of an initial guess for the
solution. For our application, after each time step, $\bphi$, and hence
the right hand side of eq.~\eqref{static2}, has changed very little, so we
have access to an excellent approximation to $f(t+\delta t)$, namely
$f(t)$.  
To exploit this feature, we reinterpret eq.~\eqref{static2} as the
Euler-Lagrange equation for the quadratic functional 
\begin{equation}
G(f)=\int_X\left(\frac12|\d f|^2+\frac12f^2+gB_0f\right)*1,
\end{equation}
which we solve by minimising $G$ using an off-the-peg conjugate
gradient method starting at $f(t)$ (a particularly efficient choice
for quadratic functions). The first application of this method (at
$t=0$), where we have only a rough guess for $f$ (we use $f=-gB_0$) is
quite computationally costly, but after each subsequent time step very
few cycles of the conjugate gradient method (typically $0$ to $3$) are
required to correct $f$ to match the new Skyrme field $\bphi$ to within
the tolerance we require.

\begin{figure}[!th]
  \begin{center}
    \includegraphics[width=\linewidth]{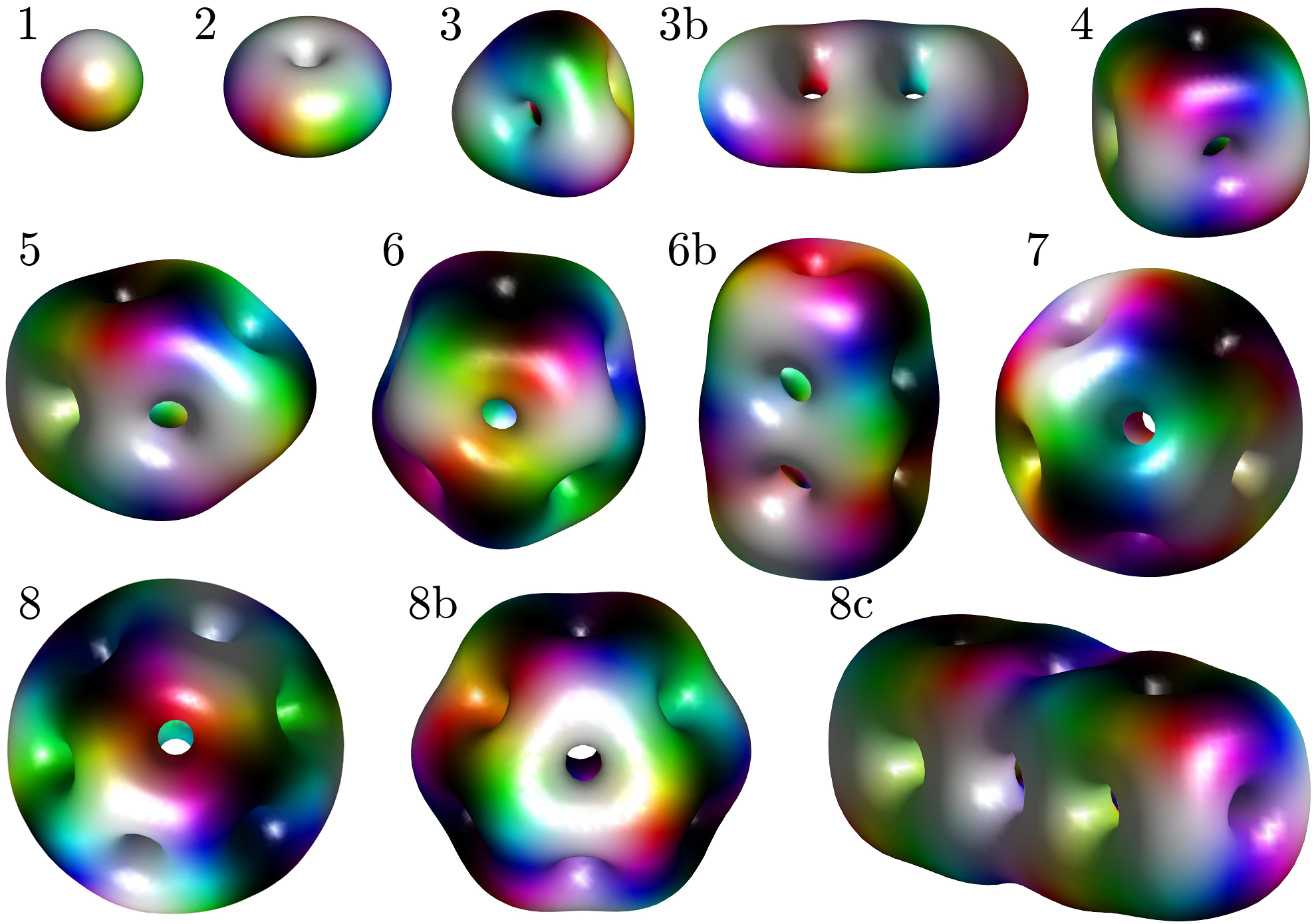}
    \caption{Numerical solutions for baryon numbers $B=1$ through
      $B=8$.
    The global minimisers (the stable solutions) are
    labelled with their topological degree, whereas the metastable
    solutions have increasing energy with letters in the Latin 
    alphabet. }
    \label{fig:g347M0176}
  \end{center}
\end{figure}

To illustrate our numerical scheme, we present classical energy
minimisers of charges $B=1,2,\ldots,8$ for the coupling 
and pion mass proposed by Sutcliffe \cite{Sutcliffe:2008sk}:
\beq
g=34.7, \qquad
m=0.176. \label{eq:Sutcliffe_calibration}
\eeq
The calibration chosen by Sutcliffe fixes $g$ by using the
experimental value for the pion decay constant and the omega mass
(hence fixing the length and energy units) and adjusting $g$ to match
the mass of the 4-Skyrmion to that of Helium-4.

Fig.~\ref{fig:g347M0176} shows coloured surfaces of constant baryon
density for these solutions. The colouring represents the value of the
normalised pion field $\bpi/|\bpi|$ using a standard colouring of the
unit sphere, which can be deduced from the picture for $B=1$.  
The 1-Skyrmion is spherically symmetric, while the
2-Skyrmion is stable and has the shape of a torus as usual in the
Skyrme-like models -- this confirms the stability of the
2-Skyrmion which was an open question in the rational map approach
with the same value of the coupling $g$ \cite{Sutcliffe:2008sk}.
The $B=3$ topological sector contains the first metastable solution
(local, but not global,  energy minimiser), 
which is a baguette-shaped solution of three 1-Skyrmions stacked
together horizontally (with the middle one flipped with respect to the
outer two), see 3b in
fig.~\ref{fig:g347M0176}\footnote{This baguette-shaped solution has
  appeared previously in the literature, i.e.~in
  ref.~\cite{Walet:1996he} where it was obtained from an instanton
  holonomy without tetrahedral symmetry. In ref.~\cite{Walet:1996he}
  the shape was referred to as ``pretzel'' shaped.}.
It has, nevertheless, higher energy compared with the tetrahedrally
symmetric ``standard'' 3-Skyrmion.
The $B=4$ Skyrmion is octahedrally symmetric and the $B=5$ is
dihedrally symmetric, as usual.
The $B=6$ sector contains a global minimiser of the energy functional
with dihedral symmetry (which is the ``standard'' 6-Skyrmion solution)
as well as a metastable solution; it can be interpreted as three
2-Skyrmions (tori) that are stacked on top of each other (with the
middle one flipped); this is similar to how a cube is made of two tori
(with one of them flipped), but just with an extra torus added in, see 
fig.~\ref{fig:g347M0176}(6b). 
In the $B=7$ sector the energy functional is minimised by the
icosahedrally symmetric Skyrmion as usual. 
Finally, the $B=8$ topological sector contains three solutions. The
stable solution is the dihedrally symmetric ``standard''
8-Skyrmion with $D_{6d}$ symmetry, unlike in the standard Skyrme model
with a pion mass term (where the stable solution is composed of
two $B=4$ cubes). 
Additionally, here, there are two metastable solutions:
the first and closest in energy to the minimiser of the energy
functional in the $B=8$ sector has a slightly smaller amount of
symmetry, which we think is $D_6$. 
The last and highest-energy solution in this sector is composed of two
cubes, but unlike in the standard Skyrme model, they do not ``melt''
together, but merely attach to each other and hence look more like a
regular crystal than the ``standard'' solution of the standard Skyrme
model with a pion mass term does.

The first two $B=8$ Skyrmions depicted in figs.~\ref{fig:g347M0176}(8)
and \ref{fig:g347M0176}(8b) are both approximately described by the
rational map \cite{Houghton:1997kg}: 
\beq
R(z) = \frac{z^6 - a}{z^2(a z^6 + 1)},
\eeq
with $z=e^{\i\varphi}\tan\tfrac{\theta}{2}$ being the coordinate on the
Riemann sphere and $a\in\mathbb{C}$.
If $a$ is real, there is an enhanced symmetry, i.e.~$D_{6d}$,
otherwise it is simply $D_6$.
The Skyrme field $\bphi$ obtained by suspending this rational map is
\cite{Houghton:1997kg}
\beq
\bphi = \left(
\cos F(r),
\frac{R+\bar{R}}{1+|R|^2}\sin F(r),
\frac{-\i(R-\bar{R})}{1+|R|^2}\sin F(r),
\frac{1-|R|^2}{1+|R|^2}\sin F(r)
\right),
\eeq
where $F$ is some (so far, unspecified) profile function.
The standard Skyrme energy of this field depends on $a$
only via
\beq
\mathcal{I} = \frac{1}{4\pi}\int
\left(\frac{1+|z|^2}{1+|R|^2}\left|\frac{\d R}{\d z}\right|\right)^4
\frac{2\i\d{z}\wedge\d{\bar{z}}}{(1+|z|^2)^2},
\eeq
which is minimised independently from $F(r)$.
The $\omega$-Skyrme energy in the rational map approximation
analogously depends only on $a$ via $\mathcal{I}$
\cite{Sutcliffe:2008sk}. 
The minimum of $\mathcal{I}(a)$ is at $a=0.135$ \cite{Houghton:1997kg},
but there is a saddle point at $a=0.101\i$.
We think that in the $\omega$-Skyrme theory, this saddle point has
become a local minimum (and possibly moved a bit in the $a$-plane).
Thus we want to identify the stable and normal $B=8$ $D_{6d}$
symmetric Skyrmion of fig.~\ref{fig:g347M0176}(8) with $a=0.135$ and
the metastable (local minimum) $B=8$ Skyrmion of
fig.~\ref{fig:g347M0176}(8b) with $a=0.101\i$, which has
$D_6$ symmetry.

We have searched extensively for a solution that looks like two cubes
attached to each other with and without a twist along the axis that
joins them (i.e.~the global minimisers in the standard Skyrme model
with a pion mass term), but have found -- to our surprise -- that they
only exist as saddle points in the theory and decay into the
dihedrally symmetric  global minimiser (see the ancillary files for
videos of this decay).

To summarise, all global energy minimisers for $B=1$ to $8$ turn out
to have the same symmetries as the global minimisers in the standard
Skyrme energy \emph{without a pion mass term}. 
This model has a pion mass term and thus differs from the standard
Skyrme model with massive pions in having a dihedrally $D_{6d}$
symmetric fullerene-like $B=8$ solution as the global minimiser of the
energy functional.  

\begin{figure}[!htp]
  \begin{center}
    \mbox{\subfloat[1]{\includegraphics[width=0.49\linewidth]{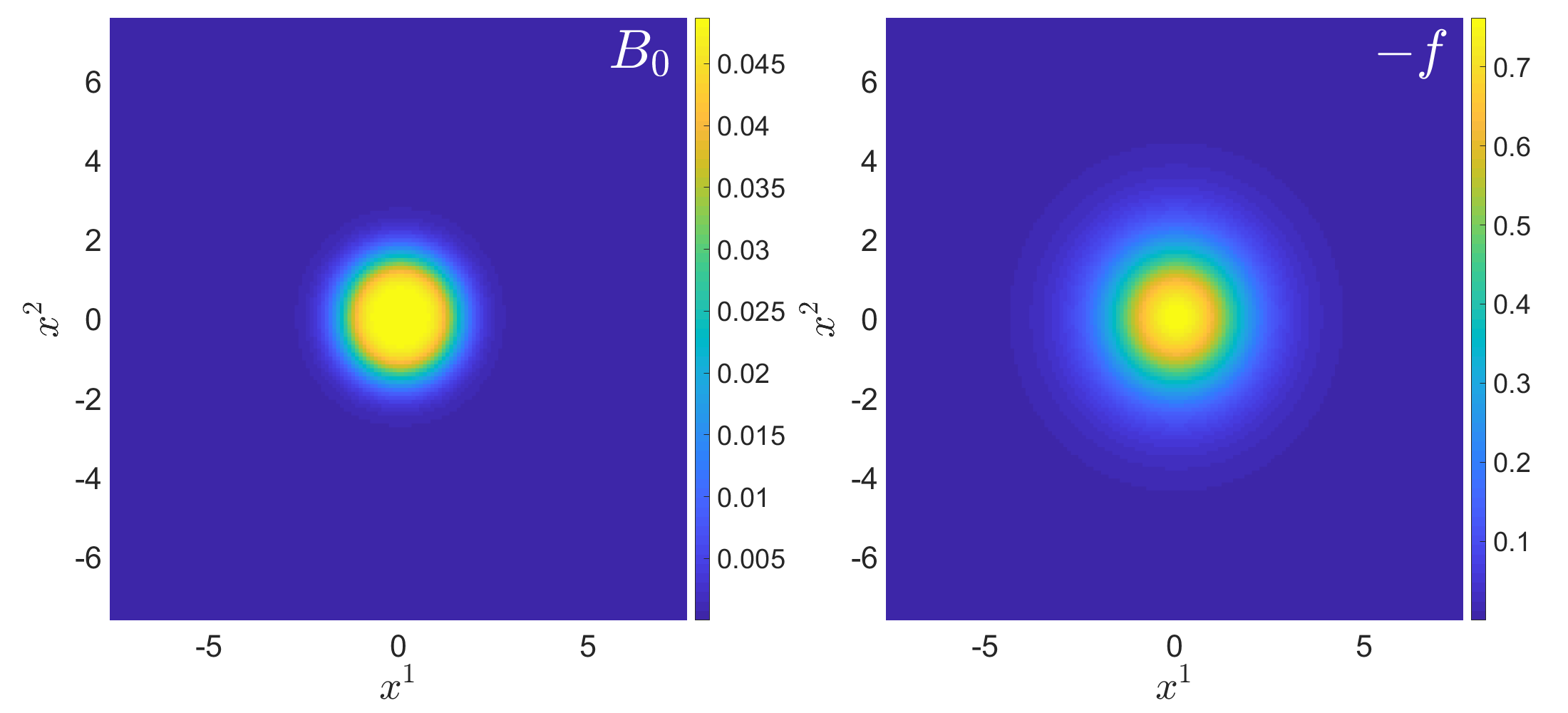}}
      \subfloat[2]{\includegraphics[width=0.49\linewidth]{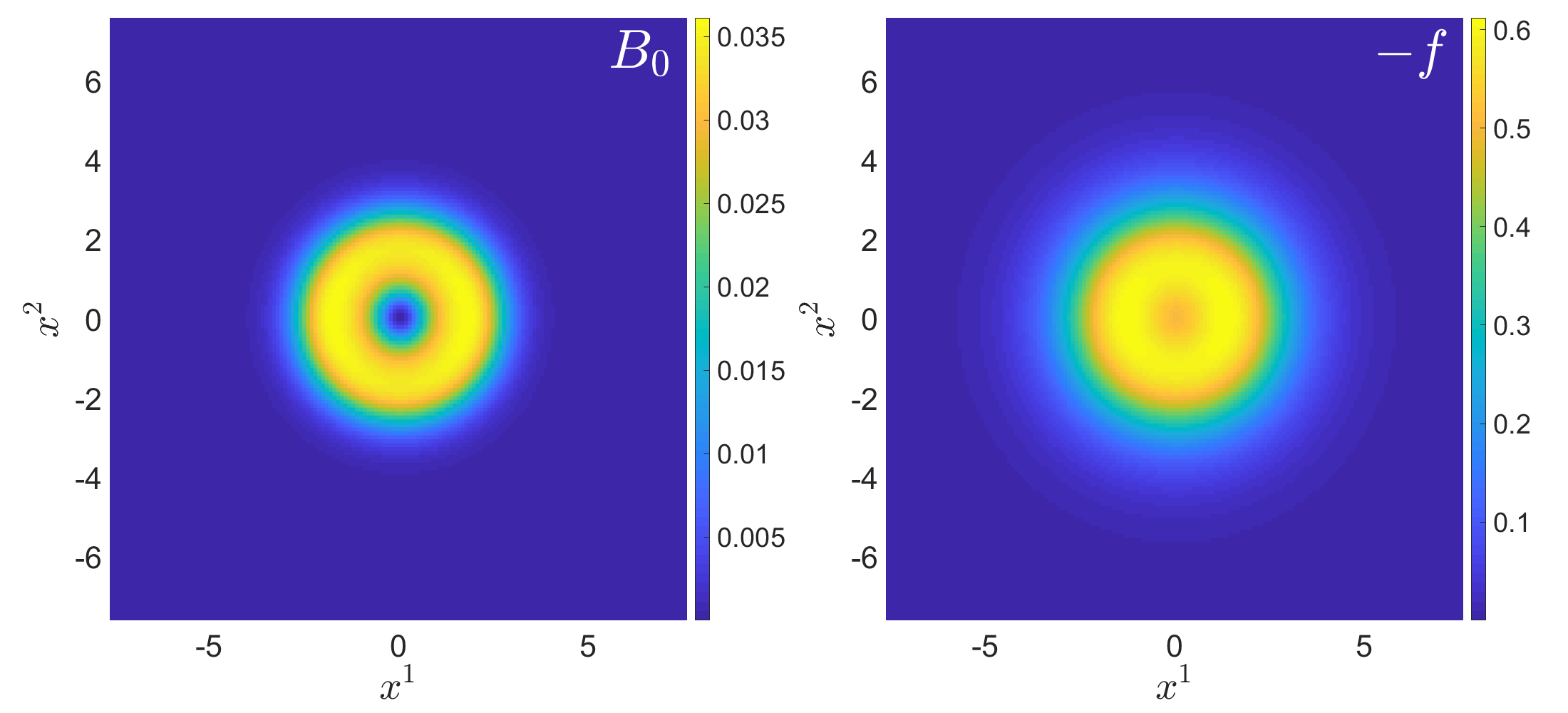}}}
    \mbox{\subfloat[3]{\includegraphics[width=0.49\linewidth]{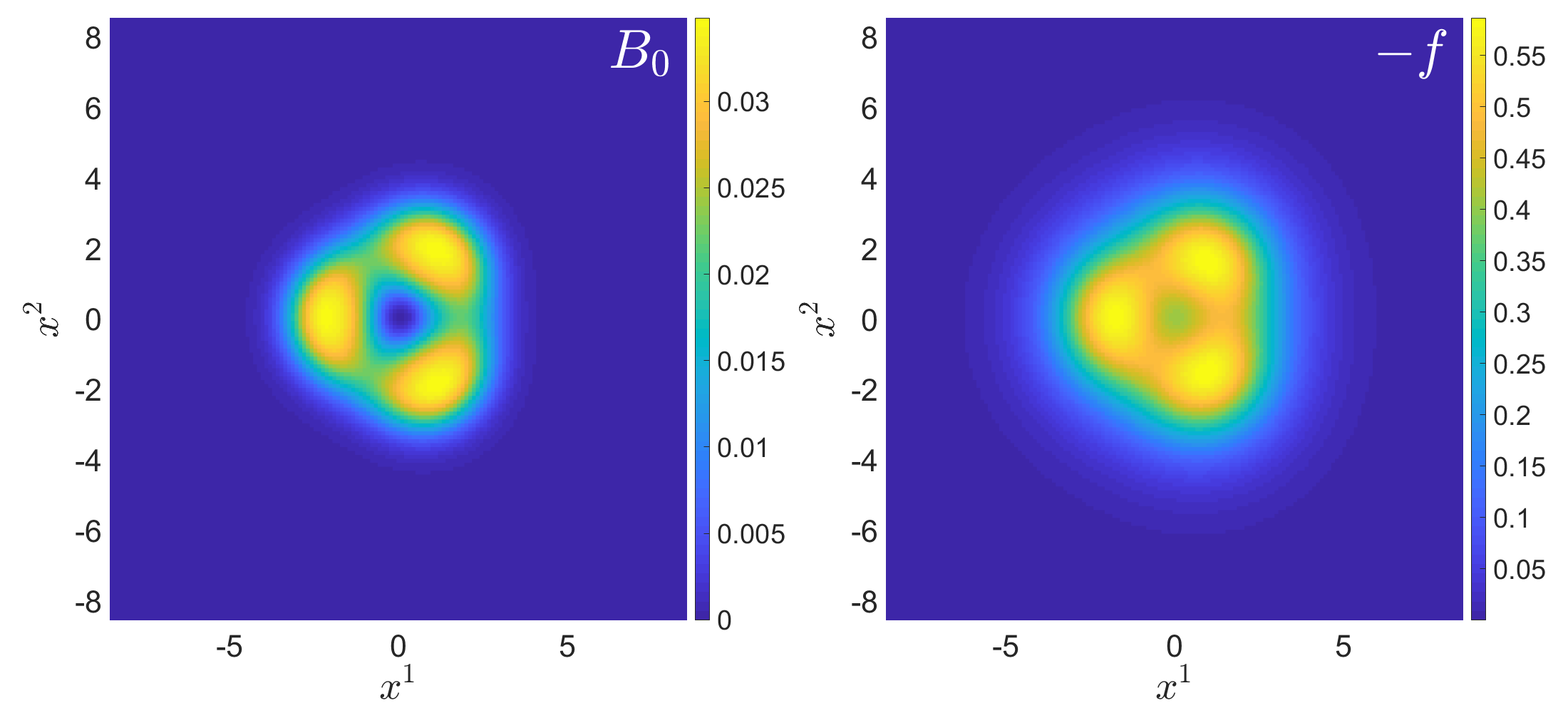}}
      \subfloat[3b]{\includegraphics[width=0.49\linewidth]{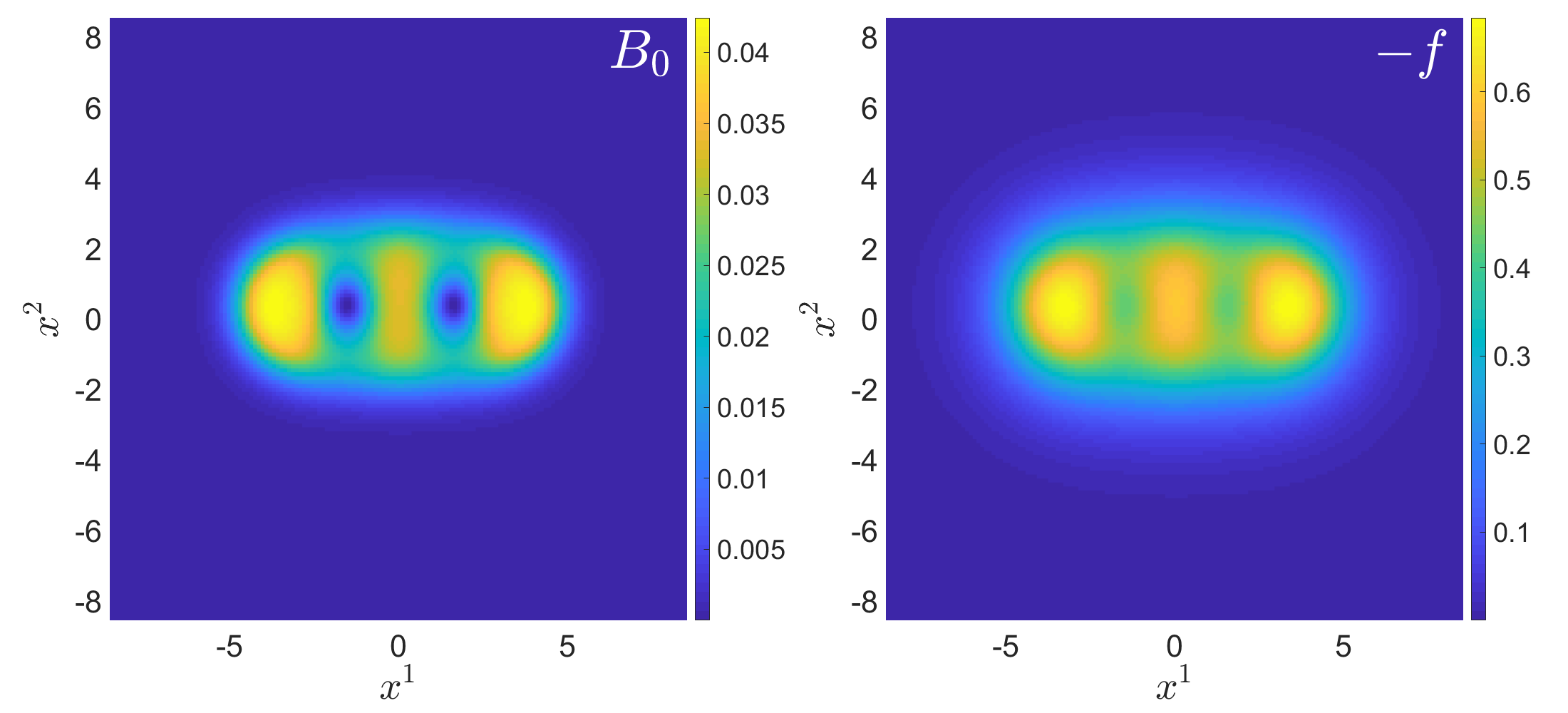}}}
    \mbox{\subfloat[4]{\includegraphics[width=0.49\linewidth]{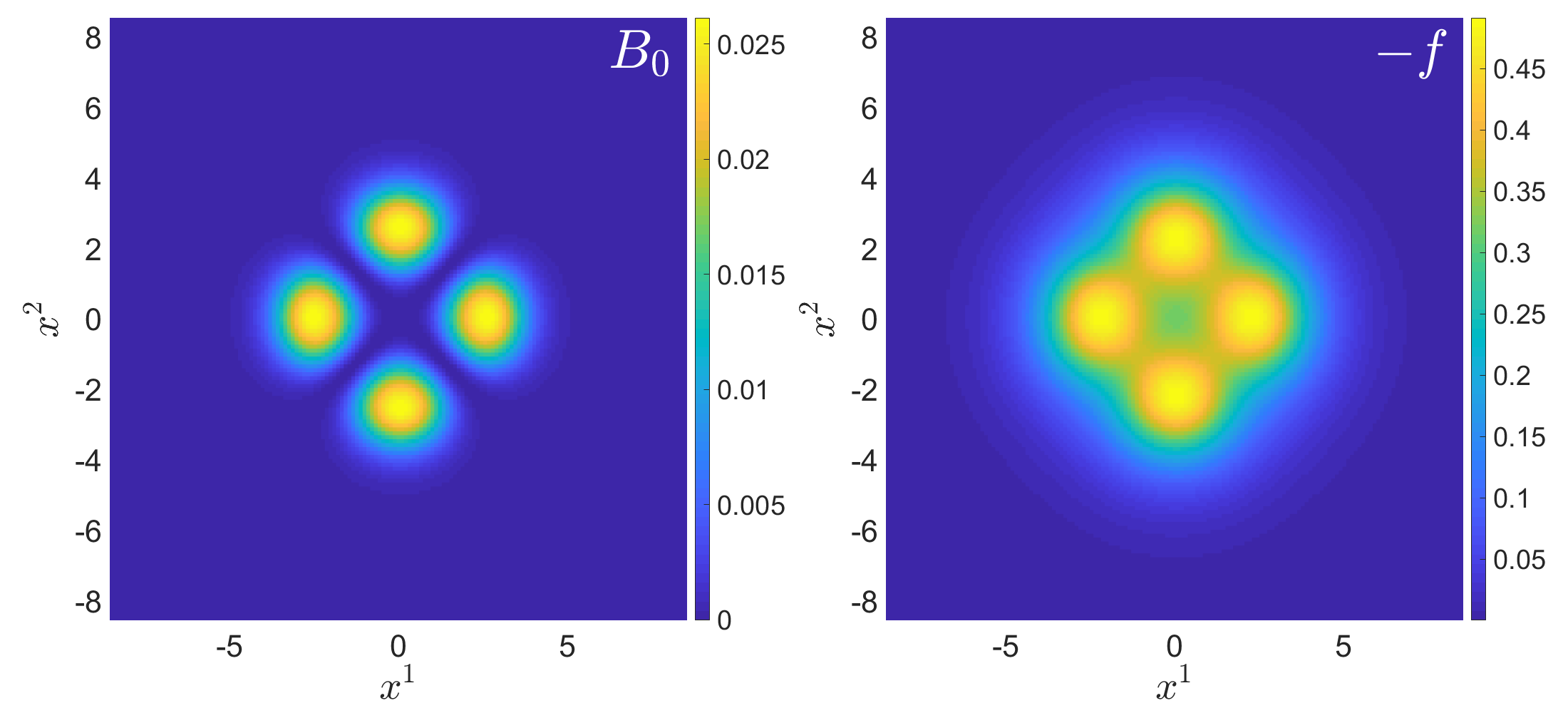}}
      \subfloat[5]{\includegraphics[width=0.49\linewidth]{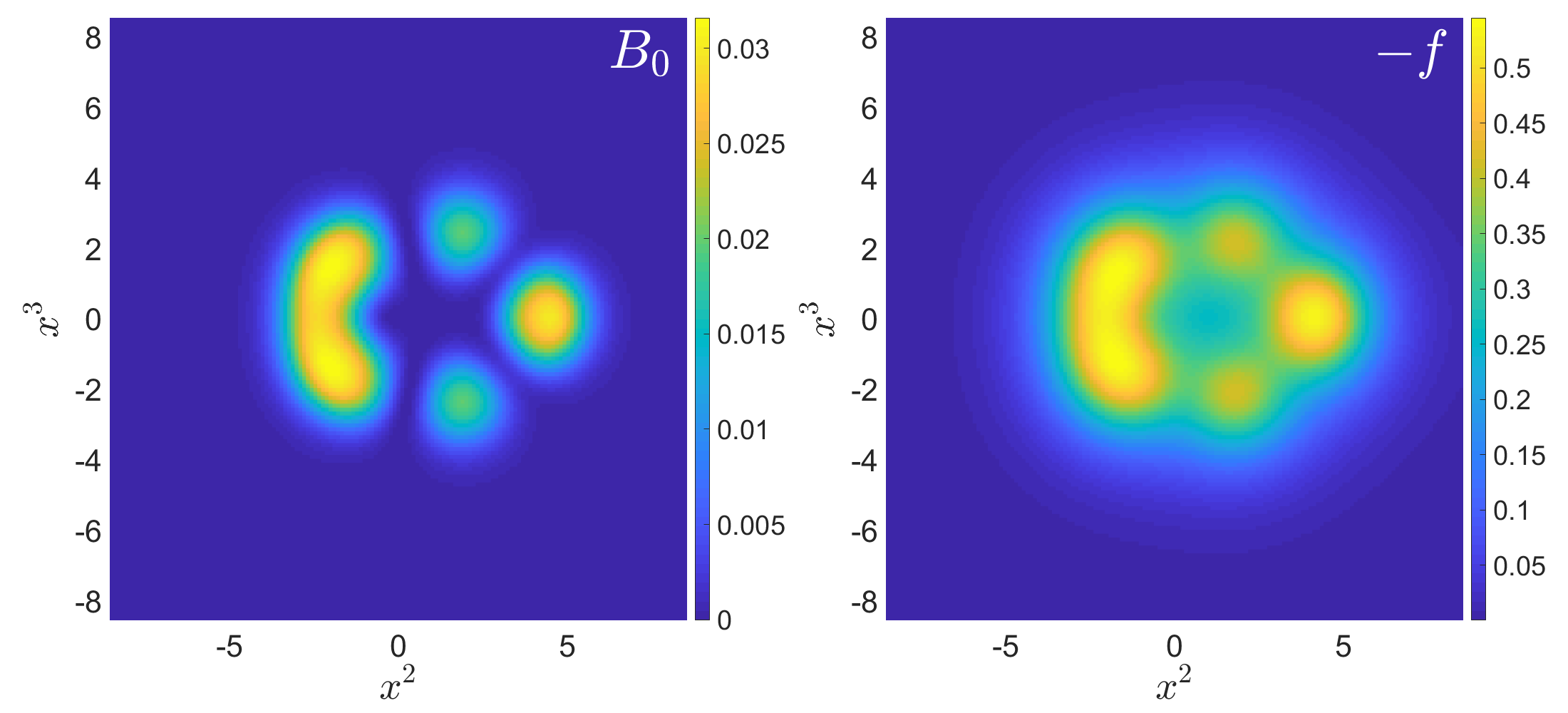}}}
    \caption{Slices of baryon charge density $B_0$ (left) compared with the
      omega meson function $f$ (right) for the Skyrmion solutions 1
      through 5 of fig.~\ref{fig:g347M0176}.}
    \label{fig:omega_baryon_slice1}
  \end{center}
\end{figure}
Although the static solutions for the pion fields $\bphi$ uniquely
determine the corresponding omega meson functions $f$ via the
constraint \eqref{static2}, it will prove instructive to look at the
difference between the baryon charge density $B_0$ and the function
$f$.
It is intuitively clear that the two quantities have some similarities
and in particular, for large enough level set, they display
surfaces of the same topology.
Of course the difference between $B_0$ and $f$ is due to the presence
of the Laplace operator in the constraint equation which
smooths out $f$ in comparison with $B_0$.
In particular, this means that the ``holes'' -- well known to reside
in Skyrmion solutions -- are filled up by said smoothing of the
Laplace operator.
This in turn has consequences for the energy density, which receives
contributions from the omega meson field $f$ and hence also is less
``hollow'' than the Skyrmion solutions in the standard Skyrme model.
Fig.~\ref{fig:omega_baryon_slice1} 
shows slices through the solutions, where each panel compares the baryon
charge density (left) and the omega meson function $f$ (right), for
all the Skyrmion solutions with $1\leq B\leq 5$. The solutions for
$B=6,7,8$ show qualitatively similar features. This ``filling in''
effect perhaps explains why the model continues to favour shell-like
fullerene structures up to values of $B$ at which such structures are
unstable in the standard Skyrme model with massive pions. 

We conclude this section by comparing our solutions, obtained by
solving the full PDE system, with the approximate solutions obtained
by Sutcliffe \cite{Sutcliffe:2008sk}. 
These latter were obtained by using
the rational map approximation for the pion field for $B=1$ through
$B=4$, where the rational maps have spherical, axial, tetrahedral and
cubic (octahedral) symmetries, respectively.
The $\omega_0=f$ field was obtained in ref.~\cite{Sutcliffe:2008sk},
by expanding it in symmetric harmonics, which are a linear combination
of the usual spherical harmonics. The expansion was further truncated
to angular quantum numbers $l\leq 10$.
Similarly, the baryon density was expanded in the same basis as the
$\omega$ meson.
This procedure led to at most 10 ODEs for the $\omega$ field and a
single ODE for the pion fields.

\begin{table}[!htp]
  \begin{center}
    \begin{tabular}{l||cc|cc}
      $B$ & $E_B$ & $E_B/E_1$ & $E_B^{\rm Sutcliffe}$ & $E_B^{\rm Sutcliffe}/E_B$\\
      \hline\hline
      1  & 22.50$\pm$0.03 & 1.000 & 22.53 & 1.001\\
      2  & 43.36$\pm$0.05 & 1.927 & 45.20 & 1.042\\
      3  & 63.53$\pm$0.08 & 2.820 & 65.88 & 1.037\\
      3b & 64.95$\pm$0.12 & 2.886 & -- & -- \\
      4  & 82.88$\pm$0.10 & 3.683 & 84.28 & 1.017
    \end{tabular}
    \caption{Comparison of the energies of the true solutions $E_B$
      for baryon numbers $B=1,2,3,4$ with the energies found in
      ref.~\cite{Sutcliffe:2008sk} using the rational map
      approximation.
    For convenience, we also display the ratio of the energies with
    respect to that of the 1-Skyrmion. }
    \label{tab:comparison}
  \end{center}
\end{table}

In the usual Skyrme model without a pion mass term, the precision of
the solutions obtained within the rational map approximation is
surprisingly good, and the energies for $B\leq 22$ are only about 1\%
higher than the energies of the true solutions (to the full PDEs), see
ref.~\cite{Houghton:1997kg}. As can be seen in tab.~\ref{tab:comparison},
the accuracy of the rational map approximation is slightly worse in
the $\omega$-Skyrme model. 
Nevertheless, for $B=1,2,3,4$ the correct symmetries were predicted
using the rational map approximation and their energies were at most
4.2\% too large compared with the true solutions. Our results should
therefore be regarded as a vindication of Sutcliffe's ingenious
approximation.

\section{Collective coordinate quantisation}\label{sec:collcoordquant}

The question remains, what value of the coupling $g$ best reproduces
the physical properties of atomic nuclei for low $B$? To answer this,
we must calibrate the model (choose its length and energy units), and
compare its data with experiment. For $B=1$, particularly, quantum
mechanical effects are an important component of these data, so we
must devise a tractable quantisation scheme for our Skyrmions. The
traditional approach is ``rigid body quantisation", in which the
action of the field theory is restricted to the spin-isospin orbit of
a degree $B$ classical energy minimiser. Recent studies of the
standard Skyrme model suggest that this is, for $B>1$, often too
restrictive: the Skyrme field should instead be restricted (for each
fixed $t$) to lie in some finite dimensional manifold $M$ of
configurations which includes the spin-isospin orbits of the global
energy minimiser and all nearby local minima, and field configurations
interpolating between these
\cite{Halcrow:2015rvz,Halcrow:2016spb,Rawlinson:2017rcq,Gudnason:2018aej,Halcrow:2019myn,Rawlinson:2019xsn}.
In general, determining $M$ is a
difficult task, more art than science at present. 
Note that by choosing $M$ to be the spin-isospin orbit of the
$B$-Skyrmion, we recover rigid body quantisation from the more general
picture.  

Let us assume that a finite dimensional manifold $M$ of static degree
$B$ field configurations has been chosen, and that $\bphi(t)$ moves
slowly in $M$. As already observed, static fields produce no source
for $\omega_X=\omega_i \d{x}^i$, so each point $\bphi$ in $M$
determines a function $f=\omega_0$, but induces no $\omega_X$. Once we
allow $\bphi(t)$ to move slowly in $M$, it produces a source for
$\omega_X$ of order $|\dot\bphi|$ so that, even in the approximation of
low velocity, the terms in $S$ involving $\omega_X$ contribute
significant terms to the Lagrangian determining slow dynamics in
$M$. This subtlety was already apparent to Adkins and Nappi
\cite{Adkins:1983nw}, although they do not give a detailed
justification of their proposed resolution of it.  

We propose the following procedure: for each $\bphi\in M$ and
$\dot\bphi\in T_\bphi M$, we take $\omega_0$ and $\omega_X$ to be the
fields determined by eq.~\eqref{eq:eomomega}. We then substitute
$\bphi$ and $\omega$ into the Lagrangian defined by $S$
(eq.~\eqref{action}), keeping only terms up to quadratic order in time
derivatives. 
This gives a Lagrangian $L|$ governing the dynamics of a point moving
in $M$ (i.e.~a slow curve of Skyrme fields) which can be quantised by
standard methods. The Lagrangian defined by $S$ of eq.~\eqref{action}
is 
\begin{equation}\label{L0}
L=\frac18\|\dot\bphi\|^2+\frac12\|\dot\omega_X\|^2-\ip{\dot\omega_X,\d\omega_0}-\frac12\|\d\omega_X\|^2-\frac12\|\omega_X\|^2-g\ip{\omega_X,B_X}-E_{\rm static},
\end{equation}
where $\ip{\cdot,\cdot}$ denotes $L^2$ inner product on $X$ and $\|\cdot\|$ the associated norm, $B_X$ denotes the spatial part of the baryon current,
$B_X=*\bphi^*\iota_{\dot\bphi}\Omega$ and
\begin{equation}
E_{\rm static}=\frac18\|\d\bphi\|^2+\int_X V(\bphi)*1-\frac12\|\d\omega_0\|^2-\frac12\|\omega_0\|^2-g\ip{\omega_0,B_0},
\end{equation}
which coincides with eq.~\eqref{Edef} in the case where
$(\bphi,\omega_0)$ is a static solution of the model. Assume now that
$\omega_X$ satisfies eq.~\eqref{eq:eomomega}. It follows immediately that
the form $\omega=\omega_0\d{t}+\omega_X$ is coclosed on $\mathcal{M}$,
and hence that 
\begin{equation}\label{q0}
\dot\omega_0+\delta\omega_X=0,
\end{equation}
where $\delta=(-1)^p*\d*$ denotes the coderivative of $p$-forms on $X$.
Furthermore, the spatial component of eq.~\eqref{eq:eomomega} reads
\begin{equation}\label{omX0}
-\ddot\omega_X+\d\dot\omega_0-\delta\d\omega_X-\omega_X=gB_X,
\end{equation}
so
\begin{equation}\label{q1}
\|\d\omega_X\|^2+\|\omega_X\|^2=\ip{\omega_X,(\delta\d+1)\omega_X}=-g\ip{\omega_X,B_X}-\ip{\omega_X,\ddot\omega_X}+\ip{\omega_X,\d\dot\omega_0}.
\end{equation}
Substituting eq.~\eqref{q1} into eq.~\eqref{L0} yields
\beq
L&=&-E_{\rm static}+\frac18\|\dot\bphi\|^2+\frac12\|\dot\omega_X\|^2-\ip{\dot\omega_X,\d\omega_0}-\frac12g\ip{\omega_X,B_X}+\frac12\ip{\omega_X,\ddot\omega_X}
-\frac12\ip{\omega_X,\d\dot\omega_0}\nonumber\\
&=&-E_{\rm static}+\frac18\|\dot\bphi\|^2+\frac12\ip{\omega_X,\d\dot\omega_0}-\frac12g\ip{\omega_X,B_X}+\frac{\d\:}{\d{t}}\ip{\omega_X,\frac12\dot\omega_X-\d\omega_0}\nonumber\\
&=&-E_{\rm static}+\frac18\|\dot\bphi|^2-\frac12\|\dot\omega_0\|^2-\frac12g\ip{\omega_X,B_X},\label{L1}
\eeq
where, in the last line, we have used eq.~\eqref{q0} and discarded the
irrelevant total time derivative.  

In principle, formula \eqref{L1} determines $L|$, the Lagrangian for
motion in $M$. Given a curve $\bphi(t)\in M$, $\omega_0(t)$ is
determined at each time $t$ by eq.~\eqref{static2}, so $\dot\omega_0$ is
determined. We work to quadratic order in time derivatives and note
that both $B_X$ and $\omega_X$ are of linear order, so only the
leading term in $\omega_X$ is required. Hence $\ddot\omega_X$ may be
discarded from eq.~\eqref{omX0} which, given eq.~\eqref{q0} reduces to 
\begin{equation}\label{omX1}
(\triangle+1)\omega_X=-gB_X=-g\bphi^*\iota_{\dot\bphi}\Omega.
\end{equation}
Then $\bphi,\dot\bphi$ uniquely determine $\omega_X$ (by solving
eq.~\eqref{omX1}), so every term in $L$ is determined by $\bphi(t)$. 

\subsection{Quantising the 1-Skyrmion}\label{sec:quantize1}

Let us apply this formalism to the motion of a $B=1$ Skyrmion, where
$M$ is its spin-isospin orbit. Then $E_{\rm static}$ is constant, and may
be discarded from $L|$. Since the unit Skyrmion is a hedgehog field,
rotation is equivalent to isorotation, and isorotation always leaves
$\omega_0$ fixed. Hence, for any curve in $M$, $\dot\omega_0=0$, and
so 
\begin{equation}
L|=\frac18\|\dot\bphi\|^2-\frac12g\ip{\omega_X,B_X},
\end{equation}
where $\omega_X$ is determined by eq.~\eqref{omX1}. To proceed
further, we must solve eq.~\eqref{omX1} approximately. For this purpose we
formally invert the operator $1+\triangle$ yielding
\begin{equation}
\omega_X=-g(1+\triangle)^{-1}B_X=-g(1-\triangle+\triangle^2-\triangle^3+\cdots)B_X.
\end{equation}
If we keep only the leading term, $\omega_X\approx-gB_X$, we obtain 
\begin{equation}\label{L3}
L|\approx\frac18\|\dot\bphi\|^2+\frac12g^2\|B_X\|^2.
\end{equation}
The curve $\bphi(t)$ takes the form
\begin{equation}
\bphi(t)=\diag(1,A(t))\bphi_{\rm H},
\end{equation} 
for some curve $A(t)\in SO(3)$, where $\bphi_{\rm H}$ is the hedgehog field 
\begin{equation}
\bphi_{\rm H}(r,\nvec)=(\cos F(r),\sin F(r)\nvec).
\label{eq:hedgehog}
\end{equation}
Hence $\dot\bphi=(0,\dot{A}\nvec)\sin F$, so
\begin{equation}\label{q4}
\|\dot\bphi\|^2=\frac{4\pi}{3}\tr(\dot A^T\dot A)\int_0^\infty \sin^2 F(r) r^2\,\d{r}.
\end{equation}
Furthermore, at the point $r\nvec\in\R^3$, 
\begin{equation}
|B_X|^2=\Omega\big(\dot\bphi,\d\bphi(E_1),\d\bphi(E_2)\big)^2+\Omega\big(\dot\bphi,\d\bphi(E_2),\d\bphi(E_3)\big)^2+\Omega\big(\dot\bphi,\d\bphi(E_3),\d\bphi(E_1)\big)^2,
\end{equation}
where $E_1,E_2,E_3$ is any orthonormal frame for
$T_{r\nvec}\R^3$. Choosing $E_1=\cd_r$, $E_2=\Yvec/r$,
$E_3=\nvec\times\Yvec/r$ where $\Yvec$ is a unit vector in $T_\nvec S^2$
one finds, after some routine algebra,  
\begin{equation}
|B_X|^2=\frac{\sin^4 F(r)}{4\pi^4r^2}F'(r)^2|\dot{A}\nvec|^2,
\end{equation}
and hence
\begin{equation}\label{q5}
\|B_X\|^2=\frac{4\pi}{3}\tr(\dot{A}^T\dot{A})\int_0^\infty\frac{\sin^4 F(r)}{4\pi^4}F'(r)^2\, \d{r}.
\end{equation}
Substituting eqs.~\eqref{q4} and \eqref{q5} into eq.~\eqref{L3} yields
\begin{equation}
L|=\frac12\Lambda\, \frac12\tr(\dot{A}^T\dot{A}),\qquad \Lambda:=\frac{2\pi}{3}\int_0^\infty\left(r^2\sin^2F(r)+\frac{g^2}{\pi^4}\sin^4F(r)F'(r)^2\right)\d{r},
\end{equation}
where the constant $\Lambda$ is the Skyrmion's moment of inertia. 

The classical dynamics determined by $L|$ is the geodesic motion on
$M\equiv SO(3)$ with respect to the metric $\gamma=\Lambda\gamma_0$,
where $\gamma_0$ is the canonical bi-invariant metric on $SO(3)$
(which on $so(3)=T_{\I_3}SO(3)$ is
$\gamma_0(Y,Z)=\frac12\tr(Y^TZ)$). To allow for fermionic
quantisation, we must lift this to the double cover $SU(2)$ of $SO(3)$
using the usual covering map $SU(2)\ra SO(3)$ defined by the adjoint
action of $SU(2)$ on $su(2)\equiv \R^3$ induced by the identification
$i(x_1\tau_1+x_2\tau_2+x_3\tau_3)\mapsto(x_1,x_2,x_3)$. This covering
map is an isometry, so the lifted metric is
$\wt\gamma=\Lambda\wt\gamma_0$ where $\wt\gamma_0$ is the round 
metric with radius 2 on $SU(2)\equiv S^3$.  
The quantum Hamiltonian for geodesic flow is
\begin{equation}
H=\frac12\triangle_{\wt\gamma}=\frac{1}{2\Lambda}\triangle_{\wt\gamma_0}=\frac{1}{8\Lambda}\triangle_{\hat{\wt\gamma}_0},
\end{equation}
where $\triangle_{\hat{\wt\gamma}_0}$ denotes the Laplacian on the unit
$3$-sphere. The spectrum of $\triangle_{\hat{\wt\gamma}_0}$ is $l(l+2)$
where $l=0,1,2,\ldots$ is physically interpreted as twice the spin
(or, equivalently isospin) of the corresponding state. Nucleons have
$l=1$ and hence the quantum correction to their total energy is 
\begin{equation}
E_1^{\rm quantum}=\frac{3}{8\Lambda}.
\label{eq:spincontrib}  
\end{equation}

\subsection{Electric charge radius}

The final phenomenological observable that we need is the electric
charge radius. Computing this will require us to consider the Noether current associated with isospin symmetry, so it is 
convenient to revert to the Lorentz covariant setting in which the Skyrme field is regarded as a map on spacetime $\bvarphi:\mathcal{M}\ra SU(2)$ (rather than a curve $\bphi(t)$ of maps $X\ra SU(2)$).
Using the Gell-mann--Nishijima relation, the electric charge $Q$ is
given by
\beq
Q = I_3 + \frac12Y,
\eeq
where $I_3$ is the isospin and $Y$ is the hypercharge which is given by
\beq
Y = B + S,
\eeq
where $B$ is the baryon number and $S$ is the strangeness quantum
number.
Since $S=0$ for Skyrmions in $SU(2)$ models (meaning 2 light flavors of
quarks), we can write the electric charge density as
\beq
\mathcal{Q} = \mathcal{I}_3 + \frac12B_0, \qquad
Q = \int_X\mathcal{Q}*1.
\eeq
We can construct the isospin density from the vectorial (Noether)
current that is defined from the vectorial (isospin) transformation
(as opposed to the axial transformation), whose infinitesimal form can
be written as
\beq
\bvarphi + \alpha\cdot\Delta\bvarphi,
\eeq
which in component form can be written as
\beq
\varphi^i + \alpha^k (\Delta^k\varphi)^i
= \varphi^i - \alpha^k\epsilon^{kij}\varphi^j,
\eeq
with $i,j,k=1,2,3$ and $\alpha^k$ being infinitesimal parameters and
$\Delta^k$ the $k$-th isospin generator.
The Noether current corresponding to the above infinitesimal
transformation is given by the 1-form
\beq
J_V^k = \frac14\d{\bvarphi}\cdot\Delta^k\bvarphi
+ g\star(\omega\wedge\bvarphi^*\iota_{\Delta^k\bvarphi}\Omega)\cdot\Delta^k\bvarphi.
\eeq
As usual with Noether currents, the time component contains the
Noether charge, once integrated.
The isospin charge density is thus proportional to
\beq
\mathcal{I}_3 \propto J_V^3(e_0),
\eeq
with
\beq
\Delta^3\bvarphi =
\begin{pmatrix}
  0 & 0 & 0 & 0 \\
  0 & 0 & -1 & 0 \\
  0 & 1 & 0 & 0 \\
  0 & 0 & 0 & 0 
\end{pmatrix}\bvarphi,
\eeq
which corresponds to the third isospin direction. 
We still have to find a proper normalisation of the current in order
to use it for the electric charge density.
Since we know that the nucleon with isospin $\pm\tfrac12$ has electric
charge $1$ and $0$, corresponding to the proton and the neutron,
respectively, we can normalise the vectorial Noether current such that
it integrates to $\pm\tfrac12$:
\beq
I_3 = \int_X\mathcal{I}_3 = \pm\frac12.
\eeq
The normalisation constant can thus be obtained simply as
\beq
\mathcal{I}_3 = \pm\frac{J_V^3(e_0)}{2\int_X J_V^3(e_0)*1}.
\eeq
Using the baryon charge density $B_0=*\bvarphi^*\Omega$ and inserting
the hedgehog Ansatz \eqref{eq:hedgehog}, we can finally write the
electric charge density as 
\beq
\mathcal{Q}_{\pm} = -\frac{\sin^2F(r)F'(r)}{4\pi^2r^2}
\pm \frac{\sin^2F(r) + g^2\pi^{-4}r^{-2}\sin^4F(r)F'(r)^2}{8\pi\int_0^{\infty}\left(r^2\sin^2F(r) + g^2\pi^{-4}\sin^4F(r)F'(r)^2\right)\d{r}},
\eeq
which can readily be checked to integrate to $1$ ($0$) for the upper
(lower) sign, corresponding to the electric charge of the proton
(neutron).
We can now define the electric charge radius as the weighted integral
\beq
r_{1,E}^2 = \int_Xr^2\mathcal{Q}_+*1
= 4\pi\int_0^{\infty}r^4\mathcal{Q}_+\,\d{r}.
\label{eq:rE}
\eeq

\section{Calibration}\label{sec:calibration}

An appealing point about the $\omega$-Skyrme theory that we study in
this paper is that it only contains 2 physical parameters:
$m\in(0,\infty)$ and $g\in(0,\infty)$.
$m$ is physically the ratio of the pion mass to the omega meson mass
$m=\frac{m_\pi}{m_\omega}$ and $g$ is a coupling constant $\beta$
multiplied by the ratio of the omega meson mass and the pion decay
constant $g=\frac{\beta m_\omega}{F_\pi}$.
$\beta$ is related to the decay $\omega\to 3\pi$ and is limited from
above by experimental data \cite{Adkins:1983nw}.
The reason the data give only an upper bound on $\beta$ is that the
calculation of the $\omega$ 
decay to 3 pions in the model does not include the resonance
$\omega\to\rho+\pi$ (since the rho meson is absent from this theory),
which enhances the decay rate. 
The upper bound calculated in Ref.~\cite{Adkins:1983nw} is
$\beta\leq 25.4$, whereas the same calculation with updated
experimental data reads $\beta\leq 23.9$, where we have used
the decay width $\Gamma(\omega\to 3\pi)\simeq 8.49$ MeV,
$m_\omega\simeq 782.65$ MeV $F_\pi\simeq 184.13$ MeV
\cite{Tanabashi:2018oca}.
In result, using the new data we get an upper bound for $g\leq 101.4$,
if we use the experimental values for $m_\omega$ and $F_\pi$.
The energy units of the model are $\frac{F_\pi^2}{m_\omega}$ and the
length units are $1/m_\omega$. 

Physically, the pions are pseudo-Nambu-Goldstone bosons of chiral
symmetry breaking in QCD and would be massless if the quarks were all
massless.
Nevertheless, this physical explanation for their small masses, puts
an upper bound on $m<1$.
Furthermore, the Skyrmions tend to destabilise for $m>1$.
However, the limit $m\to 1$ is theoretically interesting as it tends
to unbind the Skyrmions and hence lower their mutual binding
energies, which we shall see shortly.
Using the experimental values for the meson masses, $m=0.176$.

In the literature, two values of $g$ have been used: $g=98.7$
\cite{Adkins:1983nw} and $g=34.7$ \cite{Sutcliffe:2008sk}.
The former value is found by letting $F_\pi$ and $g$ be free
parameters and fit the rotational excitation energy of the Skyrmion to
the nucleon and Delta masses \cite{Adkins:1983nw}.
Fitting parameters to the Delta in Skyrme-type models, however, is
filled with subtleties \cite{Battye:2005nx,Adam:2016drk}. 
The latter value of $g$, on the other hand, is found by setting
$F_\pi$ to its experimental value (186 MeV) and fitting the $B=4$
Skyrmion mass to that of ${}^4$He \cite{Sutcliffe:2008sk}.

\subsection{Fitting the nucleon and helium-4 masses}

In this paper, we will consider the following calibration based on the
idea that in a minimalistic model like the $\omega$-Skyrme theory, we
cannot accurately describe all phenomena of hadronic and nuclear
physics with only 2 parameters over a large energy range.
Hence, if we allow to fit the parameters of the model in order to fit
baryonic quantities, disregarding the mesonic observables, then an
appropriate list of quantities to fit the model with contains the
masses of the nucleon and helium-4 as well as the size of the
nucleon.
The justification for doing so could either be that the model is
incomplete or somewhat equivalently, that the parameters in the
effective low-energy field theory have been renormalised.

The two equations for our calibration thus read
\beq
m_{{}^4{\rm He}} = \frac{F_\pi^2}{m_\omega} m_4,
\label{eq:massHe4}
\eeq
fitting the mass of helium-4 to that of the 4-Skyrmion and
\begin{equation}
m_{\rm N} = \frac{F_\pi^2}{m_\omega}
\left(m_1 + \frac{3}{8}\left(\frac{m_\omega}{F_\pi}\right)^4\frac{1}{\Lambda}\right), 
\label{eq:massN}
\end{equation}
fitting the mass of the nucleon to that of the 1-Skyrmion with the
spin quantum correction \eqref{eq:spincontrib},
where $m_B$ is the static energy of the $B$-Skyrmion.
Eq.~\eqref{eq:massHe4} does not have a quantum correction from the
spin, because the ground state of helium-4 is a spin 0, isospin 0
state. 
The factors of $F_\pi^2/m_\omega$ and $1/m_\omega$ have been
reinstated to convert to physical units (MeV). 
In principle, these two equations fix $(F_\pi,m_\omega)$ in terms of
$m_1(g,m)$, $m_4(g,m)$ and $\Lambda(g,m)$.
However, there is not always a solution, which we can see by taking
the ratio of the two equations
\beq
\frac{m_{\rm N}}{m_{{}^4{\rm He}}} =
\frac{m_1}{m_4} + 
\frac38\left(\frac{m_\omega}{F_\pi}\right)^4\frac{1}{\Lambda m_4}.
\eeq
If $\frac{m_1}{m_4}>\frac{m_{\rm N}}{m_{{}^4{\rm He}}}$ then there is no
solution because the last term in the above equation is positive
definite.  
However, if $\frac{m_1}{m_4}<\frac{m_{\rm N}}{m_{{}^4{\rm He}}}$, then
we can write this equation as 
\beq
\frac{g}{\beta}
= \frac{m_\omega}{F_\pi}
= \sqrt[4]{\frac83\Lambda m_4\left(\frac{m_{\rm N}}{m_{{}^4{\rm He}}} - \frac{m_1}{m_4}\right)},
\eeq
where we have used the definition of $g$.
Substituting back into eqs.~\eqref{eq:massHe4}-\eqref{eq:massN}, we get
\begin{align}
  F_\pi &= \frac{g}{\beta}\frac{m_{{}^4{\rm He}}}{m_4}
  = \frac{m_{{}^4{\rm He}}}{m_4}\sqrt[4]{\frac83\Lambda_1m_4\left(\frac{m_{\rm N}}{m_{{}^4{\rm He}}} - \frac{m_1}{m_4}\right)},\label{eq:fitFpi}\\
  m_\omega &= \left(\frac{g}{\beta}\right)^2\frac{m_{{}^4{\rm He}}}{m_4}
  = \frac{m_{{}^4{\rm He}}}{m_4}\sqrt{\frac83\Lambda_1m_4\left(\frac{m_{\rm N}}{m_{{}^4{\rm He}}} - \frac{m_1}{m_4}\right)}.\label{eq:fitmomega}
\end{align}
There is always a solution if
$\frac{m_1}{m_4}<\frac{m_{\rm N}}{m_{{}^4{\rm He}}}$, however, we would
additionally like the size of the nucleon to fit experimental data as
well
\beq
r_{{\rm N},E} = \frac{\hbar c}{m_\omega} r_{1,E},
\eeq
where $\hbar c\simeq 197.3$ fm MeV and the radius of the nucleon as
perceived by an electron in scattering experiments, is the electric
charge radius given in eq.~\eqref{eq:rE}.

\begin{figure}[!th]
  \begin{center}
    \includegraphics[width=0.7\linewidth]{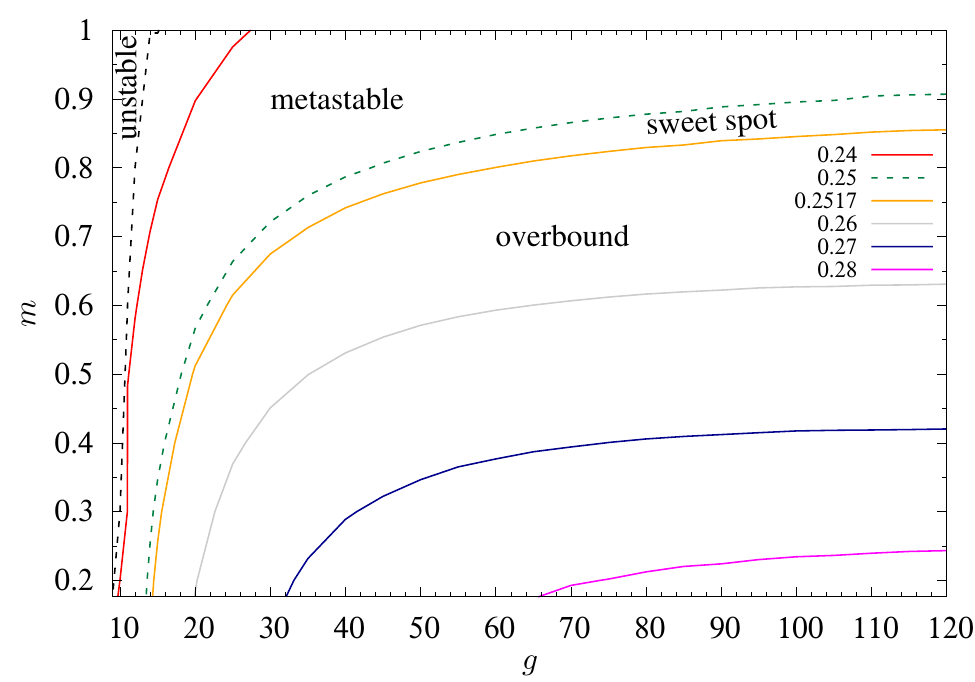}
    \caption{The ratio $\frac{m_1}{m_4}$ of the static energies of the
    1-Skyrmion to the 4-Skyrmion in the $(g,m)$ parameter space. The
    overbound region (from the orange line and below) means that the
    classical binding energy is already bigger than the physical data
    and will only be exacerbated by including the spin quantum
    correction. The metastable region (between the black and the green
    dashed lines) means that the 4-Skyrmion could gain energy
    from breaking up into 4 individual 1-Skyrmions.
    In the unstable region, the 4-Skyrmion breaks up into two
    2-Skyrmions or four 1-Skyrmions without a perturbation.
    The level sets show the value of the ratio $\frac{m_1}{m_4}$. }
    \label{fig:m1m4gM}
  \end{center}
\end{figure}

\begin{figure}[!th]
  \begin{center}
    \mbox{\includegraphics[width=0.49\linewidth]{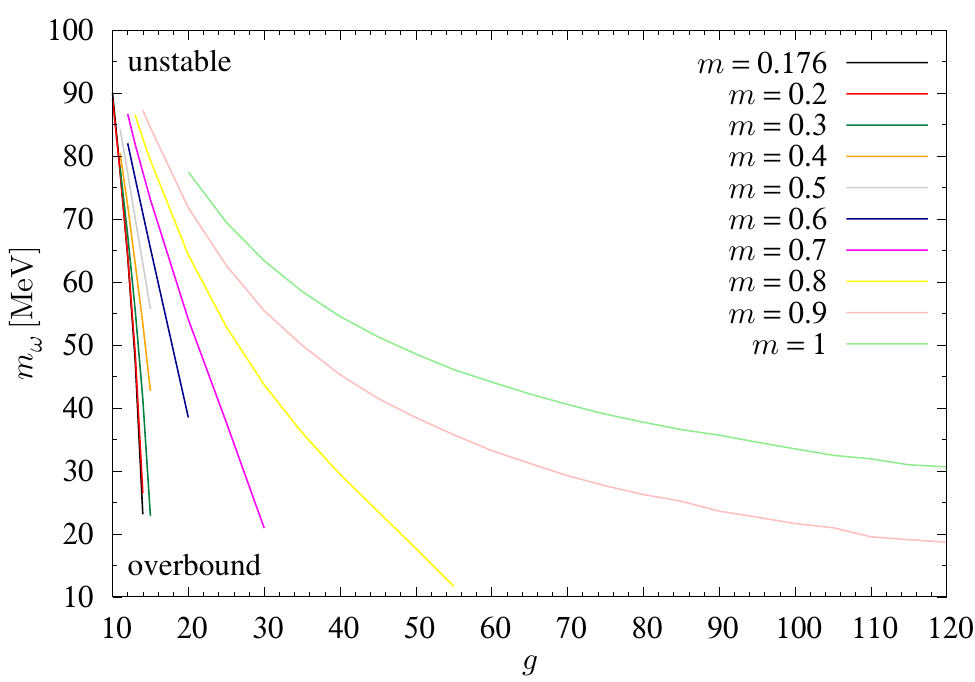}
      \includegraphics[width=0.49\linewidth]{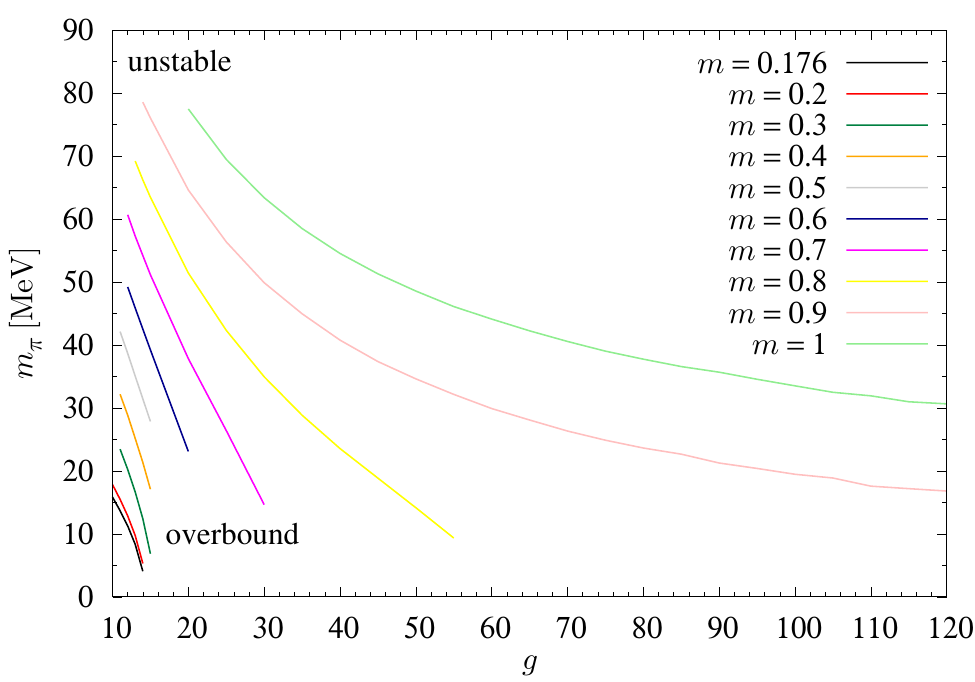}}
    \mbox{\includegraphics[width=0.49\linewidth]{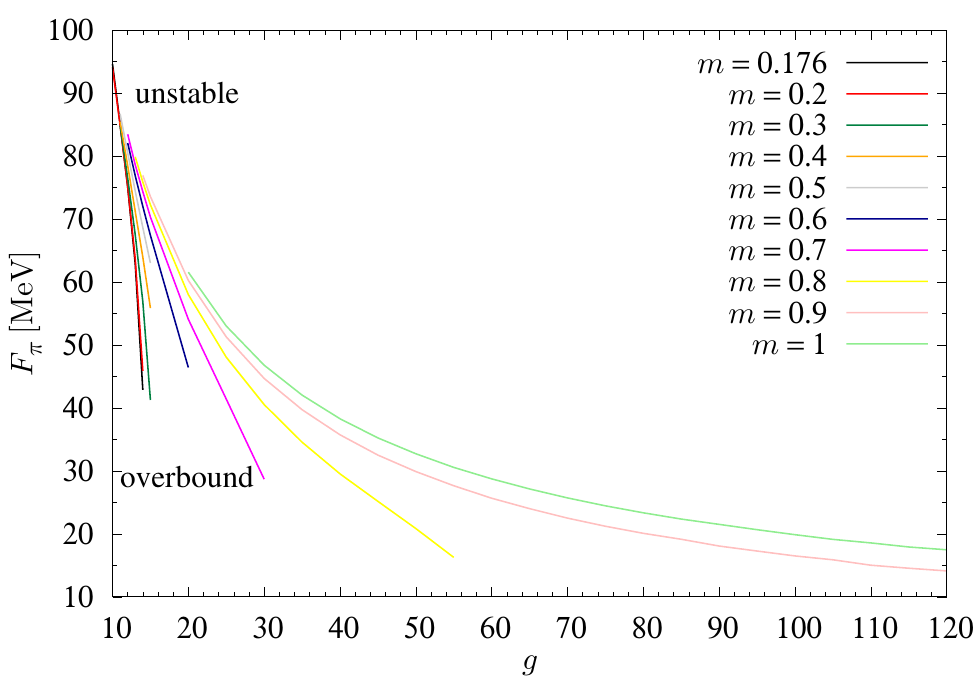}
      \includegraphics[width=0.49\linewidth]{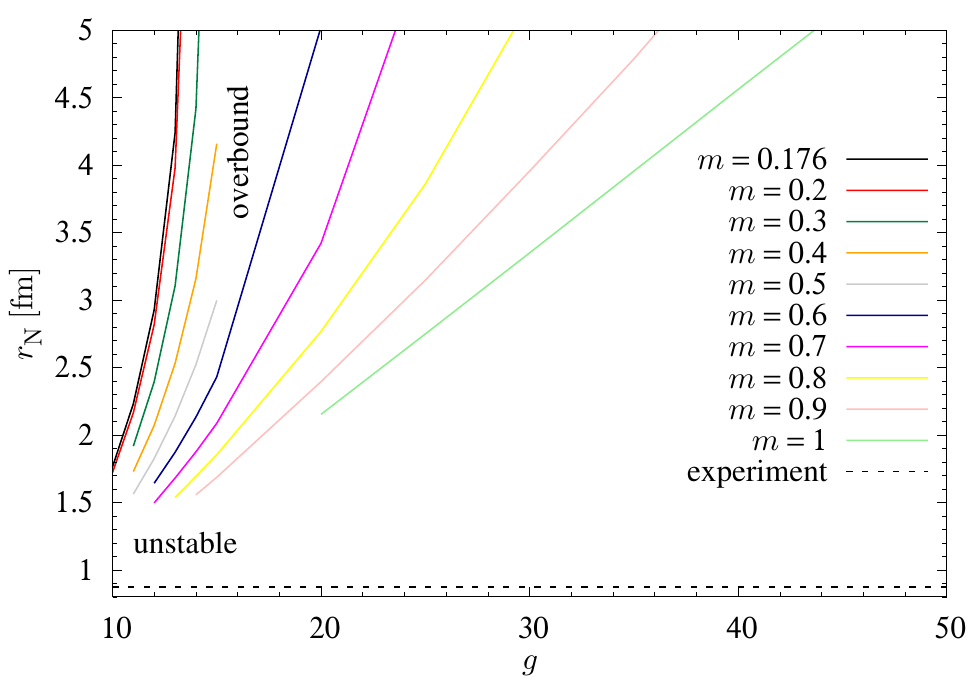}}
    \includegraphics[width=0.49\linewidth]{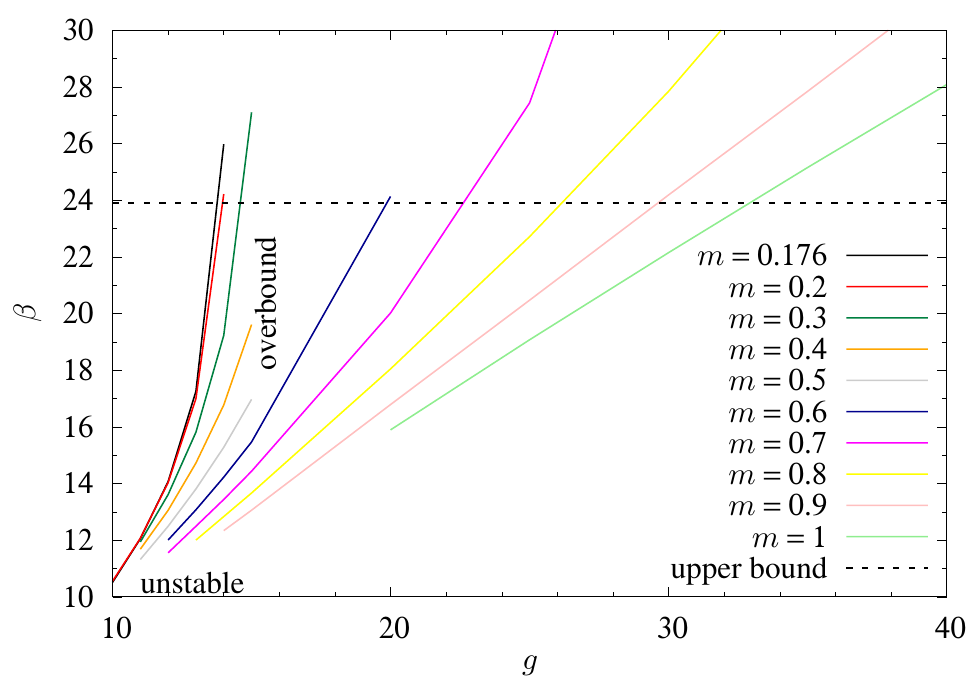}
    \caption{Solutions that fit to the nucleon mass and the helium-4
      mass.
      The panels show the omega mass $m_\omega$, the pion mass
      $m_\pi$, the pion decay constant $F_\pi$, the nucleon radius
      $r_{\rm N}$ and the coupling constant $\beta$.
      The figures for $r_{\rm N}$ and $\beta$ have been cropped so as to
      better see the viable content.
    }
    \label{fig:gM1details}
  \end{center}
\end{figure}

In order to see where we can get a solution in parameter space, we
first plot the ratio $\frac{m_1}{m_4}$ in fig.~\ref{fig:m1m4gM}.
It is possible to find a solution to
eqs.~\eqref{eq:fitFpi}-\eqref{eq:fitmomega} in the region over the
orange line in the figure.
Solutions of this type are shown in fig.~\ref{fig:gM1details}.

Fig.~\ref{fig:gM1details} shows the omega mass $m_\omega$, the pion
mass $m_\pi$, the pion decay constant $F_\pi$, the nucleon radius
$r_{\rm N}$ and the coupling constant $\beta$ between the omega meson
and the baryon current as functions of the dimensionless coupling
constant $g$ for various values of the mass ratio $m$.
First we can see that this fitting procedure yields omega masses in
the range $\sim(10,90)$ MeV, which is between 1 and 2 orders of
magnitude too small.
The largest values of the omega mass tend to prefer small values of
$g$.
The pion masses are in the range $\sim(3,80)$ MeV, which is also too
small compared with data.
The pion decay constant is in the range of $\sim(14,95)$ MeV, which is
not much worse than in many other Skyrme-like models, but still at
least a factor of 2 too small compared with data.
The nucleon radii are in the range $\sim(1.5,41)$ fm, which is at
least 71\% too large compared with data; this is the Achilles heel of
this fitting procedure.
The coupling constant $\beta$ is in the range $\sim(0.48,91)$; the
experimental upper bound is at about $23.9$ and there are many
solutions that obey this bound for $g\lesssim33$. 

The biggest issue here is that the nucleon radius (electric charge
radius) is at least 71\% too large compared with experimental data.

\subsection{Fitting the nucleon radius and the helium-4 mass}

\begin{figure}[!th]
  \begin{center}
    \mbox{\includegraphics[width=0.49\linewidth]{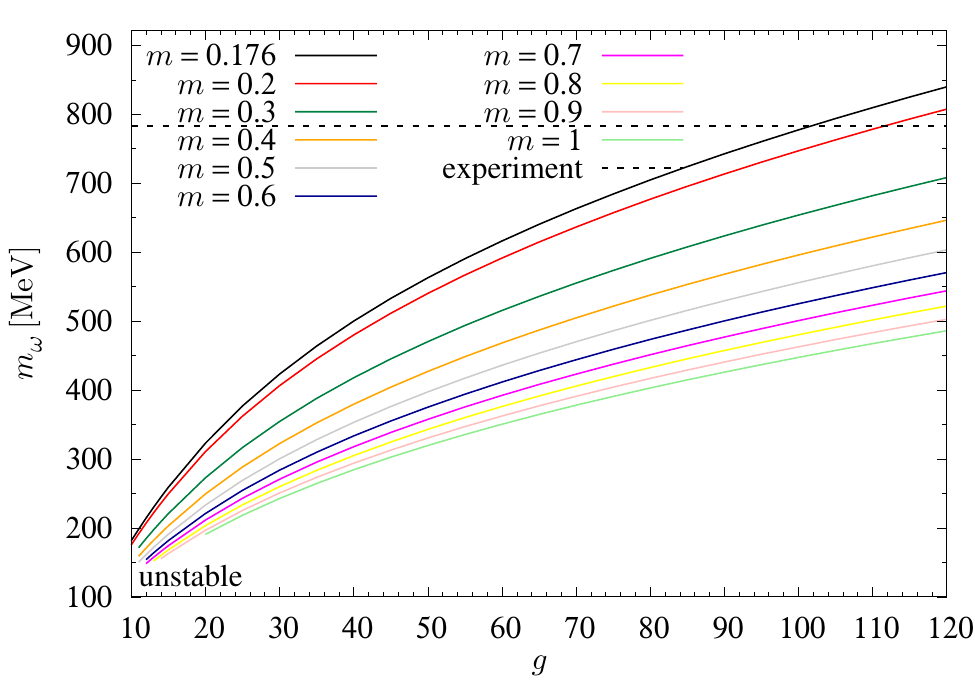}
      \includegraphics[width=0.49\linewidth]{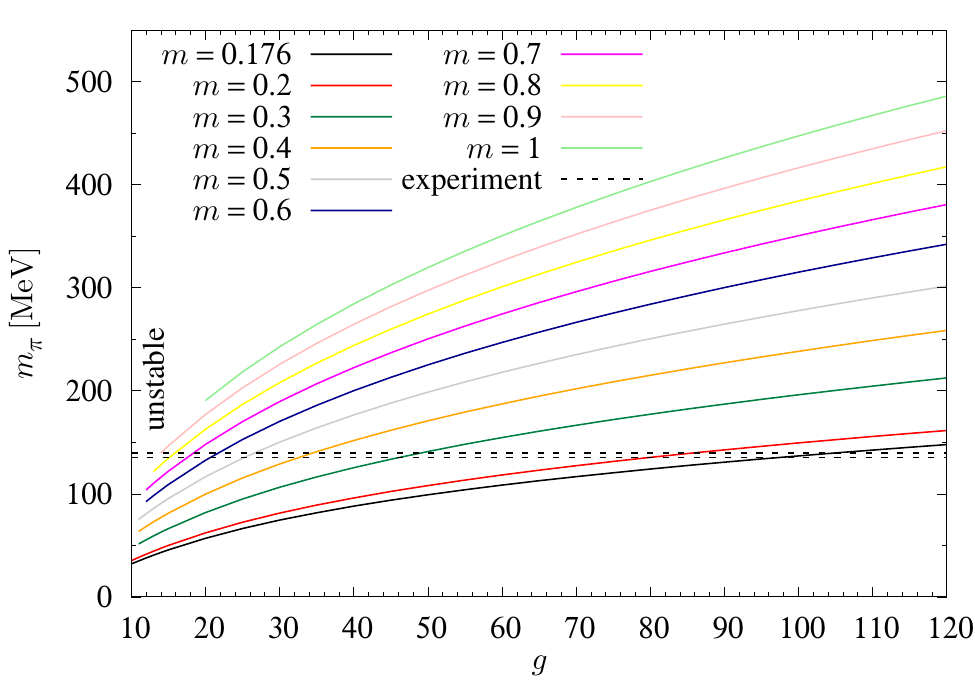}}
    \mbox{\includegraphics[width=0.49\linewidth]{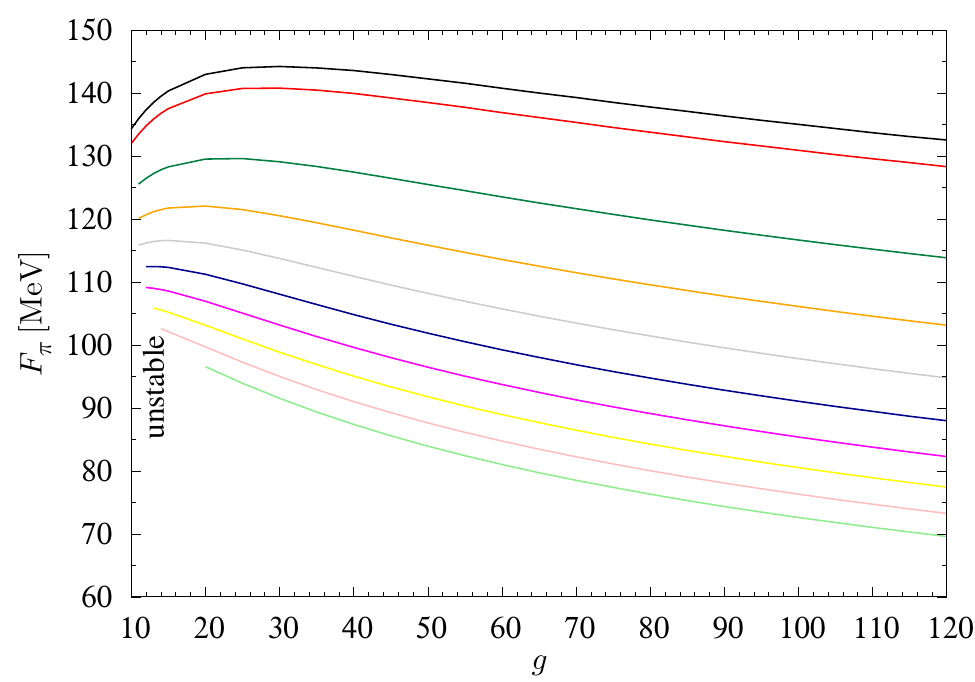}
      \includegraphics[width=0.49\linewidth]{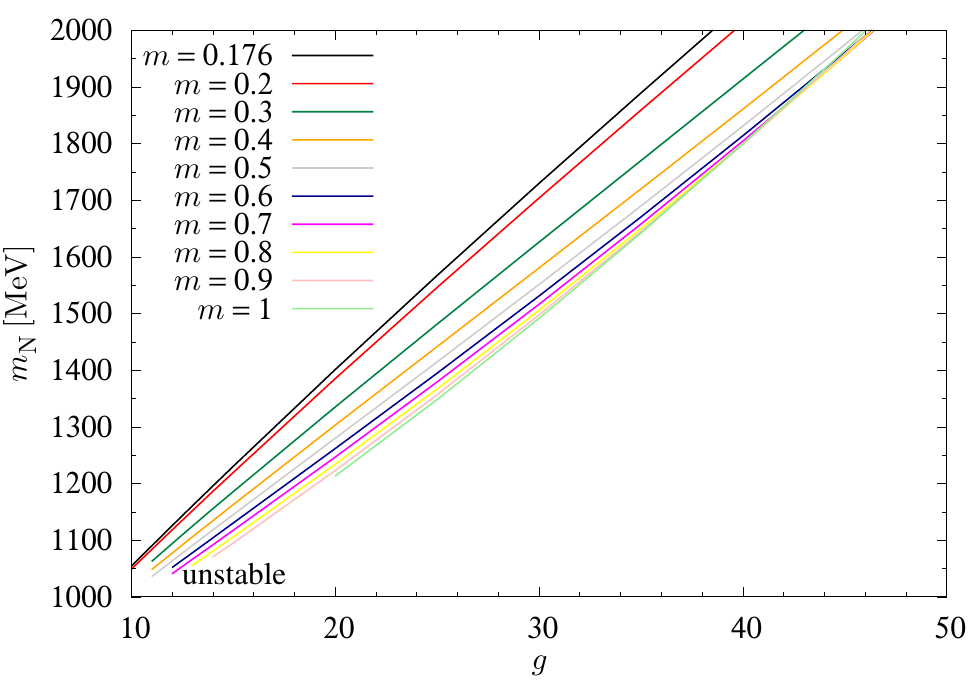}}
    \includegraphics[width=0.49\linewidth]{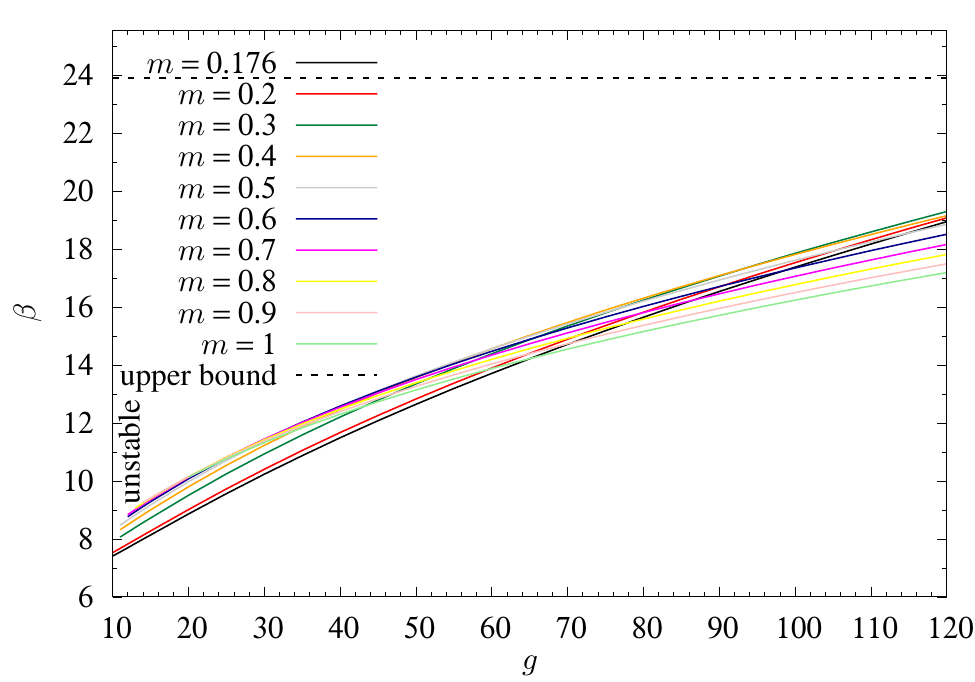}
    \caption{Solutions that fit to the nucleon radius and the helium-4
      mass.
      The panels show the omega mass $m_\omega$, the pion mass
      $m_\pi$, the pion decay constant $F_\pi$, the nucleon mass
      $m_{\rm N}$ and the coupling constant $\beta$.
      The figure for $m_{\rm N}$ has been cropped so as to better see the
      viable content. 
    }
    \label{fig:gM2details}
  \end{center}
\end{figure}

In this subsection, we will fit the size of the nucleon and the 
mass of helium-4 to experimental data. 
The mismatch that naturally will happen now is that the nucleon mass
will be larger than its experimental value.
Fig.~\ref{fig:gM2details} shows the omega mass $m_\omega$, the pion
mass $m_\pi$, the pion decay constant $F_\pi$, the nucleon mass
including the spin quantum correction $m_{\rm N}$ and finally the coupling
constant $\beta$ as functions of the dimensionless coupling constant
$g$ for various mass ratios $m$.
The omega mass is generally too small in this fitting scheme, but for
$m\lesssim2.5$ and large $g$, its experimental value can be
reproduced, but at the price of the nucleon mass being more than 5
times heavier than it should be.
The pion mass can be reproduced in this fitting procedure for
$g\lesssim 100$ for various mass ratios $m<0.9$.
The pion decay constant is generally larger in this fitting procedure
than in the latter and is in the range $\sim(70,145)$ MeV and hence
always smaller than its experimental value.
The nucleon mass is too large and in the range $\sim(1035,5045)$ MeV.
An issue is that the best values for the nucleon mass is just before
the $B=4$ Skyrmion becomes unstable; this is problematic because it is
one of the most tightly bound Skyrmions.
Finally, the coupling constant $\beta$ is in the range $\sim(7,19)$
and hence is everywhere smaller than the upper bound from pion
scattering.

Ideally we would choose a point in the model parameter space where the
nucleon mass -- including the spin quantum correction -- fits
experimental data.
Since such a point is absent from the set of solutions, we could
consider a less ambitious calibration scheme: we could continue to fit
the 4-Skyrmion mass to that of helium-4 and the size of the 1-Skyrmion
to that of the nucleon.
If we set the classical mass ratio $m_1/m_4\sim1/4$, then we are in the
right ballpark for a model with small binding energies -- provided
that the quantum corrections to each of the Skyrmions are roughly
proportional to the topological degree.
This choice corresponds to the green-dashed and the orange lines in 
fig.~\ref{fig:m1m4gM}. 
Then the nucleon mass with the spin quantum correction is off and
generally (always) too large compared with data. 
The justification of this lowering of ambition is that we do not
really expect the spin quantisation to be the only quantum correction
to the Skyrmion energies -- especially in a regime where the binding
energy is small \cite{Gudnason:2018jia}.
The latter is due to the expectation of small binding energies
yielding small vibrational frequencies \cite{Gudnason:2018ysx}.

\begin{figure}[!th]
  \begin{center}
    \mbox{\includegraphics[width=0.49\linewidth]{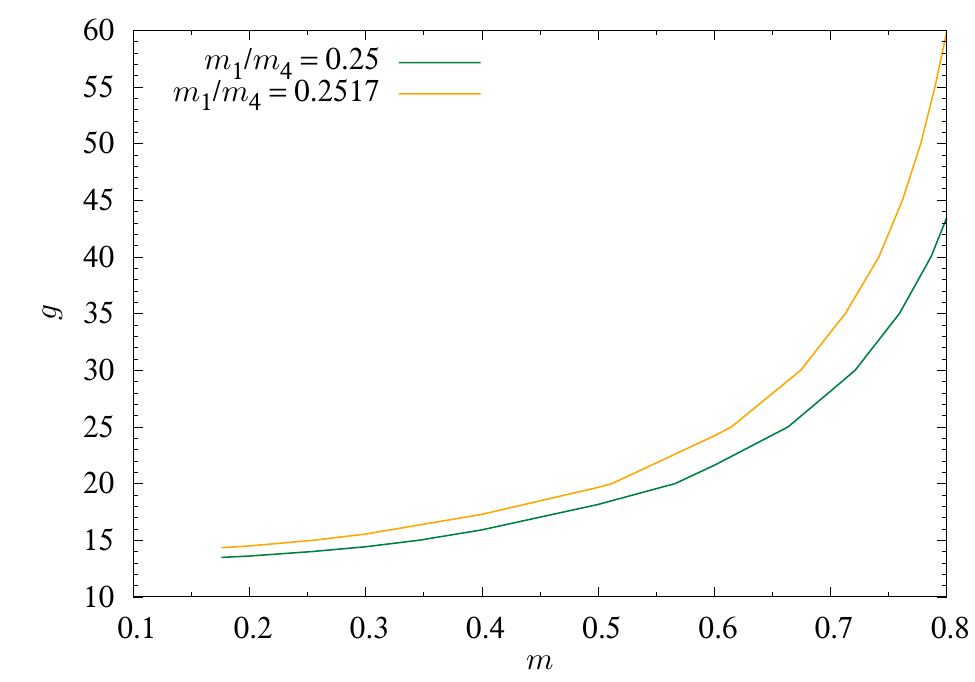}
      \includegraphics[width=0.49\linewidth]{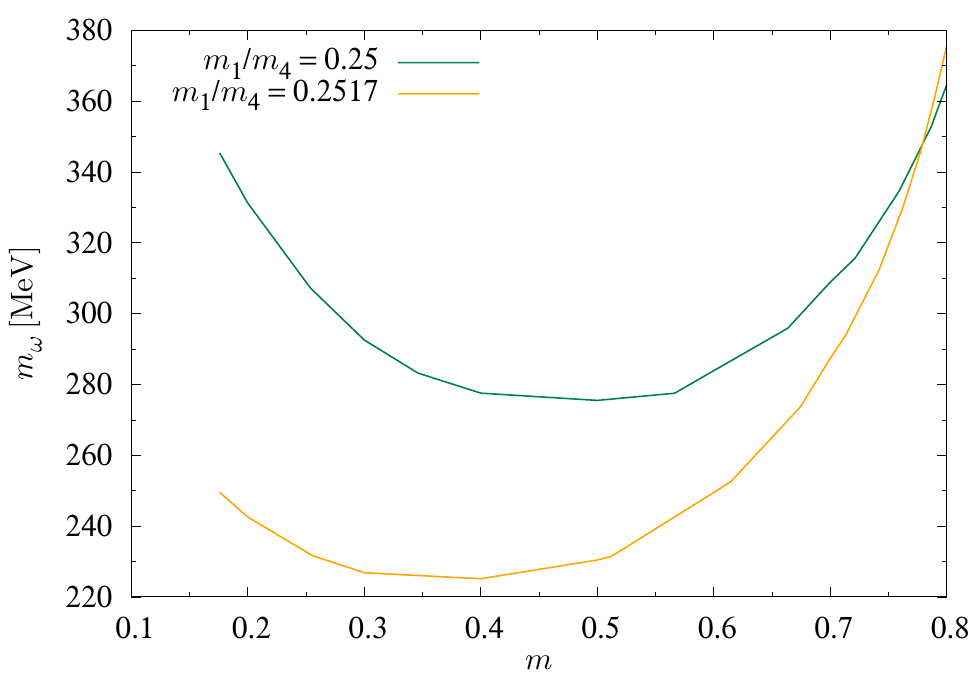}}
    \mbox{\includegraphics[width=0.49\linewidth]{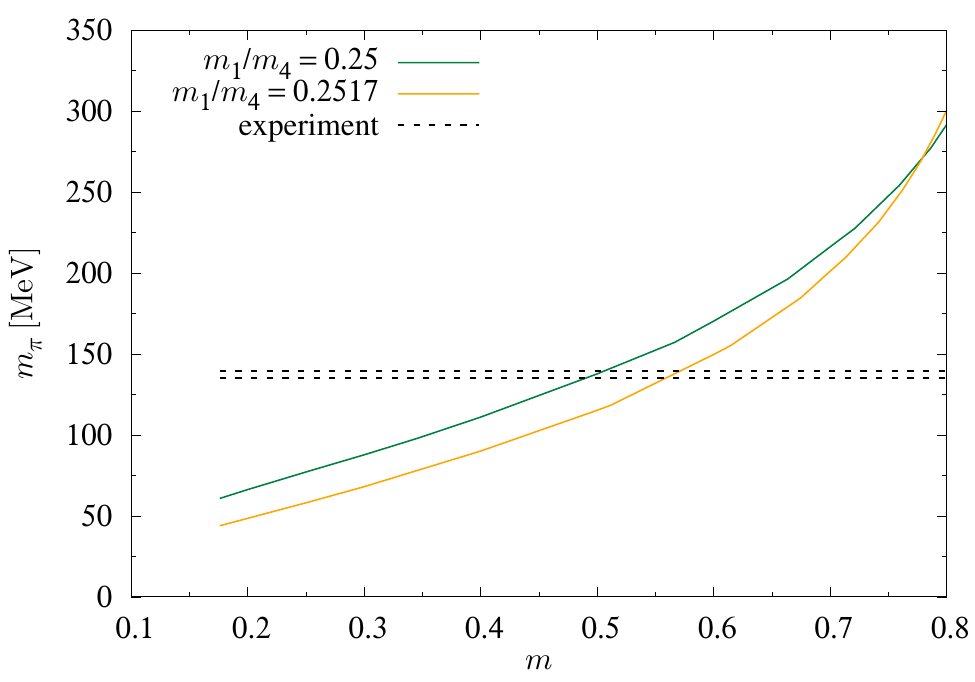}
      \includegraphics[width=0.49\linewidth]{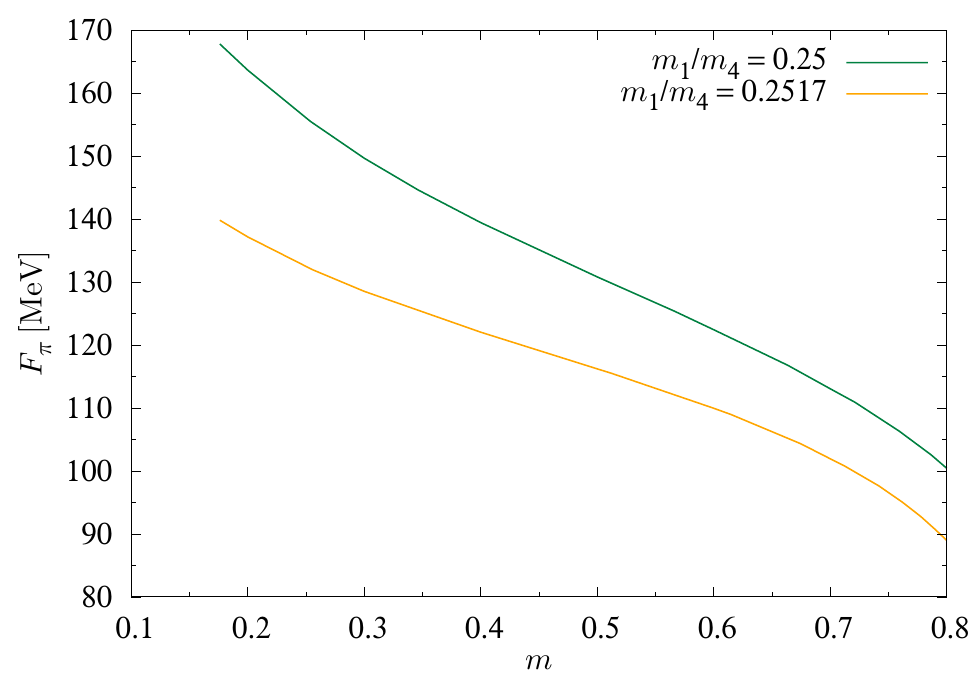}}
    \mbox{\includegraphics[width=0.49\linewidth]{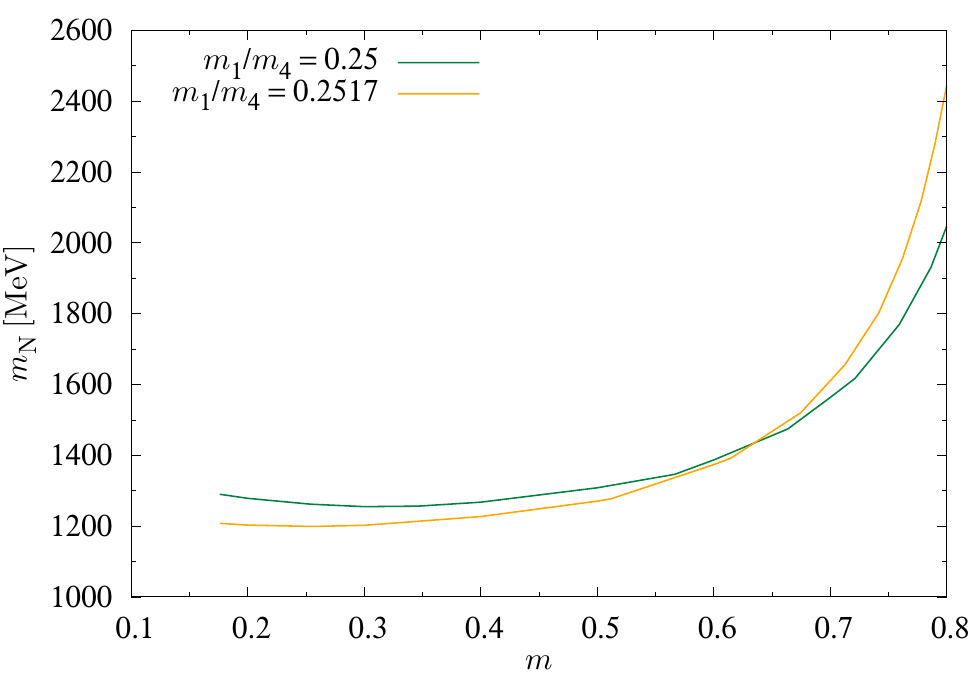}
      \includegraphics[width=0.49\linewidth]{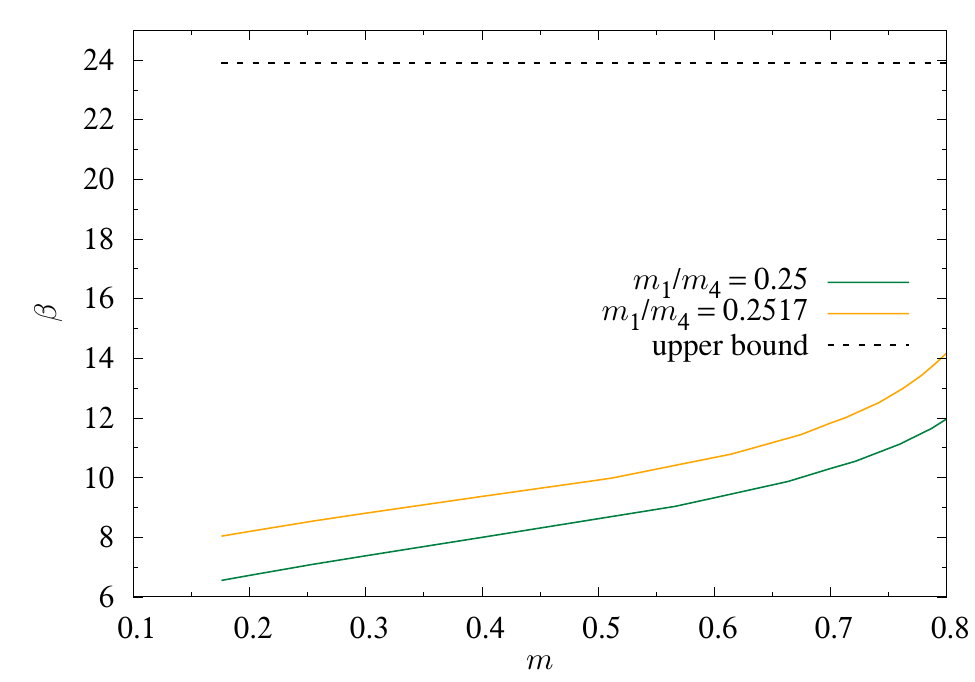}}
    \caption{Solutions that fit to the nucleon radius and the helium-4
      mass with classical mass ratios $m_1/m_4=0.25,0.2517$. 
      The panels show the coupling constant $g$, the omega mass
      $m_\omega$, the pion mass $m_\pi$, the pion decay constant
      $F_\pi$, the nucleon mass $m_{\rm N}$ and the physical coupling
      constant $\beta$.
    }
    \label{fig:gM3details}
  \end{center}
\end{figure}

Fig.~\ref{fig:gM3details} shows the coupling constant $g$, the omega
mass $m_\omega$, the pion mass $m_\pi$, the pion decay constant
$F_\pi$, the nucleon mass $m_{\rm N}$ and finally, the physical
coupling constant $\beta$ as functions of the mass ratio $m$.
The omega mass is generally too small and is smallest near
$m\sim0.5$ ($m\sim0.4$) for $m_1/m_4=0.25$ ($m_1/m_4=0.2517$). 
The pion mass naturally grows with $m$ and passes through its
experimental value(s) (there are two, because due to isospin breaking,
the charged pions are heavier than the neutral one).
The pion decay constant is always too small (but not too much for
small $m$ and $m_1/m_4=0.25$).
The nucleon mass is too large throughout the series of solutions and
has a minimum at $m\sim0.3$ ($m\sim0.25$) for $m_1/m_4=0.25$
($m_1/m_4=0.2517$).
The physical coupling constant $\beta$, is quite a bit smaller than
its upper bound and it grows monotonically with $m$.

Since there is no perfect data point (because the nucleon mass with
the quantum spin correction is always too large), we will select a
point in the parameter space as follows.
We notice that although the minimum of the nucleon mass is around
$m\sim 0.25$, there is a plateau in the curve for $m\lesssim 0.4$,
whereas both the omega mass and the pion decay constant are improved
with respect to their experimental data by lowering $m$ to $m=0.176$.
This data point is thus at $m=0.176$ and $g=14.34$ for the
$m_1/m_4=0.2517$ series of solutions.
For this point in parameter space, we have: the omega mass
$m_\omega=249.5$ MeV, the pion mass $m_\pi=43.91$ MeV, the pion decay
constant $F_\pi=139.8$ MeV, the nucleon mass $m_{\rm N}=1207$ MeV and
finally the physical coupling $\beta=8.036$.
Of course, by the definition of the fitting scheme, we also have
$r_{{\rm N},E}=0.875$ fm and $m_{{}^4{\rm He}}=3727$ MeV, which are the
experimental values for the electric charge radius and the
4-Skyrmion's mass. 

We will present numerical solutions for $g=14.37$, $m=0.176$ in the
next section. As we will see, they exhibit some striking differences
from those obtained previously for Sutcliffe's coupling $g=34.7$. (The
situation for the Adkins-Nappi coupling $g=98.7$ is rather similar to
$g=34.7$).

\section{Numerical solutions}\label{sec:numsols}

We present numerical solutions for the $\omega$-Skyrme model with
topological degrees 1 through 8, corresponding to the light nuclei.
The solutions are shown for $g=14.34$ and $m=0.176$ and the detailed 
observables are given at the end of the last section. 

For the multi-Skyrmion solutions, we begin the numerical calculations
with initial configurations which are all made up of 1-Skyrmions
placed in various random spatial patterns -- generally rotated so as
to attract each other. The existence of an attractive channel for
$m<1$ follows from a point source analysis whose details we postpone
until next section. 
The numerical method described in sec.~\ref{sec:nummet} then evolves
the initial configuration using the arrested Newton flow until a
local minimum of the energy functional has been obtained. 

\begin{figure}[!th]
  \begin{center}
    \includegraphics[width=\linewidth]{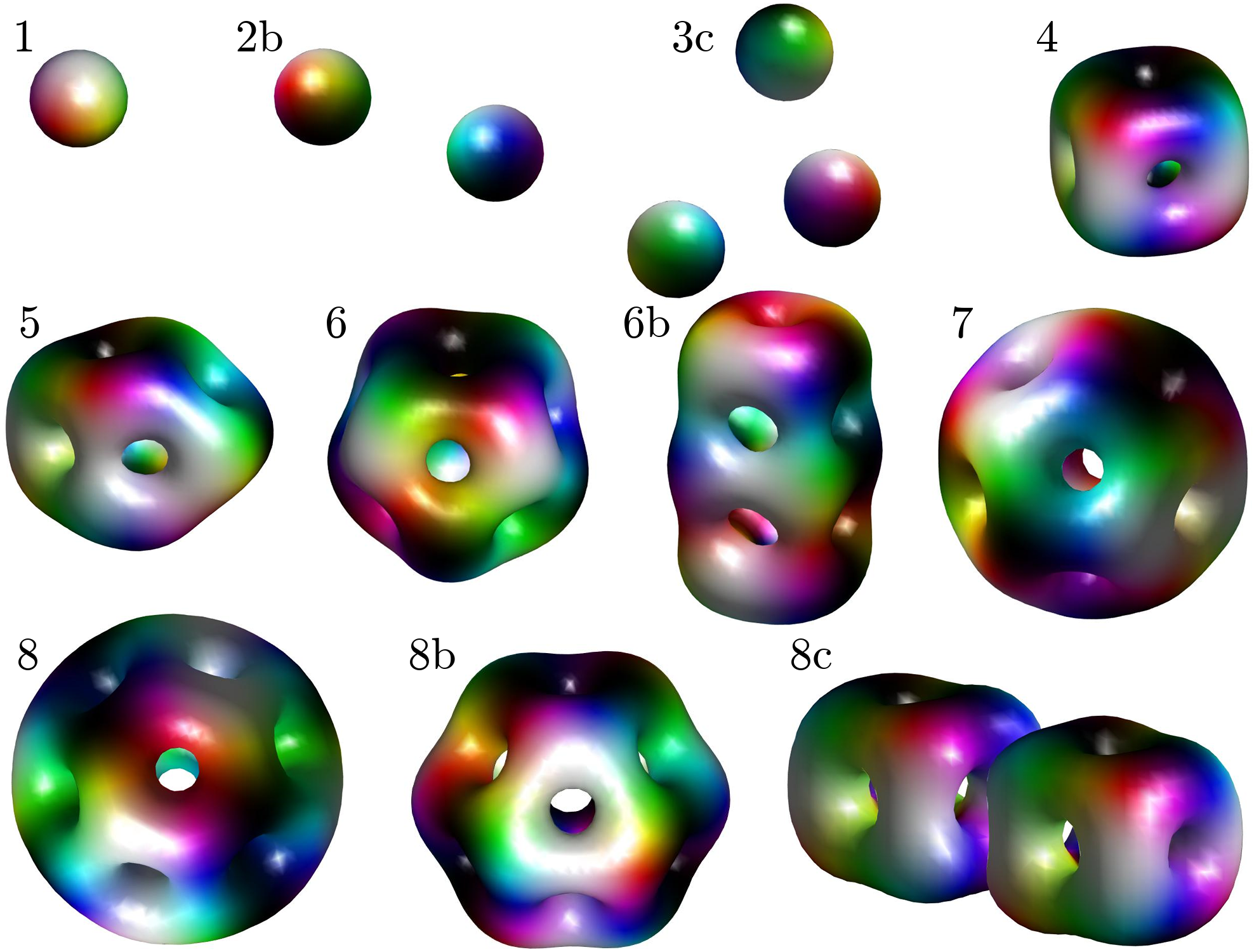}
    \caption{Numerical solutions for baryon numbers $B=1$ through
      $B=8$.
      The $B=2,3,8{\rm c}$ solutions are delocalised but bound states.
      The labels are kept the same as used in
      sec.~\ref{sec:nummet}.
      The stable solutions appear first (left-most) and the metastable
      solutions have increasing energy in order of appearance. 
    }
    \label{fig:g1434M0176}
  \end{center}
\end{figure}

Fig.~\ref{fig:g1434M0176} shows the numerically obtained multi-Skyrmion
solutions for $B=1$ through $B=8$.
Obviously the $B=1$ Skyrmion is a spherically symmetric solution. 
The first surprise is that the $B=2$ and $B=3$ solutions are
delocalised bound states for the chosen calibration. Some insight into
this phenomenon will be gained  
from a study of the inter-Skyrmion interaction energy. 
The obtained solutions are similar to those found in the
point-particle model
\cite{Gillard:2015eia,Gillard:2016esy}\footnote{The point-particle
  Skyrmion solutions also appear naturally in the holographic
  Witten-Sakai-Sugimoto model in the limit of strong 't Hooft coupling
  \cite{Baldino:2017mqq}. }.
The remaining Skyrmion solutions with $B=4$ through $B=8$ are very
similar to those found in
sec.~\ref{sec:nummet} for $g=34.7$ (the
Sutcliffe coupling), showing some universal features of the solutions.
Briefly, the $B=4$ Skyrmion has octahedral symmetry, the $B=5$
Skyrmion has dihedral symmetry, the $B=6$ Skyrmion has dihedral
symmetry, the $B=6b$ Skyrmion is metastable and composed of three
tori, the $B=7$ Skyrmion has icosahedral symmetry, the $B=8$ Skyrmion
is $D_{6d}$ symmetric whereas the $B=8b$ is only $D_6$ symmetric.
Finally the $B=8c$ Skyrmion is similar to that of
sec.~\ref{sec:nummet}, i.e.~composed by two
cubes sitting next to each other.
However, for this value of the coupling, $g=14.34$, the two cubes have
repelled themselves to become a bound state of separated $B=4$
cubes. 

To summarise, the solutions for $B=2,3$ are like in the point-particle
models, whereas the remaining solutions are qualitatively similar to solutions of the 
standard Skyrme model without pion mass.

\begin{table}[!htp]
\begin{center}
\begin{tabular}{l||l|llll|llll}
&& \multicolumn{4}{c}{$g=14.34$} & \multicolumn{4}{c}{$g=34.7$}\\
\hline  
$B$ & Sym    & $E$   & $E$   & BEPN  & QBEPN & $E$   & $E$   & BEPN  & QBEPN\\
    &        &       & [MeV] & [MeV] & [MeV] &       & [MeV] & [MeV] & [MeV]\\
\hline\hline  
1   & $O(3)$ & 11.98 & 938   & 0     & 22.5  & 22.50 & 1012  & 0     & 36.7\\
2   & $T^2$  & --    & --    & --    & --    & 43.36 & 1950  & 37.0  & 73.6\\
2b  & $D_2$  & 23.88 & 1871  & 2.6   & 25.1  & --    & --    & --    & --\\
3   & $T_d$  & --    & --    & --    & --    & 63.53 & 2857  & 59.6  & 96.3\\
3b  & ?      & --    & --    & --    & --    & 64.95 & 2921  & 38.5  & 75.1\\
3c  & $C_3$  & 35.74 & 2800  & 5.0   & 27.5  & --    & --    & --    & --\\
4   & $O_h$  & 47.57 & 3727  & 6.5   & 29.0  & 82.88 & 3727  & 80.2  & 116.9\\
5   & $D_{2d}$& 59.49 & 4661  & 6.0   & 28.5  & 103.25& 4643  & 83.4  & 120.1\\
6   & $D_{4d}$& 71.05 & 5567  & 10.4  & 32.9  & 122.71& 5518  & 92.3  & 129.0\\
6b  & ?      & 71.26 & 5583  & 7.7   & 30.2  & 123.18& 5539  & 88.8  & 125.4\\
7   & $Y_h$  & 82.39 & 6455  & 16.1  & 38.6  & 141.77& 6375  & 101.2 & 137.9\\
8   & $D_{6d}$& 94.22 & 7382  & 15.5  & 38.0  & 161.94& 7282  & 101.7 & 138.4\\
8b  & $D_6$  & 94.26 & 7385  & 15.1  & 37.6  & 162.40& 7303  & 99.1  & 135.8\\
8c  & ?      & 94.92 & 7437  & 8.7   & 31.2  & 163.60& 7357  & 92.3  & 129.0
\end{tabular}
\caption{Energies of the numerical solutions for two values of the
  coupling, $g=14.34$ and $g=34.7$. 
  The column 'Sym' shows the symmetry group of the Skyrmion solution,
  if known.
  The columns for each value of the coupling represent the energy in
  Skyrme units, the energy in MeV, the binding energy per nucleon
  (BEPN) in MeV and the quantum binding energy per nucleon (QBEPN) in
  MeV.
  The mass ratio is $m=0.176$.
}
\label{tab:energies}
\end{center}
\end{table}

\begin{figure}[!ht]
  \begin{center}
    \includegraphics[width=0.7\linewidth]{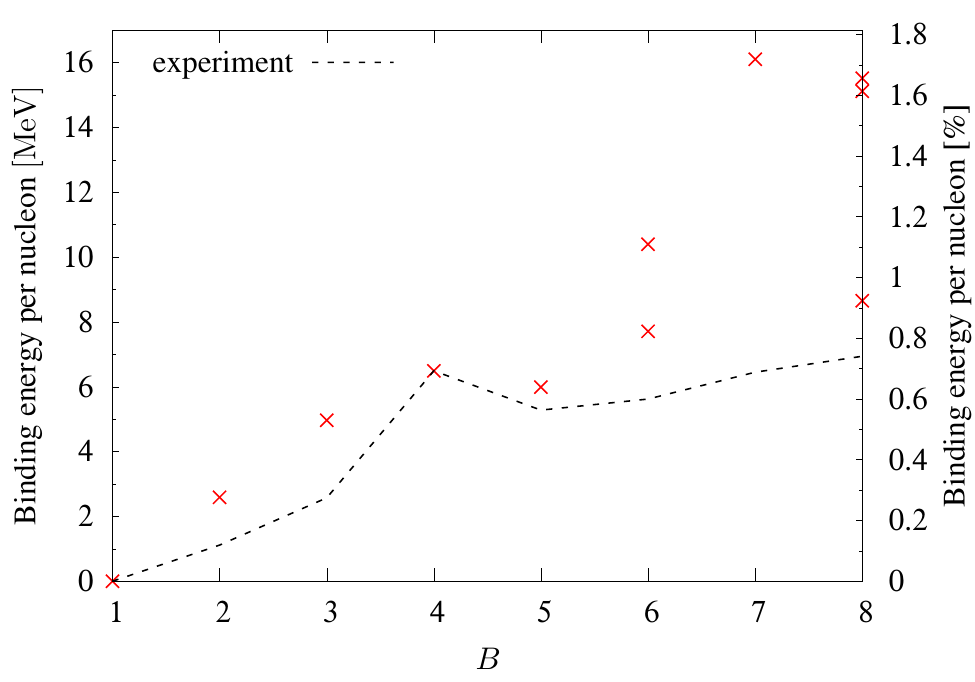}
    \caption{The classical binding energies of the numerical
      multi-Skyrmion solutions for the (first) calibration point,
      compared with experimental data.
    }
    \label{fig:bepn}
  \end{center}
\end{figure}

We provide the energies in Skyrme units and in physical units for
all solutions for $g=14.34$ and $g=34.7$ (see
sec.~\ref{sec:nummet}) in
tab.~\ref{tab:energies}. 
Finally, we illustrate the classical binding energies for our
calibration (i.e.~with $g=14.34$) compared with experimental data in 
fig.~\ref{fig:bepn}.

\section{Inter-Skyrmion forces}\label{sec:scatpot}

In this section we will compute the forces between widely separated
1-Skyrmions using a point-source formalism. This formalism was
developed for the conventional massless Skyrme model by Schroers
\cite{Schroers:1993yk}, and will require two modifications to deal
with the omega-meson version of the model studied here: the pion field
is massive, and we must introduce point sources to replicate the
Skyrmion's asymptotic $\omega$ field. Although the extra forces
induced by this field are subleading if the pion to $\omega$ mass
ratio is given its physical value, $m=0.176$, it is instructive to
include them, and to consider the (unphysical) regime where
$m\approx 1$, so, to begin with, we keep $m$ general. 

The starting point is to observe that the 1-Skyrmion takes hedgehog form
\beq
\bphi(\xvec)=\left(\cos F(r),\sin F(r)\frac{\xvec}{r}\right), \qquad
\omega_0(\xvec)=f(r),\qquad \omega_i=0,
\eeq
where $r=|\xvec|$ and the profile functions $F,f$ satisfy the coupled ODE system
\beq
-F''(r)
-\frac{2}{r}F'(r)
+\frac{\sin 2F(r)}{r^2}
+m^2\sin F(r)
&=& \frac{2g}{\pi^2}\frac{f'(r)\sin^2 F(r)}{r^2}, \nonumber \\
\label{hhnonlin}
-f''(r)-\frac{2}{r}f'+f&=&\frac{g}{2\pi^2}\frac{F'(r)\sin^2 F(r)}{r^2},
\eeq
subject to the boundary conditions $F(0)=\pi$, $f'(0)=0$, $F(\infty)=0$, $f(\infty)=0$. Of particular interest is its asymptotic form for large $r$. Since $F,f$ are small at large $r$, we assume they are close to solutions of the linearisation of this ODE system about $(F,f)=(0,0)$,
\beq
-F''(r)
-\frac{2}{r}F'(r)
+\frac{2F(r)}{r^2}
+m^2F(r)
&=& 0, \nonumber \\
-f''(r)-\frac{2}{r}f'+f&=&0,
\eeq
from which we deduce that
\beq
F(r)\sim -p\frac{\d{\:}}{\d{r}}\left(\frac{e^{-mr}}{4\pi r}\right),\qquad f(r)\sim q\frac{e^{-r}}{4\pi r},
\eeq
where $p,q$ are some unknown constants which can be determined by solving the nonlinear system \eqref{hhnonlin} numerically. The factors of $4\pi$ are introduced for later convenience. 

The corresponding asymptotic pion and $\omega$ fields are
\beq
\pi_a=-p\frac{\cd\: }{\cd x_a}\left(\frac{e^{-mr}}{4\pi r}\right),\qquad (\omega_0,\omega_i)=\left(q\frac{e^{-r}}{4\pi r},0\right).
\eeq
These coincide precisely with the solution of the {\em linearisation} of our model about the vacuum $\bphi=(\sigma,\bpi)=(1,\zerovec)$, $\omega_\mu=0$,
\beq\label{Laglin}
{\cal L}_{\rm lin}=\frac18\cd_\mu\pi_a\cd^\mu\pi_a-\frac18m^2\pi_a\pi_a+\frac14\rho_a\pi_a-\frac14\omega_{\mu\nu}\omega^{\mu\nu}+\frac12\omega_\mu\omega^\mu-j_\mu\omega^\mu,
\eeq
in the presence of the external point sources
\beq
\rho_a=-p\cd_a\delta^{(3)}(\xvec),\qquad
(j_0,j_i)=\big(q\delta^{(3)}(\xvec),0\big).
\eeq
Viewed from afar, therefore, the 1-Skyrmion looks like a point particle in a linear field theory consisting of three
uncoupled scalar boson fields of mass $m$ (the pions) and a single vector boson of mass $1$ (the $\omega$). This point particle carries three orthogonal scalar dipole moments $\pvec_a=p\evec_a$, inducing the pion fields, and a vector monopole charge $q$ inducing the $\omega_0$ field. It has no vector current density ($j_i=0$) so (or rather, because) the point Skyrmion has no $\omega_i$ field. This is the point Skyrmion in standard position (located at $\xvec=\zerovec$) and orientation. We may obtain the general point Skyrmion by translation and rotation (or isorotation, since these coincide within the
hedgehog Ansatz).

Since the 1-Skyrmion is asymptotically indistinguishable from a point
Skyrmion inducing fields in the linearised model \eqref{Laglin}, and
physics should be model independent, we assume that the {\em forces}
between well-separated 1-Skyrmions approach those between
well-separated point Skyrmions interacting via the Lagrangian
\eqref{Laglin}, as their separation grows. Consider the case where the
1-Skyrmions are static and located at $\Xvec^{(1)}$, $\Xvec^{(2)}$ and
have been (iso)rotated through $\Rrot^{(1)},\Rrot^{(2)}\in SO(3)$
respectively. Then the corresponding sources are ($\alpha=1,2$), 
\beq
\rho_a^{(\alpha)}=-p\Rrot^{(\alpha)}_{ab}\cd_b\delta^{(3)}\big(\xvec-\Xvec^{(\alpha)}\big),\qquad
\big(j_0^{(\alpha)},j_i^{(\alpha)}\big)=\left(q\delta^{(3)}\big(\xvec-\Xvec^{(\alpha)}\big),0\right),
\eeq
which induce fields
\beq
\pi_a^{(\alpha)}=-p\Rrot^{(\alpha)}_{ab}\cd_b\left(\frac{e^{-m|\xvec-\Xvec^{(\alpha)}|}}{4\pi|\xvec-\Xvec^{(\alpha)}|}\right),\qquad
\big(\omega_0^{(\alpha)},\omega_i^{(\alpha)}\big)=\left(q\frac{e^{-|\xvec-\Xvec^{(\alpha)}|}}{4\pi|\xvec-\Xvec^{(\alpha)}|},0\right).
\eeq
The interaction Lagrangian corresponding to this configuration of fields is
\beq
L_{\rm int}=\int_{\R^3}\left({\cal L}_{\rm lin}^{(1)+(2)}-{\cal L}_{\rm lin}^{(1)}-{\cal L}_{\rm lin}^{(2)}\right),
\eeq
where ${\cal L}_{\rm lin}^{(\alpha)}$ is the Lagrangian density
evaluated for field and source $\alpha$, and
${\cal L}_{\rm lin}^{(1)+(2)}$ is evaluated for their linear
superposition. Since $(\pi_a^{(\alpha)},\omega_\mu^{(\alpha)})$
satisfies the Euler-Lagrange equation for ${\cal L}_{\rm lin}$ with
source $(\rho_a^{(\alpha)},j_\mu^{(\alpha)})$ we find, after an
integration by parts, that 
\beq
L_{\rm int}&=&\int_{\R^3}\left(\frac14\rho_a^{(1)}\pi_a^{(2)}-j_\mu^{(1)}\omega^\mu_{(2)}\right)\nonumber \\
&=&\frac{p^2}{4}\Rrot_{ab}^{(1)}\Rrot_{ac}^{(2)}\frac{\cd\: }{\cd X_c^{(2)}}\frac{\cd\: }{\cd X_b^{(1)}}
\frac{e^{-m|\Xvec^{(1)}-\Xvec^{(2)}|}}{4\pi|\Xvec^{(1)}-\Xvec^{(2)}|}
-q^2\frac{e^{-|\Xvec^{(1)}-\Xvec^{(2)}|}}{4\pi|\Xvec^{(1)}-\Xvec^{(2)}|}
\eeq
Let us define the relative position $\Rvec$ and orientation $\Orot$ of Skyrmion 2 with respect to Skyrmion 1,
\beq
\Rvec=\Xvec^{(2)}-\Xvec^{(1)},\qquad \Orot=[\Rrot^{(1)}]^T\Rrot^{(2)}.
\eeq
Then the interaction {\em potential}, according to our point source model, is
\beq
V_{\rm int}=-L_{\rm int}
=\frac{p^2}{4}\Orot_{bc}\frac{\cd^2\: }{\cd R_b\cd R_c}\left(\frac{e^{-mR}}{4\pi R}\right)+q^2\frac{e^{-R}}{4\pi R},
\eeq
which can be written explicitly as
\beq
V_{\rm int}=
\frac{p^2 e^{-m R}}{16\pi R}\left[
  \left(m^2 + \frac{3m}{R} + \frac{3}{R^2}\right)
  \wh{\Rvec}\cdot\Orot\wh{\Rvec}
  -\left(\frac{m}{R} + \frac{1}{R^2}\right)\tr\Orot
  \right]
+\frac{q^2 e^{-R}}{4\pi R},
\eeq
where $\wh{\Rvec}\equiv\Rvec/R$.

\begin{figure}[!htp]
  \begin{center}
    \mbox{\subfloat[]{\includegraphics[width=0.49\linewidth]{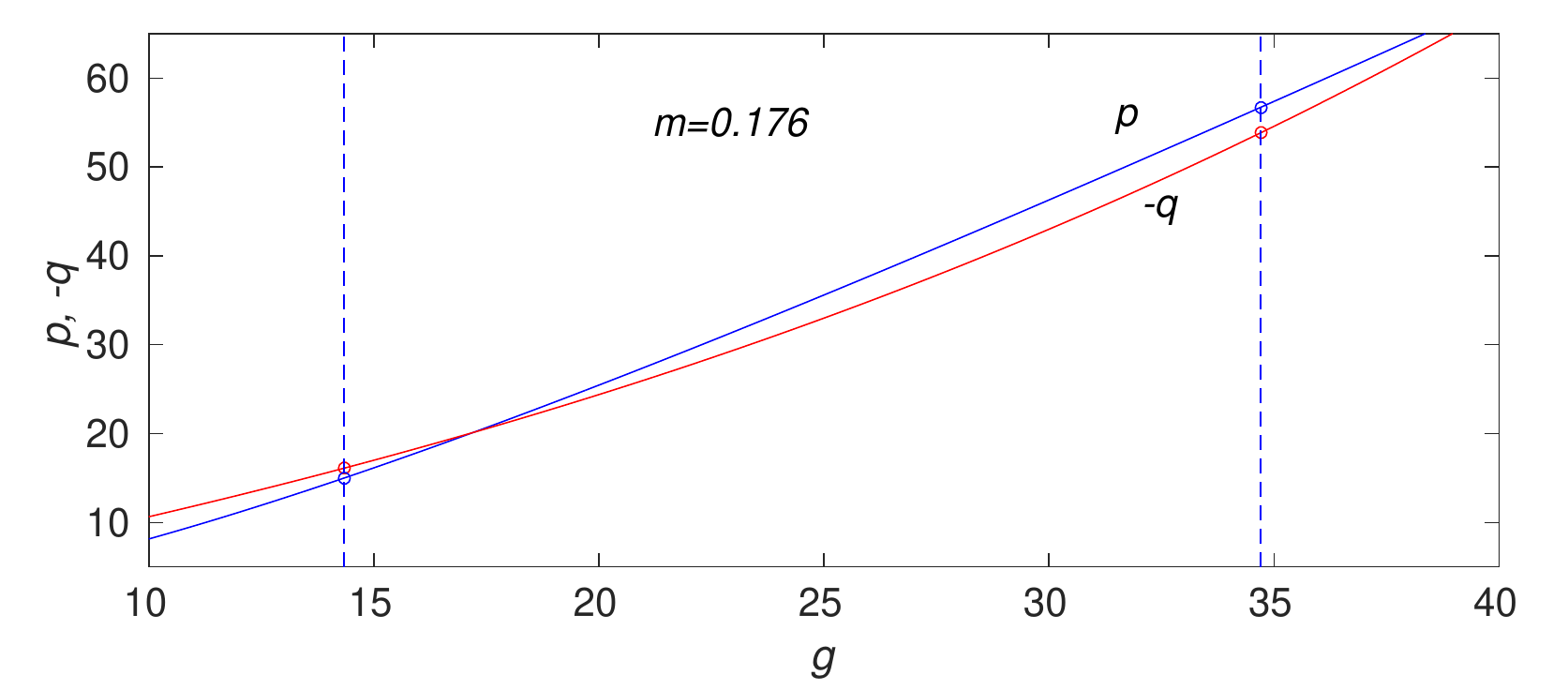}}
      \subfloat[]{\includegraphics[width=0.49\linewidth]{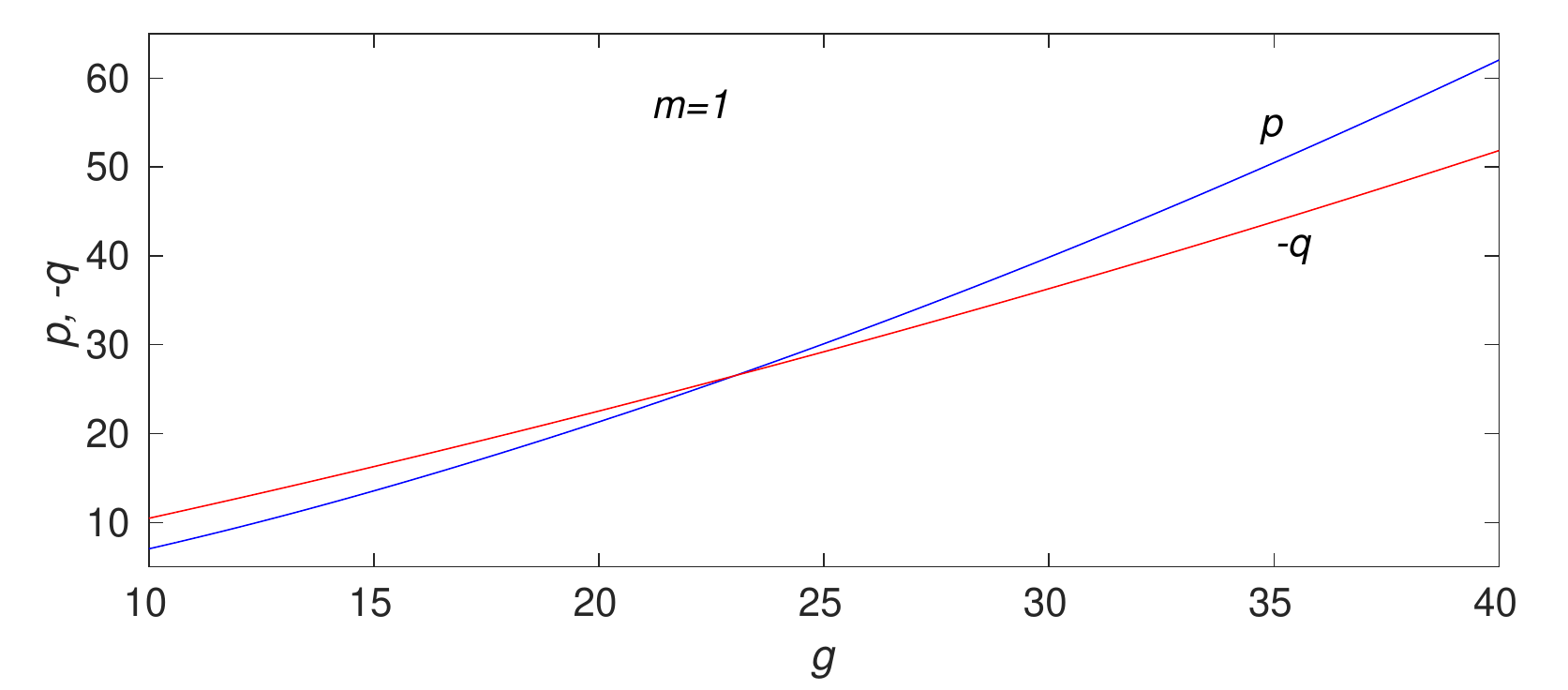}}}
    \caption{Scalar dipole moment $p$ and vector monopole charge $q$ of a 1-Skyrmion as a function of coupling $g$ for pion mass (a) $m=0.176$, and
(b) $m=1$. The dashed lines on (a) mark the coupling values studied in detail via scattering simulations.}
    \label{fig:charges}
  \end{center}
\end{figure}

If $m<1$ (for example, $m=0.176$), the leading term in $V_{\rm int}$
at large $R$ is 
\beq
V_{\rm int}=\frac{m^2 p^2 e^{-mR}}{16\pi R}\wh{\Rvec}\cdot \Orot\wh{\Rvec}+\cdots
\eeq
with corrections of order $m e^{-mR}/R^2$. Hence, the two-Skyrmion
interaction is maximally attractive if $\Orot\wh\Rvec=-\wh\Rvec$, that
is, $\Orot$ represents a rotation by $\pi$ about some axis orthogonal
to the line joining the two Skyrmions. This is the usual prediction of
an attractive channel for appropriately oriented Skyrmions, leading to
the expectation that Skyrmions can coalesce and form bound states.  
Note, however, that if $m>1$, the uniformly repulsive interaction
mediated by the $\omega$ mesons dominates at large separation, so we
expect no bound states in this regime. The case $m=1$ is
interesting. Now the (potentially) attractive scalar dipole
interaction and the repulsive vector monopole interaction have exactly
equal range, and which one dominates depends on the relative sizes of
the dipole moment $p$ and monopole charge $q$. 
These quantities depend on the coupling $g$ as well as the mass $m$,
see Figure \ref{fig:charges}. In fact, for $m=1$, $p^2/4<q^2$ for all
$10\leq g\leq 40$, so vector repulsion dominates when $m=1$ and we
expect no bound states. Of course, the physical pion mass, $m=0.176$,
is rather far from this regime. Nonetheless, the fact that the vector
monopole interaction is uniformly repulsive leads one to expect that
binding energies in this model may be unexpectedly small, at least for
some choices of $g$.  

We will now perform a numerical calculation of the interaction
potential in the full nonlinear model by sending two 1-Skyrmions towards each other in the
attractive (meaning $\Orot\wh\Rvec=-\wh\Rvec$, one of them is rotated by 180 degrees around an axis
perpendicular to the line joining them) and the maximally repulsive channels
(meaning $\Orot\wh\Rvec=\wh\Rvec$, so one is a translated copy of the other).
We treat the problem adiabatically and scatter the Skyrmions at small
velocity compared to that of light. This way we can calculate the
static energy functional at each step, neglecting the kinetic energy
contribution.
The final ingredient in this calculation is to track the position of
the Skyrmions.
We define the position of the 1-Skyrmion to be the position of the
anti-vacuum, meaning $\phi_0=-1$. It is numerically difficult to be
precise about this point using only $\phi_0$, which is why our scheme
is based on finding the simultaneous zero in $\phi_1=\phi_2=\phi_3=0$
for $\phi_0<0$. The zero can be found by determining the sign change
from one lattice point to another.

\begin{figure}[!htp]
  \begin{center}
    \mbox{\subfloat[$g=34.7$]{\includegraphics[width=0.49\linewidth]{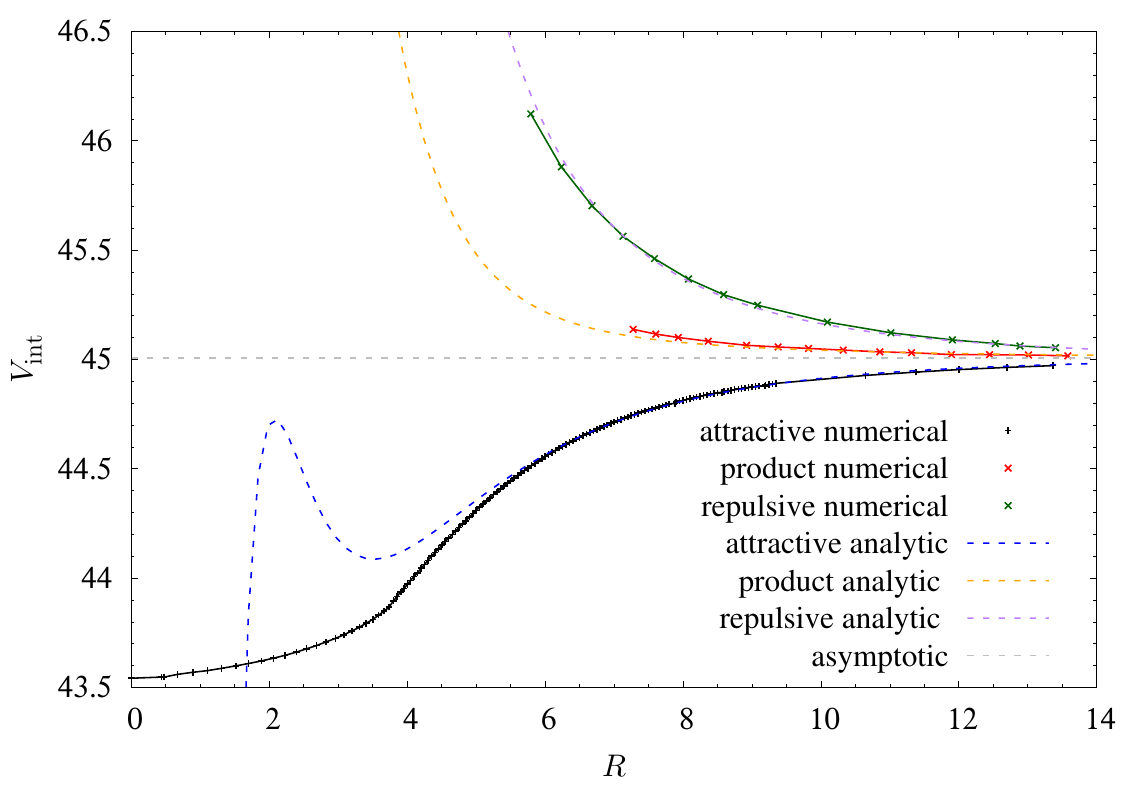}}
      \subfloat[$g=14.34$]{\includegraphics[width=0.49\linewidth]{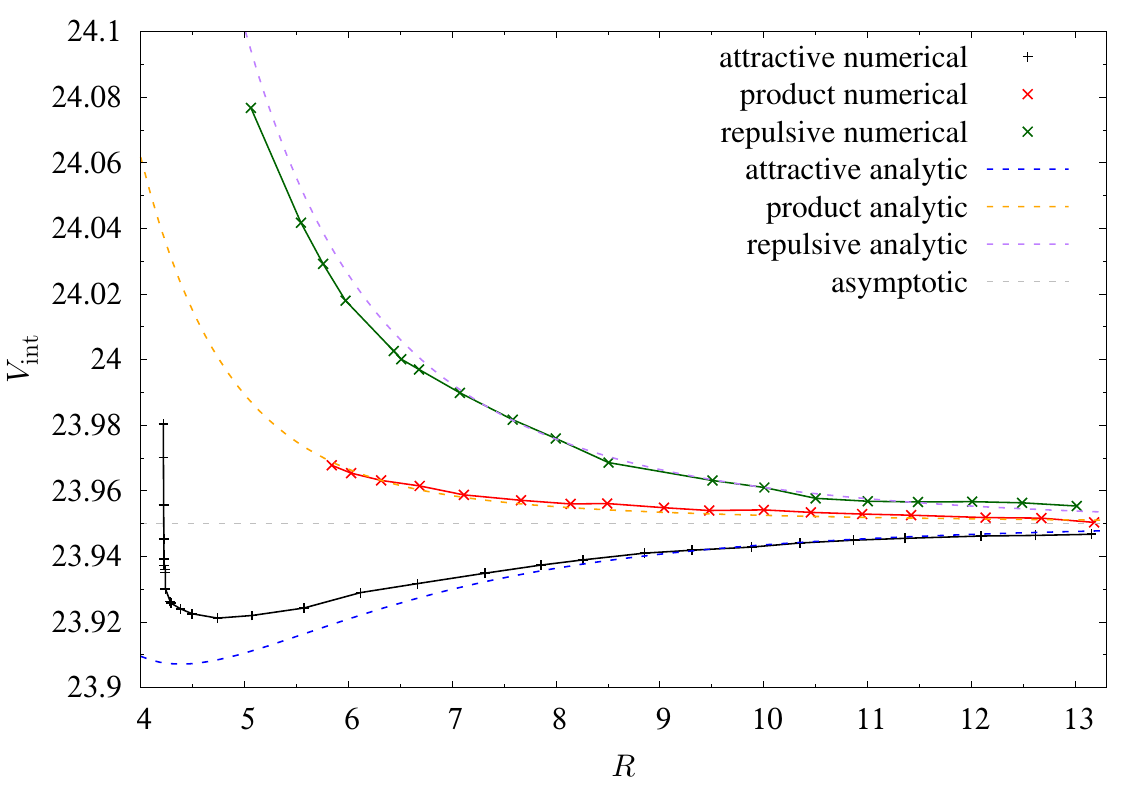}}}
    \caption{Interaction potential extracted from numerical simulations
      for (a) $g=34.7$ and (b) $g=14.34$.
      The product channel is made by translating a copy of one
      Skyrmion by $R$ in some direction.
      The attractive channel takes the translated Skyrmion and rotates
      the it by $\pi$ around and axis perpendicular to the
      axis separating them.
      The repulsive channel takes instead the translated Skyrmion and
      rotates it by $\pi$ around the axis that separates them.
      The mass parameter is $m=0.176$. }
    \label{fig:scatpot}
  \end{center}
\end{figure}
Fig.~\ref{fig:scatpot} shows the result of the numerical calculation
of the scattering potential. 
We display the scattering potential for two different values of the
coupling $g=34.7$ and $g=14.34$.

In both cases, the repulsive channel displays a growth in the energy
until it becomes difficult to continue the simulation adiabatically;
at the point we stop the curve, one of the two Skyrmions
either strays away or rotates into a different orientation.

For the attractive channel in the case of $g=34.7$, the asymptotic
energy corresponds to twice the energy of the 1-Skyrmion and as the
separation is shortened, the total energy drops monotonically to the
level of the 2-Skyrmion, which takes the shape of a torus.

For the attractive channel in the case of $g=14.34$, on the other
hand, asymptotically everything is similar. However, at short
distances where the asymptotic approximation breaks down, the
attraction (which is very weak for this value of the coupling $g$) is
overcome by some nonlinear repulsion and the bound state is not a
torus, but two 1-Skyrmions at a distance bound extremely weakly by
their soliton tails. This is reflected in the classical energy minimisers
for $B=2$ and $B=3$ for this coupling, which resemble lightly bound
clusters of spherical 1-Skyrmions, rather than fully merged bound
states.

\section{Conclusion}\label{sec:conc}

In this paper we have studied the omega extension of the chiral
Lagrangian, which gives stable topological solitons -- known as
Skyrmions -- without the use of the Skyrme term.
The stabilisation is provided by the interaction between the omega
vector meson and the baryon current, which is a topological current --
whose zeroth component measures the topological degree of the field.

Although the model has been discussed in one of the seminal papers by
Adkins and Nappi, numerical solutions have not been obtained
from the full PDEs -- until now.
Our method of solving the model entails rewriting the energy
functional in terms of the pion field and a scalar (the 0-th component
of the omega vector meson) field. In addition to this we implement a
constraint equation that is itself also a PDE, but it is linear and
can readily and quickly be solved by the use of the conjugate gradients
method.
We check the omega field at each time step in our code and improve it
iteratively once it is needed.
The pion field instead is evolved by means of a second-order method
which we denote arrested Newton flow.
In order to settle on a minimum of the energy functional, we remove
the kinetic energy once in a while and every time that the potential
energy increases.

Interestingly, we find that although the model only contains 2
parameters that we can dial, it has a large parameter space which
includes a line with zero classical binding energy and even negatively
bound metastable classical multi-Skyrmion solutions.
This happens when the mass ratio parameter $m$ is large (but still
less than one) and the coupling to the omega meson is small
($g\lesssim20$).
Due to the possibility of extremely lightly bound Skyrmions, there is
in turn an emergence of a large number of metastable solutions (local
minimisers of the energy functional) and hence a large potential for
nuclear clustering in the model. The model at low coupling exhibits
some similarities with the lightly bound Skyrme model studied by
Harland et. al. \cite{Gillard:2015eia}. These dissociated point-like
Skyrmion solutions are also found in the Witten-Sakai-Sugimoto model
\cite{Witten:1998zw,Sakai:2004cn} at strong 't Hooft coupling
\cite{Bartolini:2017sxi}, see Ref.~\cite{Baldino:2017mqq}.

This model, the omega extension of the chiral Lagrangian, is somewhat
similar to a generalised Skyrme model with a kinetic term and a
sixth-order derivative term, where the latter is made of the squared
baryon charge current
\cite{Jackson:1985yz,Zenkin:1987zs,Kopeliovich:2004pd,Ding:2007xi}.
In fact, our approximate Lagrangian \eqref{L3} for
calculating the spin contribution to the $B=1$ Skyrmion is exactly the
kinetic (time-dependent) part of the latter theory.
The quickest way to realise this, is to disregard the Laplacian in the
constraint equation \eqref{static2} and insert the expression for
$\omega_0=f$ into the static energy \eqref{Edef}, which yields the
kinetic term and the sixth-order derivative term to leading order.
By Lorentz invariance, the time-dependent part naturally follows as
well.
Although this approximation was useful for the quantisation of the
1-Skyrmion (the nucleon), it is a rather crude approximation and loses
important aspects of the solution.
The difference can be seen visually in
fig.~\ref{fig:omega_baryon_slice1}, which shows both $f$ and $B_0$,
which without the above-mentioned Laplacian in the constraint equation
\eqref{static2} would be locally proportional to each other.

The ability to accommodate very low classical binding energies is a
somewhat unexpected feature of the $\omega$-Skyrme model. 
Another interesting feature is that the model can reproduce, in a very
elementary manner, the mass splitting between protons and neutrons
\cite{Speight:2018zgc}. It would be interesting to see what effect the
isospin symmetry breaking perturbation proposed in
\cite{Speight:2018zgc} has on the Skyrmions presented here.

\section*{Acknowledgements}

We thank Calum Ross for discussions. 
S.~B.~G.~thanks the Outstanding Talent Program of Henan University for
partial support.
The work of S.~B.~G.~is supported by the National Natural Science
Foundation of China (Grant No.~11675223).

\begin{appendices}
\section{Proof of Proposition \ref{prop1}}\label{app:secondvar}

We make extensive use of the definitions and calculations presented in
ref.~\cite[ch.\ 5]{Urakawa:1993}. Given a two-parameter variation
$\bphi_{s,t}$ of a critical point $\bphi=\bphi_{0,0}$ of $E(\bphi)$ we
define the associated smooth map
$F:P=(-\delta,\delta)\times(-\delta,\delta)\times X\ra N$,
$F(s,t,x)=\bphi_{s,t}(x)$ and denote by $\nabla^F$ the pullback of the
Levi-Civita connexion on $TN$ to 
$F^{-1}TN$, and by $F_*$ the push-forward of vector fields on $P$. The
infinitesimal generators of the variation are $\eps=F_*\cd/\cd
s|_{s=t=0}$ and $\wh\eps=F_*\cd/\cd t|_{s=t=0}$. We will also
encounter $\dot\eps:=\nabla^F_{\cd/\cd t}F_*\cd/\cd s|_{s=t=0}$ which,
like $\eps$ and $\wh{\eps}$, is a section of $\bphi^{-1}TN$. Let
$\{e_i\}$ denote a local orthonormal frame on $(X,\zeta)$. Then the
energy of $\bphi_{s,t}$ is 
\begin{equation}
E(\bphi_{s,t})=\int_X\left(\frac18\sum_{i}h(F_*e_i,F_*e_i)+V\circ F+\frac12f_{s,t}(\triangle+1)f_{s,t}\right)*1,
\end{equation}
where
\begin{equation}\label{amha}
(\triangle+1)f_{s,t}=-g*\bphi_{s,t}^*\Omega.
\end{equation}
Hence \cite[p.\ 154]{Urakawa:1993},
\beq
\frac{\cd\: }{\cd s} E(\bphi_{s,t})&=&\int_X\bigg(-\frac14h\big(F_*\cd/\cd s,\sum_{i}(\nabla^F_{e_i}F_*e_i-F_*\nabla_{e_i}e_i)\big)+h\big(F_*\cd/\cd s,(\grad V)\circ F\big)\nonumber \\
&&\qquad\mathop+f_{s,t}(\triangle+1)\cd_sf_{s,t}\bigg)*1,
\eeq
and, further,
\beq
\frac{\cd^2  E(\bphi_{s,t})}{\cd s\cd t}&=&\int_X\bigg(-\frac14h\big(F_*\cd_s,\sum_{i}(\nabla^F_{e_i}\nabla^F_{e_i}F_*\cd_t-\nabla^F_{\nabla_{e_i}e_i}F_*\cd_t+R(F_*\cd_s,F_*e_i)F_*e_i\big)\nonumber\\
&&\mathop-\frac14h\big(\nabla^F_{\cd_t}F_*\cd_s,\sum_{i}(\nabla^F_{e_i}F_*e_i-F_*\nabla_{e_i}e_i)\big)
+h\big(F_*\cd_s,\nabla^F_{\cd_t}(\grad V\circ F)\big) \\
&&\mathop+h\big(\nabla^F_{\cd_t}F_*\cd_s,(\grad V)\circ F\big)+\cd_t f_{s,t}(\triangle+1)\cd_s f_{s,t}
+f_{s,t}(\triangle+1)\cd^2_{t,s}f_{s,t}\bigg)*1.\nonumber
\eeq
Evaluating this at $s=t=0$ yields
\beq
\frac{\cd^2 E(\bphi_{s,t}) }{\cd s\cd t} \bigg|_{s=t=0}&=&
\int_X\bigg(\frac14 h(\eps,J_\bphi\wh\eps)-\frac14h(\dot\eps,\tau(\bphi))+h\big(\eps,(\nabla^N_{\wh\eps}\grad V)\circ\bphi\big)\nonumber \\
&&\mathop+h\big(\dot\eps,(\grad V)\circ\bphi\big) \\
&&\mathop+\cd_tf_{s,t}|_{s=t=0}(\triangle+1)\cd_sf_{s,t}|_{s=t=0}+f(\triangle+1)\cd^2_{s,t}f_{s,t}|_{s=t=0}\bigg)*1,\nonumber
\eeq
where we have used the fact that, by the definition of $\nabla^F$, for any vector fields $Y$ on $N$, and $u$ on $P$, $\nabla^F_u(Y\circ F)=(\nabla^N_{F_*u}Y)\circ F$.  
Differentiating eq.~\eqref{amha} with respect to $s$ (or $t$), setting
$s=t=0$ and using the Homotopy Lemma, we see that 
$\alpha:=\cd_sf_{s,t}|_{s=t=0}$ and $\wh\alpha:= \cd_tf_{s,t}|_{s=t=0}$ satisfy
\begin{equation}
(\triangle+1)\alpha=-g*\d(\bphi^*\iota_\eps\Omega),\qquad
(\triangle+1)\wh\alpha=-g*\d(\bphi^*\iota_{\wh\eps}\Omega),
\end{equation}
and hence $\alpha=g\alpha_\bphi(\eps)$, $\wh\alpha=g\alpha_\bphi(\wh\eps)$, where $\alpha_\bphi:\Gamma(\bphi^{-1}TN)\ra C^\infty(X)$ is the linear operator defined by
equation \eqref{liba}.
Recall that $\bphi$, by assumption satisfies \eqref{static1}, so
\beq
-\frac14\tau(\bphi)+(\grad V)\circ\bphi+g*(\d f\wedge\Xi_\bphi)=0.
\eeq
Hence
 \beq
\frac{\cd^2 E(\bphi_{s,t})}{\cd s\cd t} \bigg|_{s=t=0}&=&\int_X\bigg(h\left(\eps,\frac14J_\bphi\wh\eps+(\nabla^N_{\wh\eps}\grad V)\circ\bphi\right)
-h\big(\dot\eps,g*(\d f\wedge\Xi_\bphi)\big)\nonumber \\
&&\mathop+g^2\alpha_\bphi(\wh\eps)(\triangle+1)\alpha_\bphi(\eps)+f(\triangle+1)\cd^2_{s,t}f_{s,t}|_{s=t=0}\bigg)*1 \nonumber \\
&=&\ip{\eps,\frac14J_\bphi\wh\eps+(\nabla^N_{\wh\eps}\grad V)\circ\bphi}+g^2\ip{\alpha_\bphi(\eps),(\triangle+1)\alpha_\bphi(\wh\eps)}\nonumber \\
&&\mathop-g\int_X h\big(\dot\eps,*(\d f\wedge\Xi_\bphi)\big)*1+\cd^2_{s,t}\ip{f,(\triangle+1)f_{s,t}}|_{s=t=0},\label{tebi}
\eeq
where, as usual, $\ip{\cdot,\cdot}$ denotes the $L^2$ inner product,
and we have used the self-adjointness of $\triangle+1$. It remains to
compute $\cd^2_{s,t}\ip{f,(\triangle+1)f_{s,t}}$. Now 
\beq
&&\cd_s\ip{f,(\triangle+1)f_{s,t}}=-g\cd_s\ip{f,*\bphi_{s,t}^*\Omega}
=-g\int_Xf\d(F^*(\iota_{F_*\cd_s}\Omega))
=g\int_X\d f\wedge F^*(\iota_{F_*\cd_s}\Omega)\nonumber \\
&&\qquad =g\int_X\sum_{i=1}^d(-1)^{i+1}e_i(f)\Omega(F_*\cd_s,F_*e_1,F_*e_2,\ldots,\wh{F_*e_i},\ldots,F_*e_d)*1,
\eeq
where $\wh{\cdots}$ denotes an omitted term. Hence,
\beq
&&\cd^2_{s,t}\ip{f,(\triangle+1)f_{s,t}}=g\int_X\sum_{i=1}^d(-1)^{i+1}\big(\nabla^N_{F_*\cd_t}\Omega\big)(F_*\cd_s,F_*e_1,\ldots,\wh{F_*e_i},\ldots,F_*e_d)*1\nonumber \\
&&\mathop+g\int_X\sum_{i=1}^d(-1)^{i+1}e_i(f)\Omega(\nabla^F_{\cd_t}F_*\cd_s,F_*e_1,\ldots,\wh{F_*e_i},\ldots,F_*e_d)*1 \\
&&\mathop+g\int_X\sum_{i=1}^d\sum_{j<i}(-1)^{i+j}e_i(f)\Omega(F_*\cd_s,\nabla^F_{\cd_t}F_*e_j,F_*e_1,\ldots,\wh{F_*e_j},\ldots,\wh{F_*e_i},\ldots,F_*e_d)*1\nonumber \\
&&\mathop+g\int_X\sum_{i=1}^d\sum_{j>i}(-1)^{i+j+1}e_i(f)\Omega(F_*\cd_s,\nabla^F_{\cd_t}F_*e_j,F_*e_1,\ldots,\wh{F_*e_i},\ldots,\wh{F_*e_j},\ldots,F_*e_d)*1.\nonumber 
\eeq
The pullback connexion satisfies the identity $\nabla^F_uF_*v-\nabla^F_vF_*u-F_*[u,v]$ for all vector fields $u,v$ on $P$, so $\nabla^F_{\cd/\cd t}F_*e_j=\nabla^F_{e_j}F_*\cd/\cd t$. Hence
\beq
&&\cd^2_{s,t}\ip{f,(\triangle+1)f_{s,t}}|_{s=t=0}=g\int_X\sum_{i=1}^d(-1)^{i+1}\big(\nabla^N_{\wh\eps}\Omega\big)(\eps,\d\bphi e_1,\ldots,\wh{\d\bphi e_i},\ldots,\d\bphi e_d)*1\nonumber \\
&&\mathop+g\int_X\sum_{i=1}^d(-1)^{i+1}e_i(f)\Omega(\dot\eps,\d\bphi e_1,\ldots,\wh{\d\bphi e_i},\ldots,\d\bphi e_d)*1\nonumber \\
&&\mathop+g\int_X\sum_{i=1}^d\sum_{j<i}(-1)^{i+j}e_i(f)\Omega(\eps,\nabla^\bphi_{e_j}\wh\eps,\d\bphi e_1,\ldots,\wh{\d\bphi e_j},\ldots,\wh{\d\bphi e_i},\ldots,\d\bphi e_d)*1\nonumber \\
&&\mathop+g\int_X\sum_{i=1}^d\sum_{j>i}(-1)^{i+j+1}e_i(f)\Omega(\eps,\nabla^\bphi_{e_j}\wh\eps,\d\bphi e_1,\ldots,\wh{\d\bphi e_i},\ldots,\wh{\d\bphi e_j},\ldots,\d\bphi e_d)*1
\nonumber \\
\label{tb}
&&\qquad=g\int_X\d f\wedge\bphi^*\big(\iota_\eps\nabla^N_{\wh{\eps}}\Omega\big)+g\ip{\dot\eps,*(\d f\wedge\Xi_\bphi)}+g\ip{\eps,\dot{\Xi}_\bphi(\wh\eps)},
\eeq
where $\dot\Xi_\bphi(\wh\eps)$ is the $\bphi^{-1}TN$ valued $(d-1)$-form
defined in eq.~\eqref{linbar}. Substituting eq.~\eqref{tb} into
eq.~\eqref{tebi}, one sees that 
\beq
\frac{\cd^2 E(\bphi_{s,t}) }{\cd s\cd t} \bigg|_{s=t=0}&=&\ip{\eps,\frac14J_\bphi\wh\eps+(\nabla^N_{\wh\eps}\grad V)\circ\bphi}+g^2\ip{\alpha_\bphi(\eps),(\triangle+1)\alpha_\bphi(\wh\eps)}\nonumber \\
&&\mathop+g\int_X\d f\wedge\bphi^*\big(\iota_\eps\nabla^N_{\wh{\eps}}\Omega\big)+g\ip{\eps,\dot{\Xi}_\bphi(\wh\eps)},
\eeq
as Proposition \ref{prop1} claims.

\end{appendices}

\end{document}